\documentclass{jfm}
\usepackage{graphicx}
\usepackage{natbib}
\usepackage{subfigure}
\usepackage{color}
\ifCUPmtlplainloaded \else
  \checkfont{eurm10}
  \iffontfound
    \IfFileExists{upmath.sty}
      {\typeout{^^JFound AMS Euler Roman fonts on the system,
                   using the 'upmath' package.^^J}%
       \usepackage{upmath}}
      {\typeout{^^JFound AMS Euler Roman fonts on the system, but you
                   dont seem to have the}%
       \typeout{'upmath' package installed. JFM.cls can take advantage
                 of these fonts,^^Jif you use 'upmath' package.^^J}%
      }
  \else
  \fi
\fi

\ifCUPmtlplainloaded \else
  \checkfont{msam10}
  \iffontfound
    \IfFileExists{amssymb.sty}
      {\typeout{^^JFound AMS Symbol fonts on the system, using the
                'amssymb' package.^^J}%
       \usepackage{amssymb}%
         
       \let\ge=\geqslant  
      }{}
  \fi
\fi

\ifCUPmtlplainloaded \else
  \IfFileExists{amsbsy.sty}
    {\typeout{^^JFound the 'amsbsy' package on the system, using it.^^J}%
     \usepackage{amsbsy}}
    {\providecommand\boldsymbol[1]{\mbox{\boldmath $##1$}}}
\fi

\newsavebox{\astrutbox}
\sbox{\astrutbox}{\rule[-5pt]{0pt}{20pt}}

\title[Microstructure and rheology]{Microstructure and rheology of finite inertia neutrally buoyant suspensions}

\author[Hamed Haddadi and Jeffrey F. Morris]%
{Hamed Haddadi and Jeffrey F. Morris   \thanks{Email address for correspondence: morris@ccny.cuny.edu},\ns}

\affiliation{Benjamin Levich Institute and Department of Chemical Engineering, The City College of New York, NY, NY 10031, USA \\[\affilskip]
}

\pubyear{2010}
\volume{650}
\pagerange{119--126}
\date{?; revised ?; accepted ?. - To be entered by editorial office}

\begin{document}

\maketitle
\begin{abstract}
The microstructure and rheological properties of suspensions of neutrally buoyant hard spherical particles in Newtonian fluid under conditions of finite inertia are studied using the lattice-Boltzmann method (LBM), which is based on a discrete Boltzmann model for the fluid and Newtonian dynamics for the particles.  The suspensions are subjected to simple-shear flow and the properties are studied as a function of Reynolds number and volume fraction, $\phi$.  The inertia is characterized by the particle-scale shear flow Reynolds number $Re = \frac{\rho \dot{\gamma}a^{2}}{\mu}$, where $a$ is the particle radius, $\dot{\gamma}$ is the shear rate and $\rho$ and $\mu$ are the density and viscosity of the fluid, respectively. The influences of inertia and of the volume fraction are studied for $0.005\leqslant Re \leqslant 5$ and $0.1\leqslant \phi \leqslant 0.35$. The flow-induced microstructure is studied using the pair distribution function $g(\boldsymbol{r})$.  Different stress mechanisms, including those due to surface tractions (stresslet), acceleration, and the Reynolds stress due to velocity fluctuations are computed and their influence on the first and second normal stress differences, the particle pressure and the viscosity of the suspensions are detailed. The probability density functions of particle force and torque are also presented.    
\end{abstract}

\section{Introduction}
 The inertial flow of particle-laden fluids arises in many natural and industrial applications. Examples include blood flow in arteries (Ku 1997) and suspension coating flows (Aidun \& Lu 1995), and rather surprisingly, inertial effects on particle dynamics appear in microfluidic flow settings (Humphry \emph{et al.} 2010). It is thus of fundamental and practical interest to understand the role of inertia in altering suspension properties from those of inertialess, i.e. Stokes flow, suspensions. In this work, we address the bulk flow properties of sheared suspensions at finite particle-scale inertia, obtained using numerical simulations of the discrete-particle dynamics.  From the simulations we extract also the microstructural arrangement of the particles and 
 we seek to provide insight to its role in the stress of the mixture. \newline
  
 To characterize the role of inertia relative to viscous effects in determining the stress system in the suspension, the relevant Reynolds number is the shear-flow based $Re =  \dot{\gamma} a^2/\nu$ with $\dot{\gamma}$ the shear rate, $a$ the particle size, and $\nu = \mu/\rho$ the kinematic viscosity of the suspending fluid of dynamic viscosity $\mu$ and density $\rho$.  The Stokes number, characterizing the particle inertia relative to viscous effects may be written $St = (\rho_p/\rho_f) Re$, and for particles that are neutrally-buoyant, only $Re$ is needed.  By contrast, for a suspension of heavy particles where there is a large difference in density between the particles and the fluid, particle inertia may be relevant where fluid inertia is negligible (e.g. for particles in a gas), i.e. $Re=0$ while $St>0$.  Such a system has been considered, by assuming Stokes flow of the fluid and inertial motion of the particles (Sangani \& Koch 1996; Subramanian \& Brady 2006).  For such a suspension, linearity of the fluid Stokes equations allows analytical progress as well as important simulational simplifications. However, for a suspension of neutrally buoyant particles, a similar level of inertia is carried by the particles and the fluid, and the nonlinear governing equations for the fluid motion must be considered. This limits analytical approaches to infinitely dilute suspensions (Lin, Peery \& Schowalter 1970; Subramanian \emph{et al.} 2011), and forces use of numerical methods for solving the Navier-Stokes equations.  \newline
 
 While Stokes flow suspensions have been extensively studied analytically (Brady \& Morris 1997; Bergenholtz, Brady \& Vicic 2002; Nazockdast \& Morris 2013) and using the Stokesian Dynamics or related simulation algorithms (Brady \& Bossis 1988;  Sierou \& Brady 2001; Melrose \& Ball 2004), there are limited studies addressing finite $Re$ suspensions. Single particle studies by analytical, numerical and experimental approaches have explore the streamline topology in finite inertia (Robertson \& Acrivos 1970; Kossack \& Acrivos 1974; Poe \& Acrivos 1975). Numerical simulations have been employed for extending the studies to many-body calculations.  However, the full solution of the Navier-Stokes equation together with dynamics of the moving particles is a computationally demanding task. Direct numerical solution of the Navier-Stokes equation for two dimensional (Feng \& Joseph 1994) flows around particles or finite element computations of flow around an isolated sphere or ellipsoid (Mikulencak \& Morris 2004) are examples of utilizing conventional numerical methods. An important development was that of use of discretized fluid solvers coupled to a Newtonian dynamics for particle motions, as this tool can resolve suspension motion at finite inertia with readily accessible computational resources. In particular, the Lattice Boltzmann Method (LBM) together with methods for application of solid-fluid boundary conditions allows for solution of the fluid flow in the presence of the moving particles (Ladd 1994 a,b).   Calculation of the fluid force on the particles allows the Newtonian dynamics to be employed to compute particle trajectories.  Using this method, different aspects of finite inertia suspension flows, including pair trajectories, microstructure, shear induced self-diffusion and viscosity, at varying volume fractions have been studied in two-dimensional simple shear flows (Shakib-Manesh \emph{et al.} 2001;  Kromkamp \emph{et al.} 2005).  However, the LBM model used for these studies is limited to suspended solid particles with densities larger than the fluid density. Consequently, the results do not exactly correspond to the suspension of neutrally buoyant particles. With further extension of the LBM model to neutrally-buoyant particles (Aidun, Lu \& Ding 1998), the pair trajectories of hard spherical particles with the same density as the fluid has been studied in two dimensions (Yan \emph{et al.} 2007) and three dimensions (Kulkarni \& Morris 2008a). A limited examination of the microstructure and rheological properties of suspensions at finite $Re$ in a wall bounded simple shear flow was presented by Kulkarni \& Morris (2008b). Recently, the structure of the pair trajectories, shear-induced self diffusion and rheological properties of finite-inertia suspensions in periodic computational domains have been studied by the Force Coupling Method (Yeo \& Maxey 2013), a technique based on 
 a force-moment, or multipole, description of the particles. 
 \\
 
We address the microstructure and rheological properties for suspensions of neutrally buoyant hard spherical particles in Newtonian fluid, computing the flows by a lattice-Boltzmann scheme with corrections for short-range hydrodynamic, i.e lubrication, forces (Ladd 1994 a,b; Aidun \emph{et al.} 1998; Nguyen \& Ladd 2002). We study the effect of inertia for $0.005 \leqslant Re \leqslant 5$ and volume fractions $0.1 \leqslant \phi \leqslant 0.35$. With particle scale $Re \leqslant 5$, the inertia in the bulk length scale can be significantly higher. We find that the calculations for $Re \geqslant 5$ are influenced by particle elastic collisions and imperfections of the simulation tool. By increasing the possibility of elastic collisions at $Re \geqslant 5$, their contribution on the stress transmission and rheological properties must be considered. In this work, we complement the results previously presented on the microstructure and rheology of the finite inertia suspensions (Kulkarni \& Morris 2008b; hereafter abbreviated KM08).  Employing many-processor parallel computations enabled us to evaluate the effect of $Re$ and $\phi$ on the microstructure with much greater precision.  We compute the rheological properties using the Batchelor (1970) formulation of the stress system in a suspension. Through this approach, we differentiate the effect of stress generated by surface traction forces (stresslet), accelerations and Reynolds stress (owing to velocity fluctuations arising here from particle interaction) and characterize the contribution of each mechanism to the bulk and particle-phase stresses.  We also present detailed statistical information on the stress distribution on a particle under the various conditions, as well as on the fluctuating force and torque on the particles. 
\\

We organize this paper as follows. In \S 2, we present the governing equations and give a brief outline of the LBM method and the suspension stress formulation.  We present the microstructure of the finite inertia suspensions in \S 3 and rheology of the suspensions in \S 4. The statistical distributions noted above are presented in \S5, followed by conclusions from the work. 
     
\section{Problem Formulation}

\subsection{Governing equations and computational parameters}

We study suspensions of neutrally buoyant solid spherical particles in a Newtonian fluid subjected to simple shear flow. The governing equations of the fluid phase in dimensionless form are

\begin{subequations}
\begin{equation}
\boldsymbol{\nabla.u} = 0, \\
\end{equation}
\begin{equation}
Re(\frac{\partial\boldsymbol{u}}{\partial t} + \boldsymbol{u.\nabla u}) = -\boldsymbol{\nabla}p + \boldsymbol{\nabla^2 u},
\end{equation}
\end{subequations}
where length has been made dimensionless by the sphere radius \emph{a}, time by the inverse of the shear rate $\dot{\gamma}^{-1}$, fluid velocity by $\dot{\gamma}a$, and the pressure by $\mu\dot{\gamma}$. Motion of the particles is governed by Newtonian dynamics. The translational ($\boldsymbol{U}_i$) and rotational ($\boldsymbol{\Omega}_i$) velocity of the particles are governed by

\begin{subequations}
\begin{equation}
m_i \frac{d\boldsymbol{ U}_i}{d t} = \boldsymbol{F}_i, \\
\end{equation}
\begin{equation}
I_i \frac{d\boldsymbol{ \Omega}_i}{d t} = \boldsymbol{T}_i, 
\end{equation}
\end{subequations}
where $\boldsymbol{F}_i$ and $\boldsymbol{T}_i$ are respectively, the net force and torque exerted on particle $i$, which has the mass $m_i$ and the moment of inertia $I_i$. \newline

We have performed the numerical simulation of the suspension flow in a rectangular box with size $L\times H \times W$ in the flow ($x$), velocity gradient ($y$) and vorticity ($z$) directions. The domain is bounded by walls in the gradient direction, while periodic boundary conditions are applied in the $x$ and $z$ directions. The shearing motion is imposed by moving the top and bottom walls with opposite velocities, in the $x$ direction only. The size of the particle is $6.2$ lattice nodes per radius and the dimensions of the computation box have been set at $20a \times 20a \times 20a$. The results for rheology and microstructure have been found qualitatively independent of the further increase of the particle mesh resolution and the size of the computation box, when confinement effects are accounted properly.  Specifically, confinement in a wall-bounded domain may influence the suspension properties by forming particle layers near the walls (Shakib-Manesh \emph{et al.} 2002; KM08; Yeo \& Maxey 2010) and generating flow reversal zones (Zurita-Gotor, Blawzdziewicz \& Wajnryb 2007), and thus care has been taken to avoid artifacts due to these influences as described later in regard to sampling of the pair microstructure.\newline

\subsection{Simulation method}

We utilize the lattice-Boltzmann approach developed by Ladd (1994 \emph{a},\emph{b}) with later improvements by Aidun \emph{et al.} (1998) and a lubrication correction for close particles by Nguyen and Ladd (2002). The LBM for a suspension of particles combines a discretized Boltzmann model for the fluid with Newtonian dynamics for the particles. The fluid is assumed to be made of fictitious particles, termed here LB particles, which are constrained to move only on lattice directions. The state of the fluid phase is described by a one particle distribution function $n_i(\boldsymbol{r}, t)$, which describes the mass density of particles with velocity $\boldsymbol{c_i}$ at a lattice node $\boldsymbol{r}$, at time $t$. The quantities $\boldsymbol{c_i}$ and $t$ are discrete whereas $n_i$ is continuous. In three dimensions, isotropy requires a multi-speed model. In this work we have used a \emph{D3Q19} model (implying 3 dimensions, with 19 directions including the case of no motion) where the LB particles stream along lattice links to the nearest sites, e.g [100], or diagonal next nearest neighboring nodes, e.g [110], on a cubic lattice; [xyz] indicates the direction of motion from the node of interest in the $x$, $y$, and $z$ directions. The mass density $\rho$, the momentum density $\boldsymbol{j} = \rho \boldsymbol{u}$ and the momentum flux $\boldsymbol{\Pi}$ are defined by moments of the velocity distribution function 

\begin{equation}
\rho =  \sum_i n_i,  \;  \;  \;  \;  \;  \;  \;  \;  \;  \boldsymbol{j} = \sum_i n_i \boldsymbol{c_i},  \;  \;  \;  \;  \;  \;  \;  \;  \;  \boldsymbol{\Pi} = \sum_i n_i \boldsymbol{c_i} \boldsymbol{c_i}. 
\end{equation}

The evolution of $n_i(\boldsymbol{r}, t)$ is described by the lattice-Boltzmann equation 

\begin{equation}
n_i(\boldsymbol{r} + \boldsymbol{c_i}\Delta t, t + \Delta t) = n_i(\boldsymbol{r}, t) + \sum_j \boldsymbol{\Omega_{ij}}n^{neq}_j,
\end{equation}
where $\boldsymbol{\Omega_{ij}}$ is the collision operator which represents the change in $n_j$ due to molecular collisions and $\Delta t$ is the time step. The non-equilibrium distribution function $n^{neq}_j$, is defined $n^{neq}_j = n_j - n^{eq}_j$, and $\sum_j \boldsymbol{\Omega_{ij}}n^{eq}_j = 0$. The equilibrium distribution function is required by the moment conditions to reproduce the inviscid (Euler) equations on large length and time scales; the viscous influence is discussed below. The second moment of the equilibrium distribution should be equal to the inviscid momentum flux, $\boldsymbol{\Pi}^{eq} = \sum_i n_i \boldsymbol{c_i}\boldsymbol{c_i} = p \boldsymbol{I} + \rho \boldsymbol{uu}$ where $\boldsymbol{I}$ is the identity tensor. The appropriate form of the equilibrium distribution of the 19 velocity model is written as 
\begin{equation}
n^{eq}_i = a^{c_i} \left[\rho + \frac{\boldsymbol{j}.\boldsymbol{c_i}}{c_{s}^2} + \frac{\rho \boldsymbol{uu:}(\boldsymbol{c_{i}c_{i}} - c_{s}^2\boldsymbol{I})}{2c_{s}^4} \right], 
\end{equation}
where $c_{s} = \sqrt{\frac{1}{3}}$ is the speed of sound in lattice units and $a^{c_i}$ are coefficients of speeds 0, 1, $\sqrt{2}$, corresponding to the velocity of the LB particles that remain at the same node,  or stream to the nearest node and diagonal next nearest node, respectively. The coefficients are

\begin{equation}
a^{0} = \frac{1}{3},  \;  \;  \;  \;  \;  \;  \;  \;  \;  a^{1} = \frac{1}{18},  \;  \;  \;  \;  \;  \;  \;  \;  \;  a^{\sqrt{2}} = \frac{1}{36}. 
\end{equation}
The post collision distribution $n_{i}^{*}(\boldsymbol{r},t)$ is written as a series of moments,
\begin{equation}
n^{*}_i = a^{c_i} \left[\rho + \frac{\boldsymbol{j}.\boldsymbol{c_i}}{c_{s}^2} + (\frac{\rho \boldsymbol{uu} + \boldsymbol{\Pi}^{neq,*})\boldsymbol{:}(\boldsymbol{c_{i}c_{i}} - c_{s}^2\boldsymbol{I})}{2c_{s}^4} \right], 
\end{equation}
where the first, $\rho$, and the second moments, $\boldsymbol{j}$, remain unchanged but the non-equilibrium second moment, $\boldsymbol{\Pi}^{neq, *}$, changes according to 
 \begin{equation}
 \boldsymbol{\Pi}^{neq, *} = (1 + \lambda) \boldsymbol{\Pi}^{neq} + \frac{1}{3}(1 + \lambda_{\nu})(\boldsymbol{\Pi}^{neq}\boldsymbol{:I})\boldsymbol{I}.
\end{equation}
Here, $\Pi^{neq} = \Pi - \Pi^{eq}$ and $\lambda$ and $\lambda_{\nu}$ are eigenvalues of the collision operator. The shear and bulk viscosities are related to $\lambda$ and $\lambda_{\nu}$, respectively, as
\begin{equation}
\mu = -\rho c_{s}^2 \Delta t(\frac{1}{\lambda} + \frac{1}{2}),   \;  \;  \;  \;  \;  \;  \;  \;  \; \mu_{\nu} = -\rho c_{s}^2 \Delta t(\frac{2}{3\lambda_{\nu}} + \frac{1}{3}).
\end{equation}

After the collision the population densities stream to the neighboring nodes along lattice links. \newline

The solid-fluid boundary condition is implemented by the ``link-bounce-back" method. Solid particles are defined by surfaces cutting links between the nodes, and the boundary nodes are placed halfway along the links. The LB particles streaming on the links interact with the boundary nodes, so that $n_i$ of the fluid just outside of the particle is modified in such a way that the fluid velocity is matched to the local solid velocity. Following Aidun \emph{et al.} (1998), the fluid is removed from the interior of the particle. The moving boundary condition without interior fluid is implemented by taking a set of fluid nodes $\boldsymbol{r}$ just outside the particle surface with velocities $\boldsymbol{c}_b$ such that $\boldsymbol{r} + \boldsymbol{c}_b\Delta t$ lies inside the particle surface. The distribution function of these nodes is updated according to 

\begin{equation}
n_{-i}(\boldsymbol{r}, t+\Delta t) = n^{*}_{i}(\boldsymbol{r},t) - \frac{2a^{c_i}\rho_{0}\boldsymbol{u}_{b}\boldsymbol{.c_i}}{c^2_s},
\end{equation}
where $-i$ denotes the velocity $\boldsymbol{c}_{-i} = -\boldsymbol{c}_i$. $\rho_0$ is the mean density and is used instead of the local density $\rho$, to simplify the update procedure. The local velocity of the solid boundary,

\begin{equation}
\boldsymbol{u}_b = \boldsymbol{U} + \boldsymbol{\Omega} \times (\boldsymbol{r}_b - \boldsymbol{R}),
\end{equation}
is calculated by the particle velocity $\boldsymbol{U}$, angular velocity $\boldsymbol{\Omega}$, and the center of mass $\boldsymbol{R}$; and $\boldsymbol{r}_b = \boldsymbol{r} + \frac{1}{2}\boldsymbol{c}_b\Delta t$, which is the coordinate of the boundary node. During this update scheme, the momentum is exchanged between the solid and fluid nodes but the total momentum is conserved. The force on the boundary nodes can be calculated from the transferred momentum and is given by

\begin{equation}
\boldsymbol{f}_b(\boldsymbol{r}_b, t + \frac{1}{2}\Delta t) = \frac{\Delta x^3}{\Delta t}\left[ 2n_i^{*} - \frac{2a^{c_i}\rho_{0}\boldsymbol{u}_b\boldsymbol{.c_i}}{c_s^2}\right]\boldsymbol{c_i},
\end{equation}
where $\Delta x$ is the lattice spacing. The total force and torque is calculated by summing over all the boundary nodes $\boldsymbol{r}_b$ of that particle as

\begin{equation}
\boldsymbol{F} = \sum_{\boldsymbol{r}_b} \boldsymbol{f}(\boldsymbol{r}_b),   \;  \;  \;  \;  \;  \;  \; and   \;  \;  \;  \;  \;  \;  \;  \boldsymbol{T} = \sum_{\boldsymbol{r}_b} (\boldsymbol{r}_b - \boldsymbol{R}) \times \boldsymbol{f}(\boldsymbol{r}_b) . 
\end{equation}
The particle positions and velocities are then updated for each time step. \newline

The particle covers different numbers of nodes as it moves on the lattice, and hence there is an effective hydrodynamic radius $r_h$ which is greater than the prescribed radius $a$. The $r_h$ depends on $a$ and viscosity, and the difference from $a$ becomes smaller as the particle becomes larger (Nguyen \& Ladd 2002).  Calibration of $r_h$ is performed by calculating the drag coefficient on the sphere in uniform flow. \\

The hydrodynamic interaction between hard spheres involves calculation of near-field lubrication forces. The LBM captures hydrodynamic interactions when the separation between two solid surfaces is more than one lattice spacing. For gaps smaller than one lattice unit, the method is insufficient to resolve the lubrication forces. In these small surface separations, the hydrodynamic forces are pairwise additive and can be calculated by construction of the grand resistance matrix. Nguyen \& Ladd (2002) proposed the form
\begin{equation}
\boldsymbol{F}_{lub} = -\frac{6\pi\mu(ab)^2}{(a+b)^2} \boldsymbol{\hat{\boldsymbol{r}}\hat{\boldsymbol{r}}}\boldsymbol{.U}_{21}(\frac{1}{h} - \frac{1}{h_c}),
\end{equation}
for the normal force associated with squeezing flow between two solid sphere surfaces, where $a$ and $b$ are the radii of the two spheres, $\boldsymbol{r} = \boldsymbol{x}_2 - \boldsymbol{x}_1$ for particles located at $\boldsymbol{x}_1$ and $\boldsymbol{x}_2$ with $\hat{\boldsymbol{r}} = \frac{\boldsymbol{r}}{|\boldsymbol{r}|}$, and $\boldsymbol{U}_{21} = \boldsymbol{U}_2 - \boldsymbol{U}_1$ is the pair relative velocity. The gap $h = |\boldsymbol{r}| - (a + b)$ is the distance of closest approach of the spheres, and $h_c$ is a cut off for the added lubrication force such that for $h>h_c$, $\boldsymbol{F}_{lub} = 0$ and only the LB-computed force is used. Typically, $h_c = 1.5\times10^{-7}a$ in this work. We also repeated the calculations at $h_c = 1.5\times10^{-3}a$ and observed negligible difference between results. In case of overlaps between particles, which happens when the distance between particles is less than $1.5\times10^{-7}a$, an elastic collision force is generated along the line of centers. For neutrally buoyant particles at $0.05 \leqslant Re \leqslant 5$, the elastic collision force is exerted on the particles sporadically and we were not forced to account or correct for particle-particle overlap.\newline 

The lubrication forces complicate the update of particle velocities and cause instabilities whenever the distance between surface of particles is less than $0.01 - 0.08$ lattice units, depending on $\nu$. To update velocities, a ``cluster implicit method" proposed by Nguyen \& Ladd has been used in this study. \newline

Simulation of suspensions with low kinematic viscosities may cause errors in calculation of lubrication forces (Nguyen \& Ladd 2002). At the same time, increasing $Re$ of the calculations by increasing the wall speed results in errors associated with large Mach numbers in lattice-Boltzmann simulations ($Ma = \frac{\dot{\gamma}H}{2c_s}$). This limits the maximum $Re$ that can be achieved in the simulations. However, $Re = 5$ is large enough to cause large inertia at bulk scale and at the same time avoid numerical errors.  We study suspension properties at $\phi \leqslant 0.35$, allowing us to explore rheological behavior of suspensions as it goes from a region with dominant fluid mechanical inertia  at low volume fractions to a region more similar to Stokes flow suspensions, but where inertia amplifies the effect of excluded volume, at $\phi = 0.35$.

\subsection{Stress system in a suspension}

Batchelor (1970) proposed a formulation for calculation of stress in a suspension. For statistically homogeneous suspensions, the bulk stress is computed by averaging the stress over the volume. At each point,  the velocity and stress fields are denoted as $\boldsymbol{u}(\boldsymbol{x})$ and $\boldsymbol{\sigma}(\boldsymbol{x})$, respectively. The bulk stress of the suspension is written in dimensional form as \newline

\begin{equation}
 \boldsymbol{\Sigma} = \frac{1}{V} \int_{V_{f}} -p \boldsymbol{I} dV + \mu(\boldsymbol{\nabla U} + \boldsymbol{\nabla U}^{T}) + \boldsymbol{\Sigma^p},
\end{equation}
where $p$ is the isotropic pressure and $\boldsymbol{\nabla U}$ is the average velocity gradient in the fluid.  The first two terms are fluid contributions to the suspension stress.  The influence of the particles is reflected in  $\boldsymbol{\Sigma^p}$, which is termed \emph{particle stress}. \newline

The stress generated by the particles originates from: $(1)$ the actual stress in the particles and $(2)$ velocity fluctuations caused by particles. We can split stress within the particles into the surface stress which originates from interaction of the particle with the fluid and the stress due to acceleration as

\begin{equation}
\label{eq:accel}
\int_{V_p} \boldsymbol{\sigma} dV_p = \emph{sym}(\int_{A_p} \boldsymbol{x\sigma.n} dA_p) - \emph{sym}(\int_{V_p} \boldsymbol{x \nabla. \sigma} dV_p),
\end{equation}
where $\boldsymbol{n}$ is the normal outward from the particle surface, and \emph{sym} indicates the symmetric part of the quantity which follows. The first integral on the right hand side is called \emph{stresslet} and the second the \emph{acceleration stress}, respectively. The stresslet is the symmetric first moment of surface traction applied on the particle surface, and depends on the suspension conditions ($\phi$ and $Re$ for this work). The acceleration on the particle results in a stress seen in the final term of (\ref{eq:accel}); in the absence of all non-hydrodynamic forces, $\boldsymbol{\nabla.\sigma}$ is related to the acceleration (note that $\frac{D \rho \boldsymbol{u}}{Dt} = \boldsymbol{\nabla.\sigma} $). \newline

In a dilute suspension, the velocity fluctuations are due only to the fluid disturbance caused by the essentially isolated particles. At larger $\phi$, interactions cause particles to deviate from their average paths and result in additional fluctuations.  The velocity at each point is the combination of the average and fluctuating velocities ($\boldsymbol{u} = \boldsymbol{\bar{u}} + \boldsymbol{u^{\prime}}$), and the fluctuations provide a mechanism for momentum transfer, of form $\rho \boldsymbol{u^\prime u^\prime}$, a Reynolds stress.\newline 

By taking the volume average, the particle stress can be written as

\begin{equation}
\label{eq:stresslet}
 \boldsymbol{\Sigma^{p}} = \frac{1}{V} \sum_{i} \int_{A_p} \boldsymbol{x \sigma.n}dA_i - \frac{1}{V} \sum_{i} \int_{V_p}  \boldsymbol{x \nabla . \sigma}dV_i - \frac{1}{V} \int_V \rho \boldsymbol{{u^\prime}{u^\prime}} dV,
\end{equation}
where summations are over all particles within the volume $V$. \newline

For a point inside a rigid particle, the acceleration $\boldsymbol{a} =  \boldsymbol{a_{i}} + \boldsymbol{\alpha_{i}}  \times \boldsymbol{r} + \boldsymbol{\omega_{i}}\times(\boldsymbol{\omega_{i}} \times \boldsymbol{r})$, where $\boldsymbol{a_i}$ is the linear acceleration of the center of mass, $\boldsymbol{\alpha_i}$ is the angular acceleration, $\boldsymbol{\omega_i}$ is the angular velocity and $\boldsymbol{r}$ is the distance from the point to the center of mass of the particle.  For each particle, we can find the acceleration stress by a volume integral of a first moment of acceleration with respect to $\boldsymbol{x}$, $\int_{V_p}  \frac{\rho}{2}(\boldsymbol{x a + a x})dV_i$.  Decomposing the coordinate into the center of mass and the distance from the center of mass ($\boldsymbol{x} = \boldsymbol{x_{cm}} + \boldsymbol{r}$), we rewrite the particle stress as 

\begin{equation}
 \boldsymbol{\Sigma^{p}} = \frac{1}{V} \sum_{i} \int_{A_p} \boldsymbol{(r + x_{cm}) \sigma.n}dA_i - \frac{1}{V} \sum_{i} \int_{V_p}  \boldsymbol{(r + x_{cm}) \nabla . \sigma}dV_i - \frac{1}{V} \int_V \rho \boldsymbol{{u^\prime}{u^\prime}} dV.
\end{equation}
The two center-of-mass dependent terms, i.e $ \int_{A_p} \boldsymbol{x_{cm} \sigma.n}dA_i$ and $\int_{V_p} \boldsymbol{x_{cm} \nabla. \sigma}dV_i$ will cancel each other upon application of the divergence theorem, and the suspension stress is independent of the coordinate frame. Scaling the stress by $\mu \dot{\gamma}$, the final particle stress formula is

\begin{equation}
\label{eq:stressletf}
 \frac{\boldsymbol{\Sigma^{p}}}{\mu\dot{\gamma}} = \frac{1}{V} \sum_{i} \int_{A_p} \boldsymbol{r \sigma.n}dA_i - \frac{Re}{V} \sum_{i} \int_{V_p}  \boldsymbol{r \nabla . \sigma}dV_i - \frac{Re}{V} \int_V  \boldsymbol{{u^\prime}{u^\prime}} dV.
\end{equation}

The rheological properties reported in this work are the normal stress differences,  
\begin{equation}
N_1 = \langle \Sigma_{xx}^p-\Sigma_{yy}^p\rangle   \,\,\,\,\,\,\mbox{\rm and} \,\,\,\,\,\, N_2 = \langle \Sigma_{yy}^p-\Sigma_{zz}^p\rangle  ,
\end{equation}
the particle pressure,
\begin{equation}
\Pi = -\frac{1}{3} (\langle \Sigma_{xx}^{p}\rangle + \langle \Sigma_{yy}^{p}\rangle  + \langle \Sigma_{zz}^{p}\rangle ),
\end{equation}
and the relative viscosity $\mu_r$,  defined by
\begin{equation}
\mu_r = 1 + \frac{<\boldsymbol{\Sigma}^{p}_{xy}>}{\mu \dot{\gamma}},
\end{equation}
where $\langle \,\,\rangle $ denotes the time-averaged quantity. \newline

\section{Microstructure}

We study the effect of inertia on the flow-induced structure as characterized by the pair distribution function, which is defined by \newline
\begin{equation}
  g(\boldsymbol{r})= \frac{P_{1|1}(\boldsymbol{r}|\boldsymbol{0})}{n}, 
\end{equation}
where $n=3\phi/(4\pi a^3)$ is the average number density of particles and $P_{1|1}(\boldsymbol{r|0})$ is the conditional probability of finding a particle at location $\boldsymbol{r}$ from a reference particle at the origin, $\boldsymbol{r} = 0$. To sample pair vectors, the space around a reference particle is discretized in the $r, \theta,$ and $\psi$ directions where $\theta$ is the angle from flow direction and $\psi$ is the angle from the positive $z$ direction, as illustrated for a coordinate system centered on a reference particle in figure 1(\emph{a}).  \newline

\begin{figure}
\centering
\subfigure[]{\includegraphics[totalheight=0.2\textheight,]{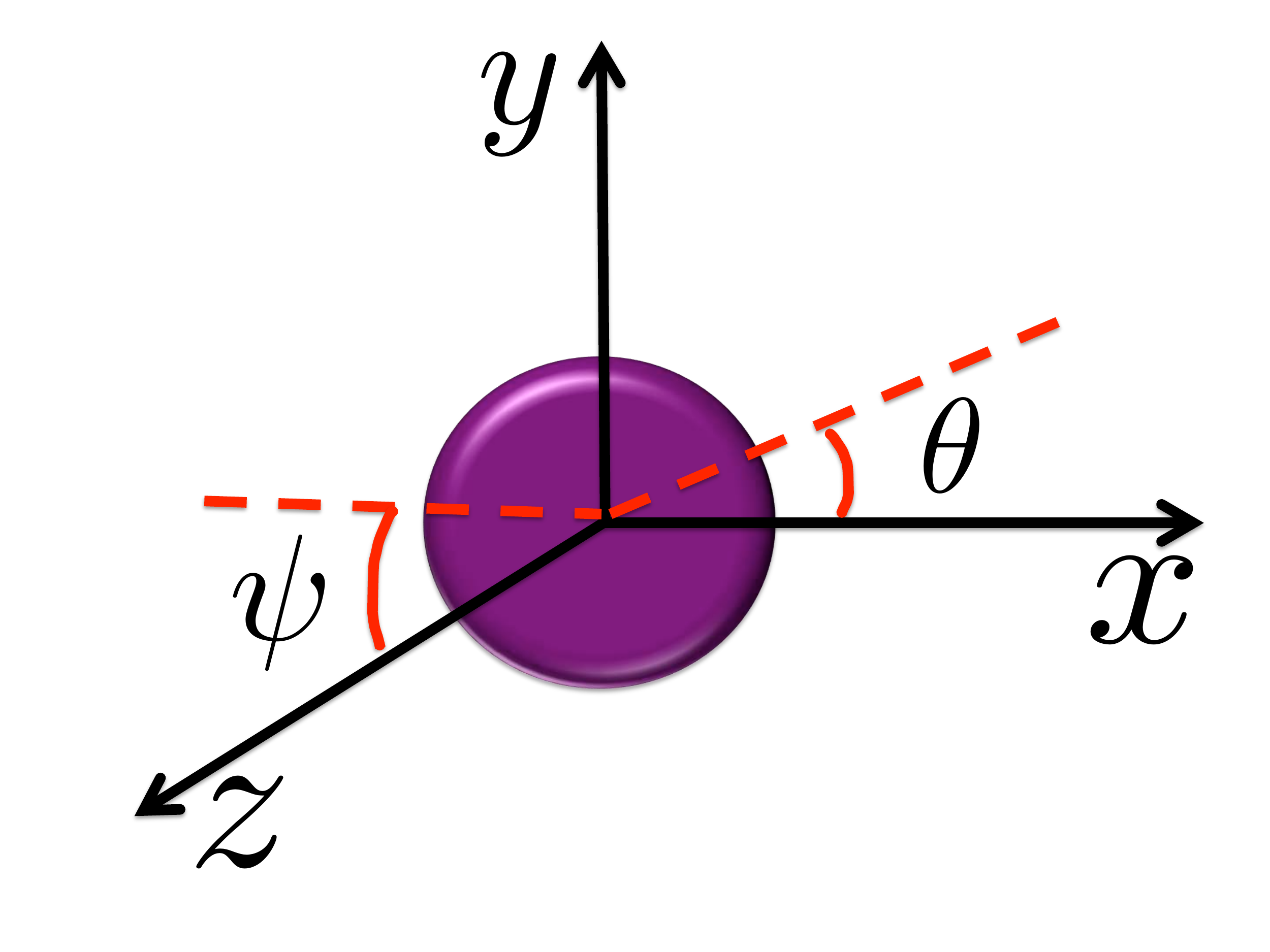}}
\subfigure[]{\includegraphics[totalheight=0.2\textheight,]{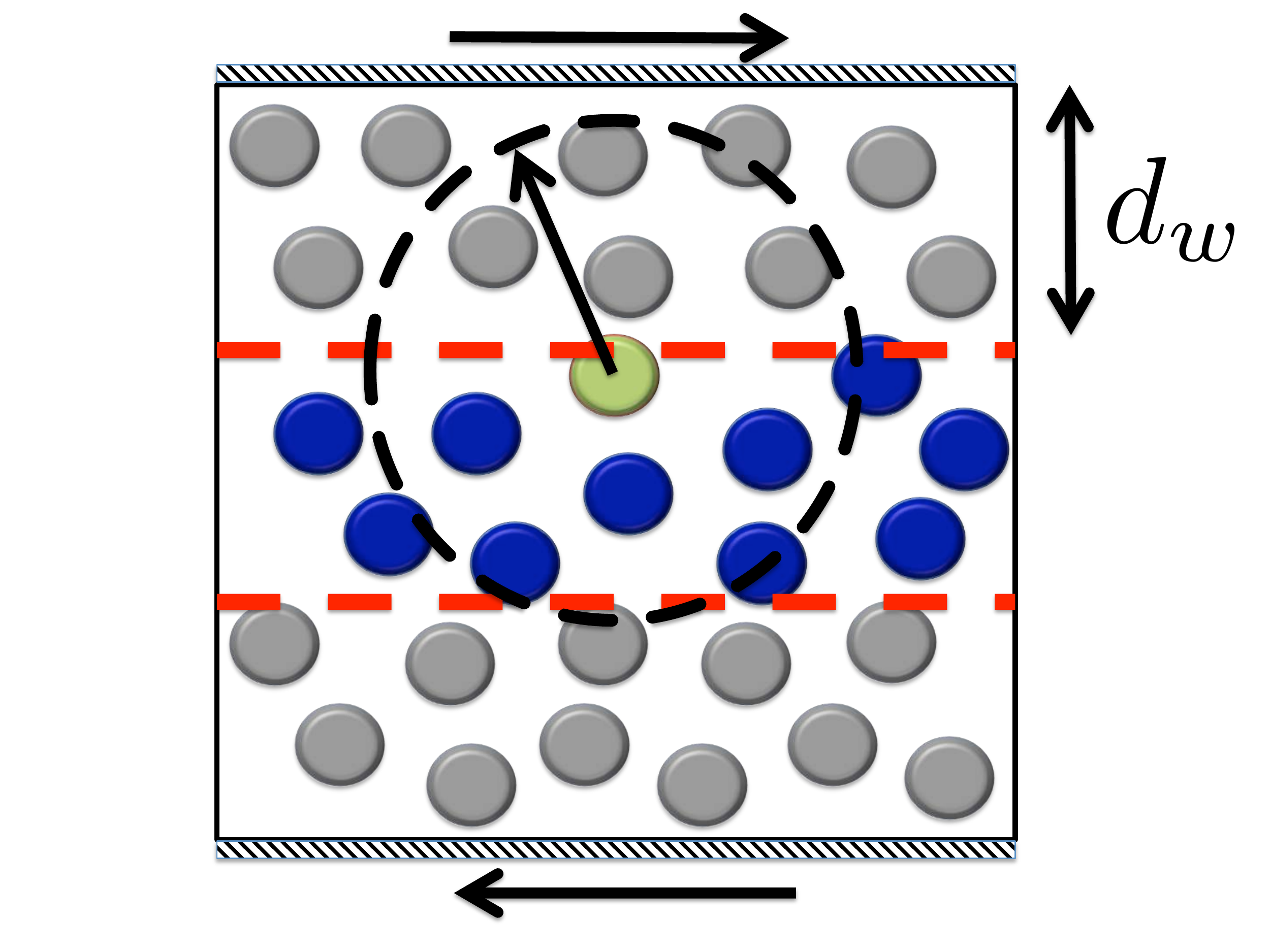}}
\caption{(\emph{a}) Convention used to define angles in a simple shear flow $u_{x}(y)$: $0^{\circ}< \theta <360^{\circ}$ is measured counter-clockwise from the positive $x$ (flow) axis, and $0^{\circ}<\psi<180^{\circ}$ is the polar angle measured from positive $z$ (vorticity) direction. (\emph{b}) The scheme for sampling of $g(\boldsymbol{r})$ in a confined domain. Reference particles lie within the boundaries of the secondary domain, displayed by straight dashed lines, which are located $d_w$ away from walls and pair separation vectors are sampled from a spherical domain (which does not intersect a wall) centered on the reference particle.}
\label{fig:crd}
\end{figure}

In a shear flow, noninteracting particles would approach each other in the compressional region, and separate in the extensional region.  The pair space in simple-shear flow is symmetric with respect to the origin. For the upper half space, the compressional region is $90^{\circ}< \theta <180^{\circ}$ ($xy < 0$) and the extensional zone covers $0^{\circ}< \theta <90^{\circ}$ ($xy > 0$). Calculation of $g(\boldsymbol{r})$ is executed by assigning the sampled pair vectors in the appropriate spatial bins.  To minimize effects due to boundaries of the simulation box in the velocity gradient ($y$) direction, sampling for $g(\boldsymbol{r})$ is taken from a secondary domain inside the main computational box with boundaries in $y$ away from the walls. The distance of the secondary domain boundaries from the walls, $d_w,$ depends on the distance from the reference particle for which the pair distribution function is calculated. A typical value for spatial distributions in this work  is $d_w = 5a$; but for instance, to compute $g(\boldsymbol{r})$ within the range $2\leqslant|\boldsymbol{r}|\leqslant8$, the boundaries must be chosen at $8a$ away from walls. The secondary domain within the computational box is indicated by straight dashed lines in figure 1(\emph{b}). The reference particle center must be within the secondary domain for our sampling; while the second of the pair may be outside the domain. Vectors of the pair separation located in a spherical zone around the reference particle are calculated and dispensed in bins of volume $\triangle V = r^2\triangle r \sin\psi\triangle\psi\triangle\theta$, and  a running histogram $H(r,\theta,\psi)$ for the population of pairs inside the bins is generated. The pair distribution function is calculated as

\begin{equation}
  g(r, \theta, \psi)= \frac{H(r, \theta, \psi)}{n_p t_s \triangle V} ,
\end{equation}
in which $n_p$ is the number density of pairs and $t_s$ is the number of samplings. \newline

\begin{figure}
\centering
\subfigure[$Re = 0.05$ and $\phi = 0.1$]{\includegraphics[totalheight=0.15\textheight,]{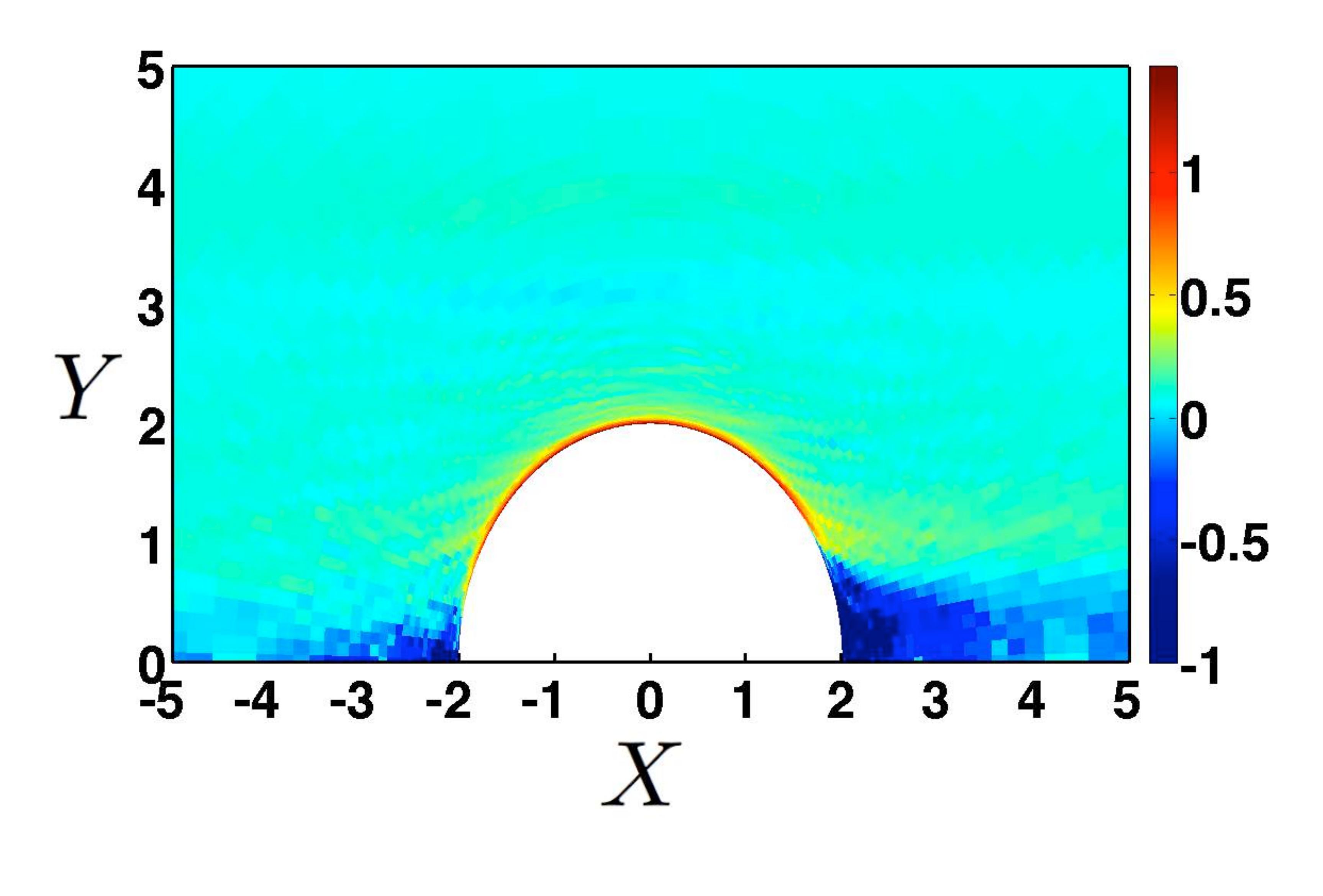}}
\subfigure[$Re = 0.05$ and $\phi = 0.25$]{\includegraphics[totalheight=0.15\textheight,]{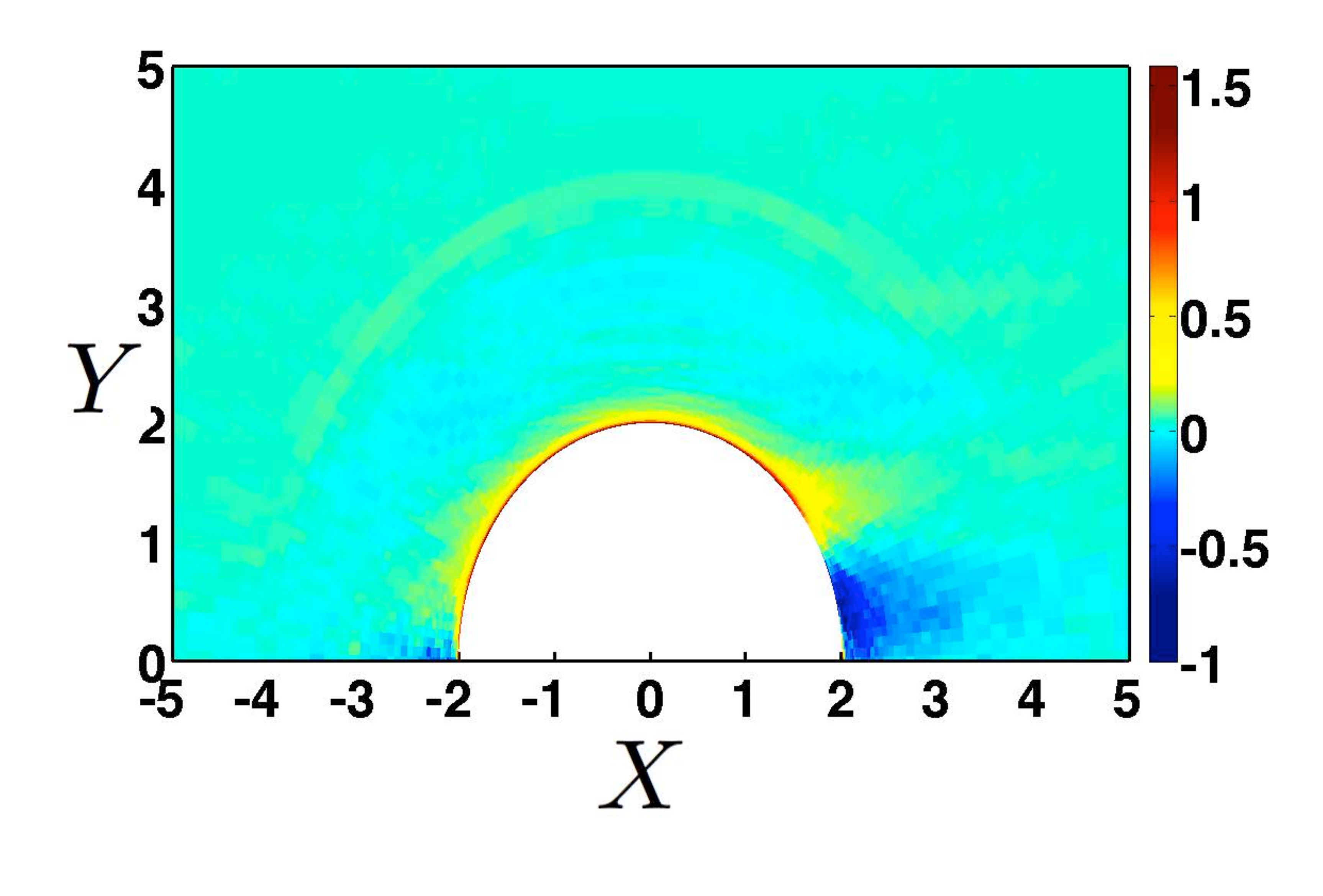}}
\subfigure[$Re = 1.2$ and $\phi = 0.1$]{\includegraphics[totalheight=0.15\textheight,]{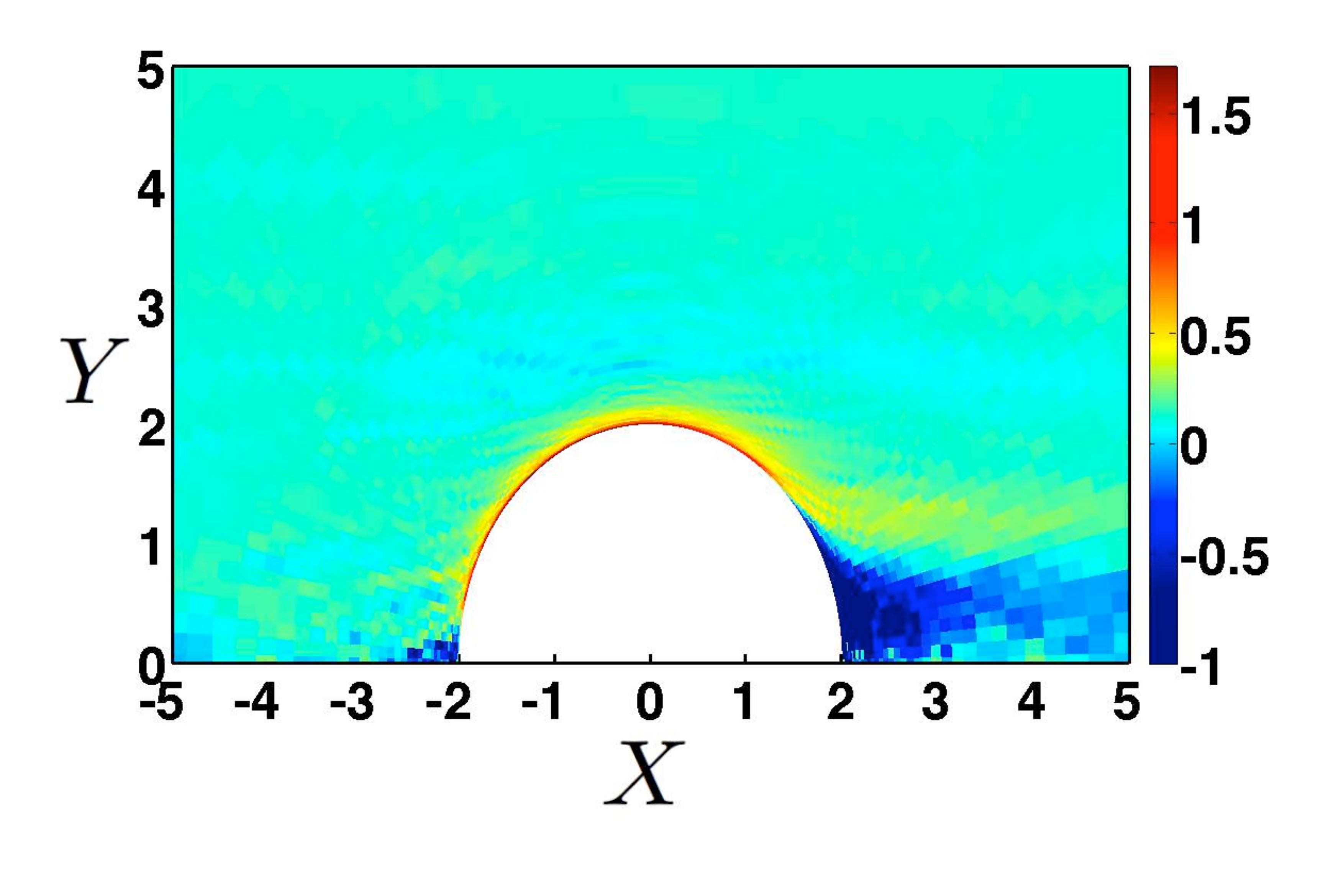}}
\subfigure[$Re = 1.2$ and $\phi = 0.25$]{\includegraphics[totalheight=0.15\textheight,]{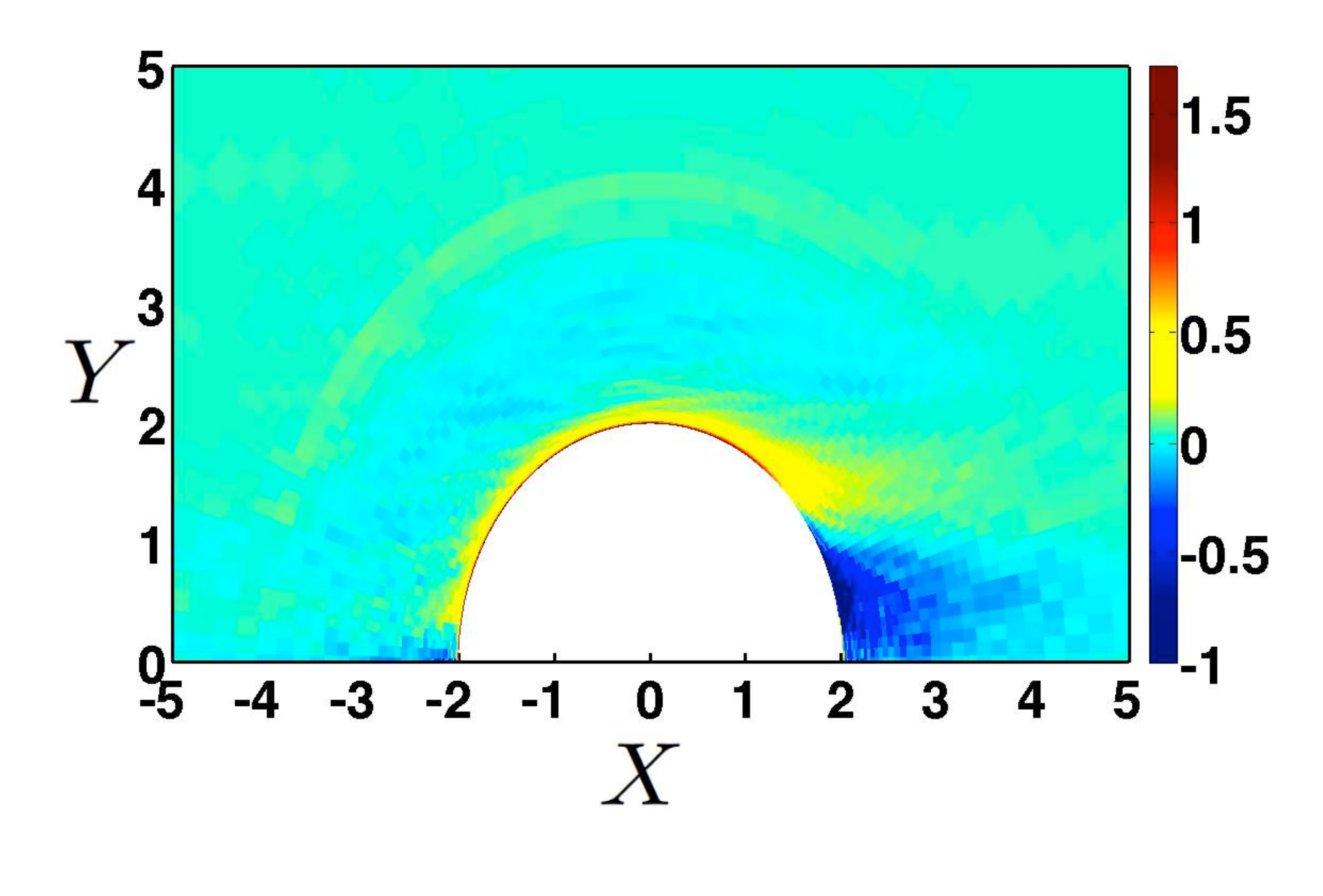}}
\subfigure[$Re = 2$ and $\phi = 0.1$]{\includegraphics[totalheight=0.15\textheight,]{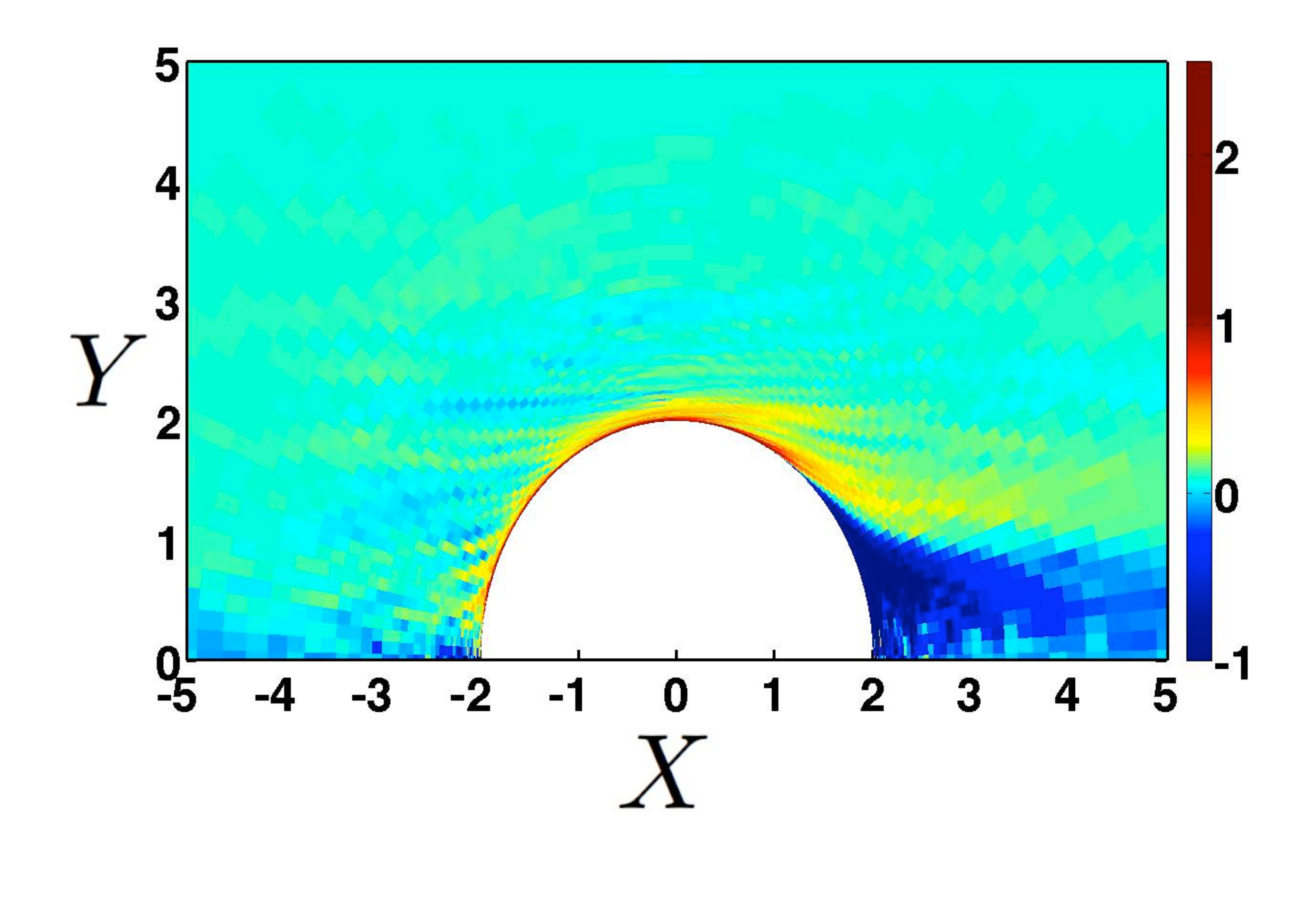}}
\subfigure[$Re = 2$ and $\phi = 0.25$]{\includegraphics[totalheight=0.148\textheight,]{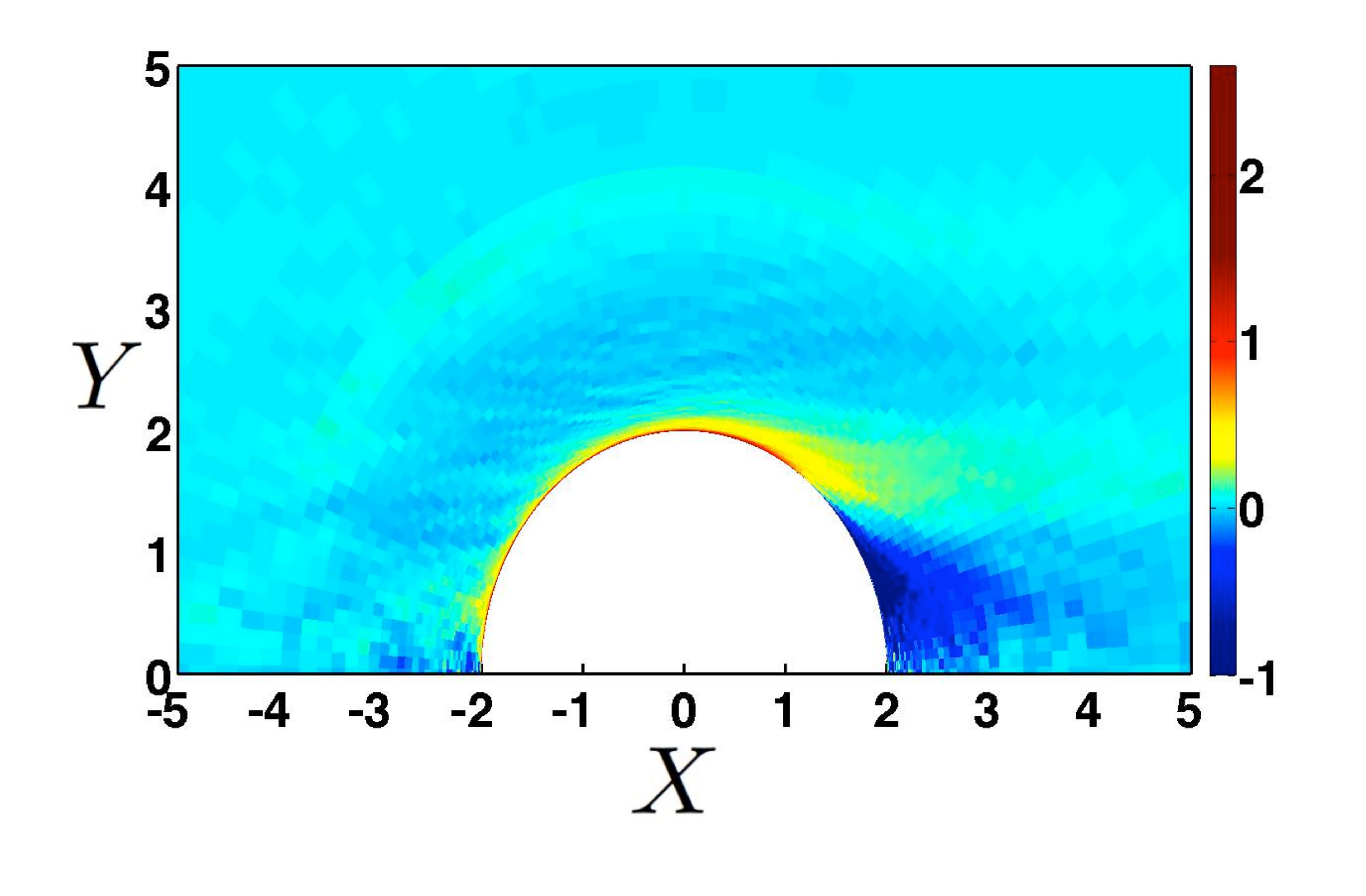}}
\subfigure[$Re = 5$ and $\phi = 0.1$]{\includegraphics[totalheight=0.15\textheight,]{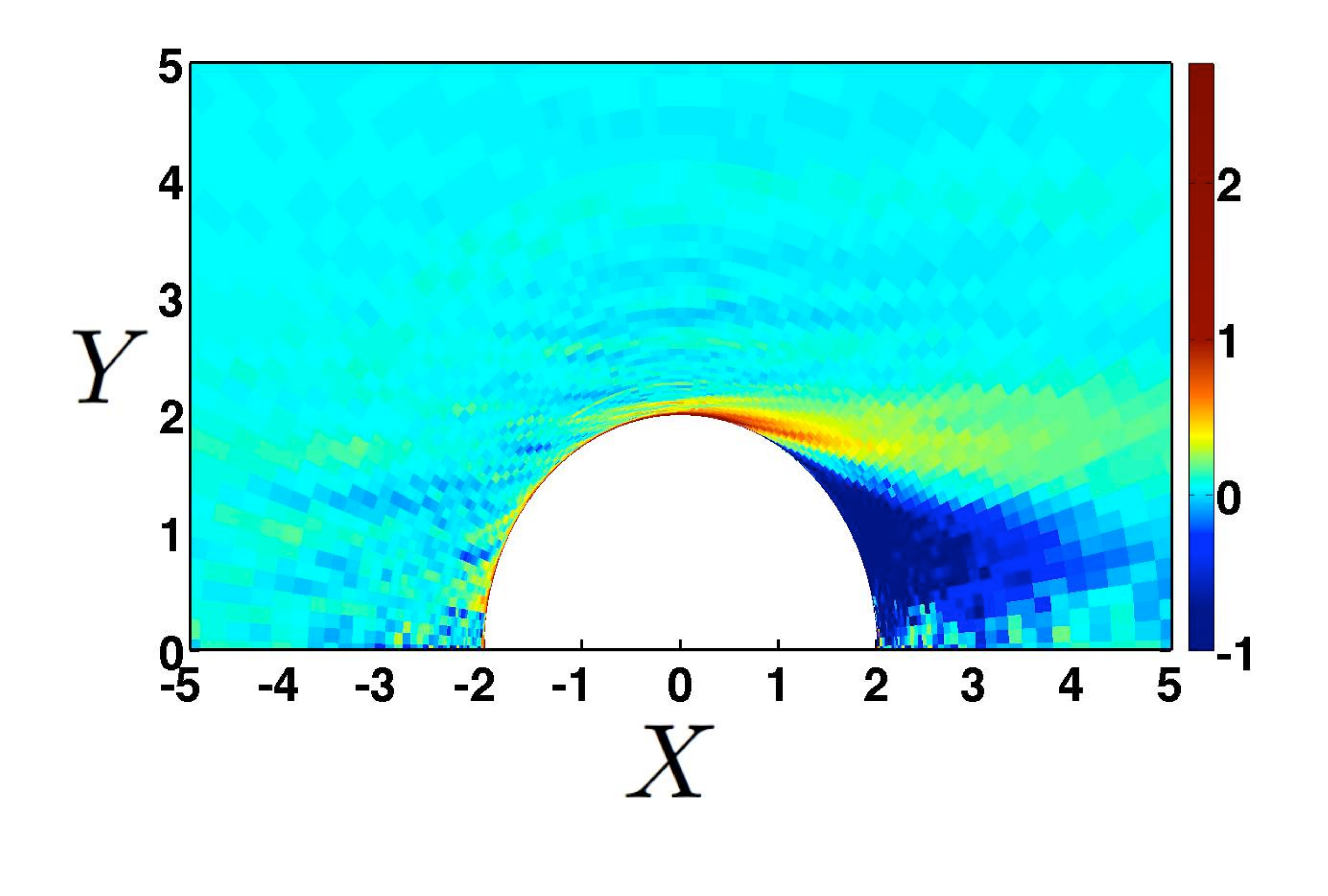}}
\subfigure[$Re = 5$ and $\phi = 0.25$]{\includegraphics[totalheight=0.148\textheight,]{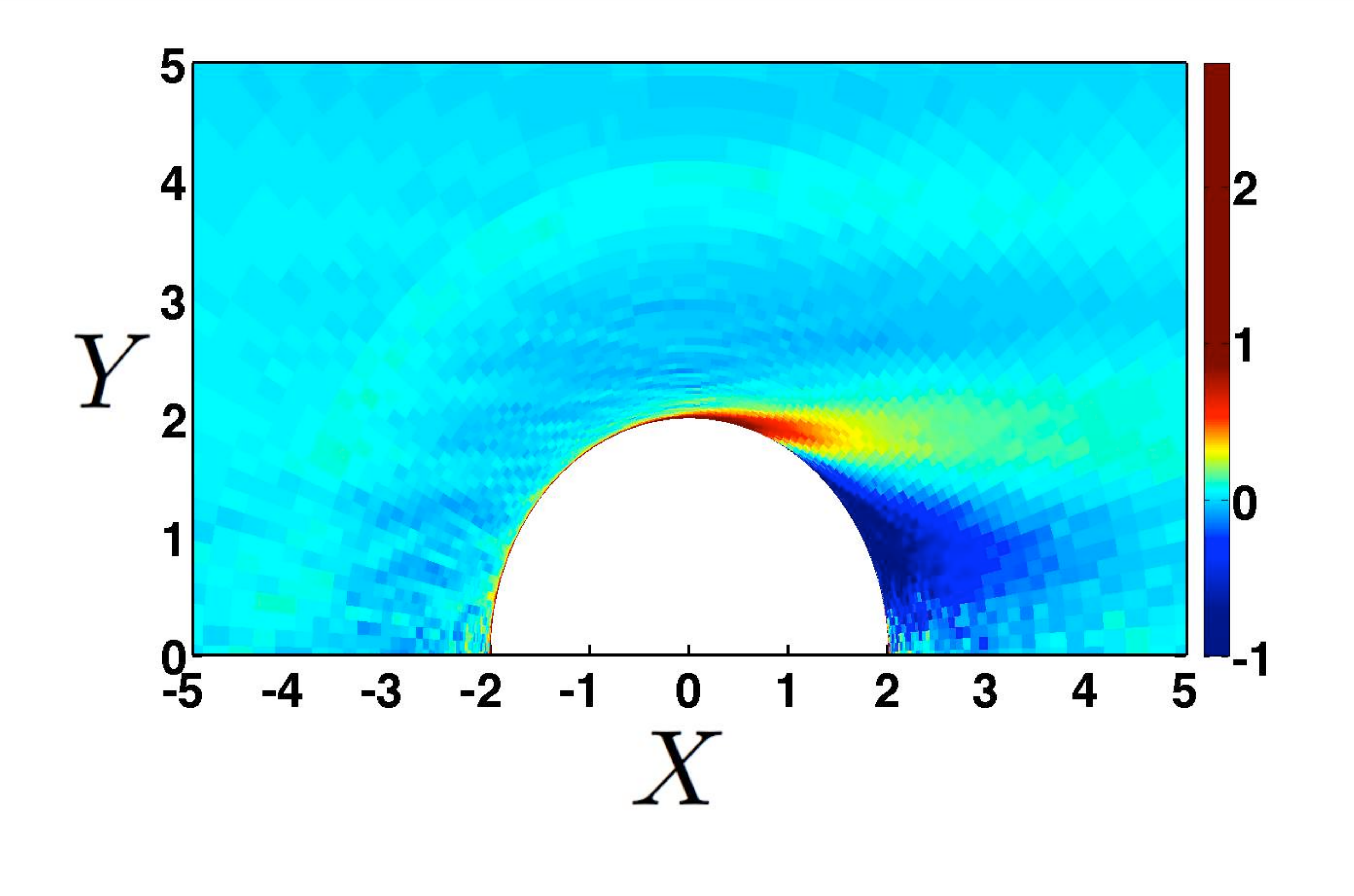}}
\caption{The flow induced microstructure of the suspension on the plane of shear.}
 \label{fig:GSP}
\end{figure}

In order to satisfy statistical accuracy of pair sampling, we performed the simulations for large total strains, typically $\dot{\gamma}t=220$ - 440, depending on $Re$ with larger simulation durations for smaller $Re$. Samples were taken at intervals of $\dot{\gamma}\Delta t= 0.003$ - 0.006.  Because the objective is to find $g(\boldsymbol{r})$ at steady state, data sampled from the initial $30$ strain are discarded from any simulation. For each $\phi$ and $Re$, we performed simulations using several initial configurations generated by a Monte Carlo algorithm, with the reported data being the average over all sampled runs at each $\phi$ and $Re$. \newline

\begin{figure}
\centering
\subfigure[$\phi$ dependence at $Re = 0.6$]{\includegraphics[totalheight=0.2\textheight,]{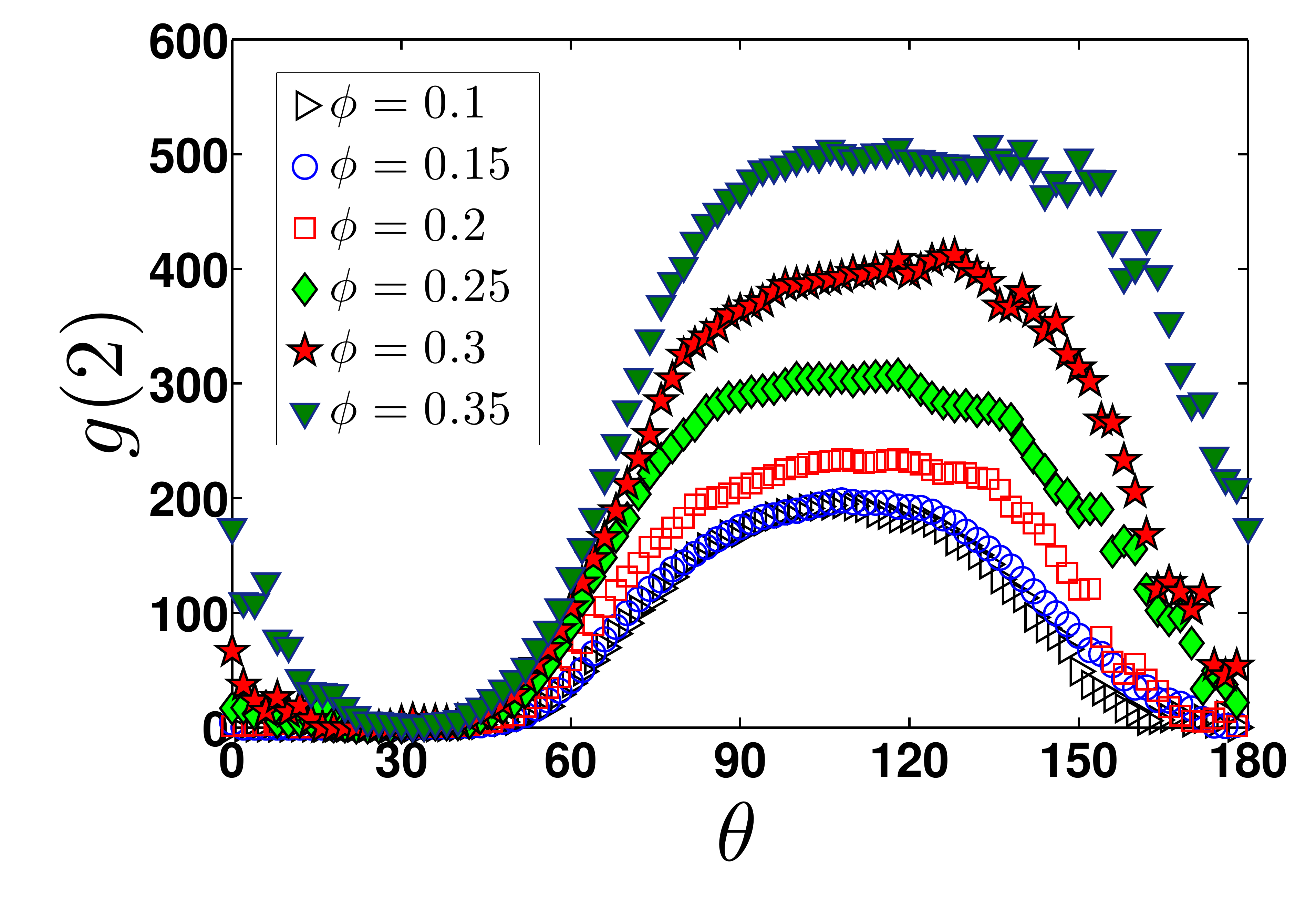}}
\subfigure[$Re$ dependence at $\phi = 0.1$]{\includegraphics[totalheight=0.202\textheight,]{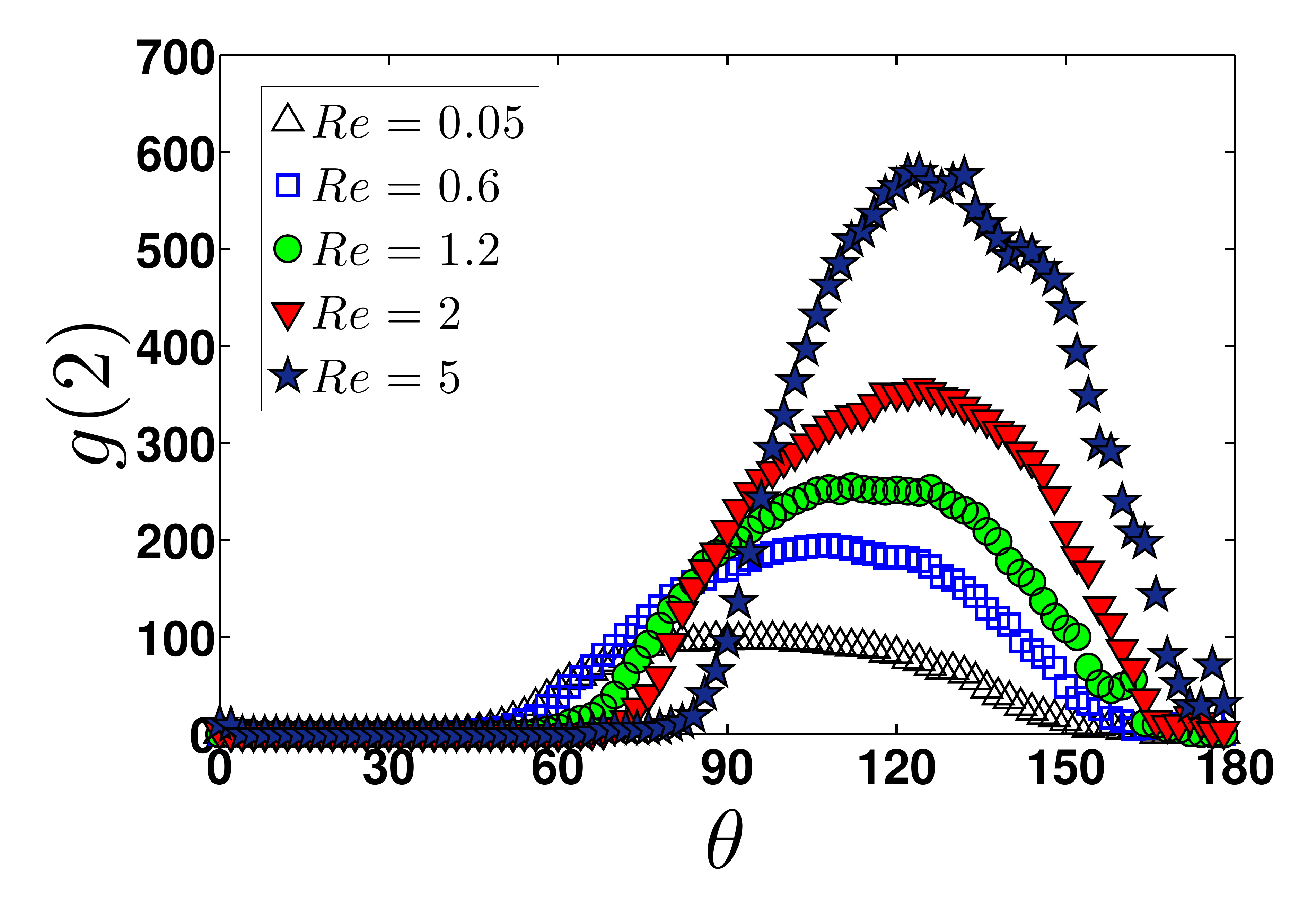}}
\subfigure[$Re$ dependence at $\phi = 0.3$]{\includegraphics[totalheight=0.2\textheight,]{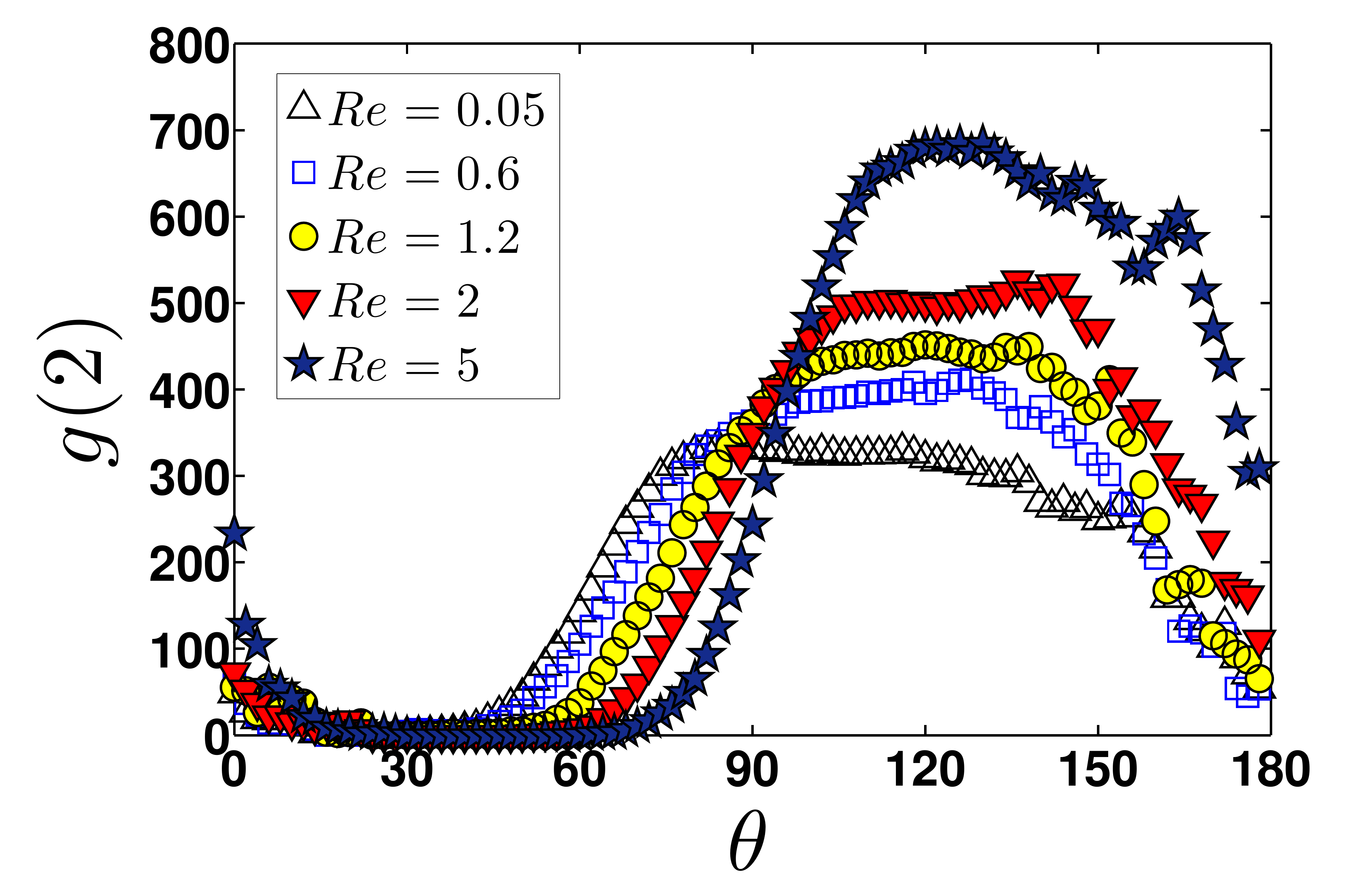}}
\caption{\emph{(a)} The influence of $\phi$ on the contact value of $g$, taken as $g(2<\boldsymbol{r}<2.001$, at $Re = 0.6$. The dependence on $Re$ is displayed in (\emph{b}) for $\phi = 0.1$ and (\emph{c}) for $\phi = 0.3$. }
 \label{fig:G2}
\end{figure}

In our determination of $g(\boldsymbol{r})$ on the plane of shear, the first bin adjacent to contact was radially centered at $r/a = 2.001$ and the angular discretization was fixed at $ \Delta\theta = 2^\circ$. For three dimensional samplings, the first bin was at $r/a  = 2.01$ and $\Delta \theta = \Delta \psi = 4^\circ$. Because the pair correlation is large at contact and declines rapidly toward unity as the pair separation increases,  we scale the radial bins starting from a dense mesh at contact as $\Delta \boldsymbol{r}^{i+1}/\Delta \boldsymbol{r}^{i} = \alpha$, where $\alpha = 1.1$ has been chosen empirically as a convenient value.\newline
 
We first present the flow-induced structure and its variation with $Re$ and $\phi$ on the shear plane, where the influence of shear on a pair interaction is strongest. The form of $g(\boldsymbol{r})$ at $Re = 0.05, 1.2, 2$ and $5$, for $\phi = 0.1$ and $0.25$, is shown in figure~\ref{fig:GSP}. We observe an asymmetric build up of $g(\boldsymbol{r})$ at contact in the shear plane; the value of $g$ is largest in compression but an elevated region extends to the extensional zone. Following the separation of the particles in the extensional zone, the highly correlated layer disappears and a depleted or wake zone develops in the downstream side.  The downstream wake zone is more pronounced at higher $Re$. Increasing $\phi$ results in a reduction of the size of the downstream wake. At $\phi = 0.25$ a secondary zone of large $g(\boldsymbol{r})$, the next-nearest neighbor ring, forms at $\boldsymbol{r}\approx 4$ in the compressional zone. Comparing the microstructures at $Re = 0.05$ and higher $Re$ shows that with increasing inertia the high correlation layer detaches at larger $\theta$, meaning pairs remain `attached' -- i.e. directly adjacent to contact -- for a shorter distance. This results in higher asymmetry of the high correlation layer at contact. With increasing $Re$ and specifically at low $\phi$ a streak of large pair correlation is observed in the extensional zone.  \newline  
 
Strong build up of pair correlation at contact has a controlling effect on the rheological properties of suspensions, as shown in Stokes-flow studies (Sierou \& Brady 2004; Nazockdast \& Morris 2012). Here we examine the influence of inertia on this contact correlation. In figure~\ref{fig:G2} we display $g(2<\boldsymbol{r}<2.001)$ on the plane of shear, with $\phi$ dependence presented in figure ~\ref{fig:G2}(\emph{a}) for suspensions at $Re = 0.6$. With increasing $\phi$, the contact value of $g$ increases. Because $g(\boldsymbol{r})$ is normalized by the pair number density, the origin of higher $g(2)$ for large $\phi$ can not be attributed to the increase of the number of pairs. The width of the zone of elevated $g(\boldsymbol{r})$ increases at larger volume fractions, i.e. $g \gg 1$ over a larger range of $\theta$. With increasing $\phi$ there is a growth of correlation in the flow direction. 
Figures  ~\ref{fig:G2} (\emph{b}) and (\emph{c}) show the effect of $Re$ on the shear plane values of $g(2<\boldsymbol{r}<2.001)$ at $\phi = 0.1$ and $0.3$. For both volume fractions, increasing $Re$ leads to increased pair correlation at contact. At $\phi = 0.1$, larger inertia results in appearance of a maximum in $g(\boldsymbol{r})$ near the compression axis at $\theta = 120^{\circ}$. At $\phi = 0.3$, the increase of $g(2<\boldsymbol{r}<2.001)$ is less pronounced mainly because the effect of inertia is weakened by relatively large excluded volume effects. We can observe in these figures that with increasing $Re$, the separation point, where $g(2<\boldsymbol{r}<2.001)$ declines rapidly, shifts to larger pair orientation angles, meaning the departure is earlier in sense of motion with the bulk flow. \newline

\begin{figure}
\centering
\subfigure[$Re = 0.05$, $\phi = 0.15$]{\includegraphics[totalheight=0.12\textheight,]{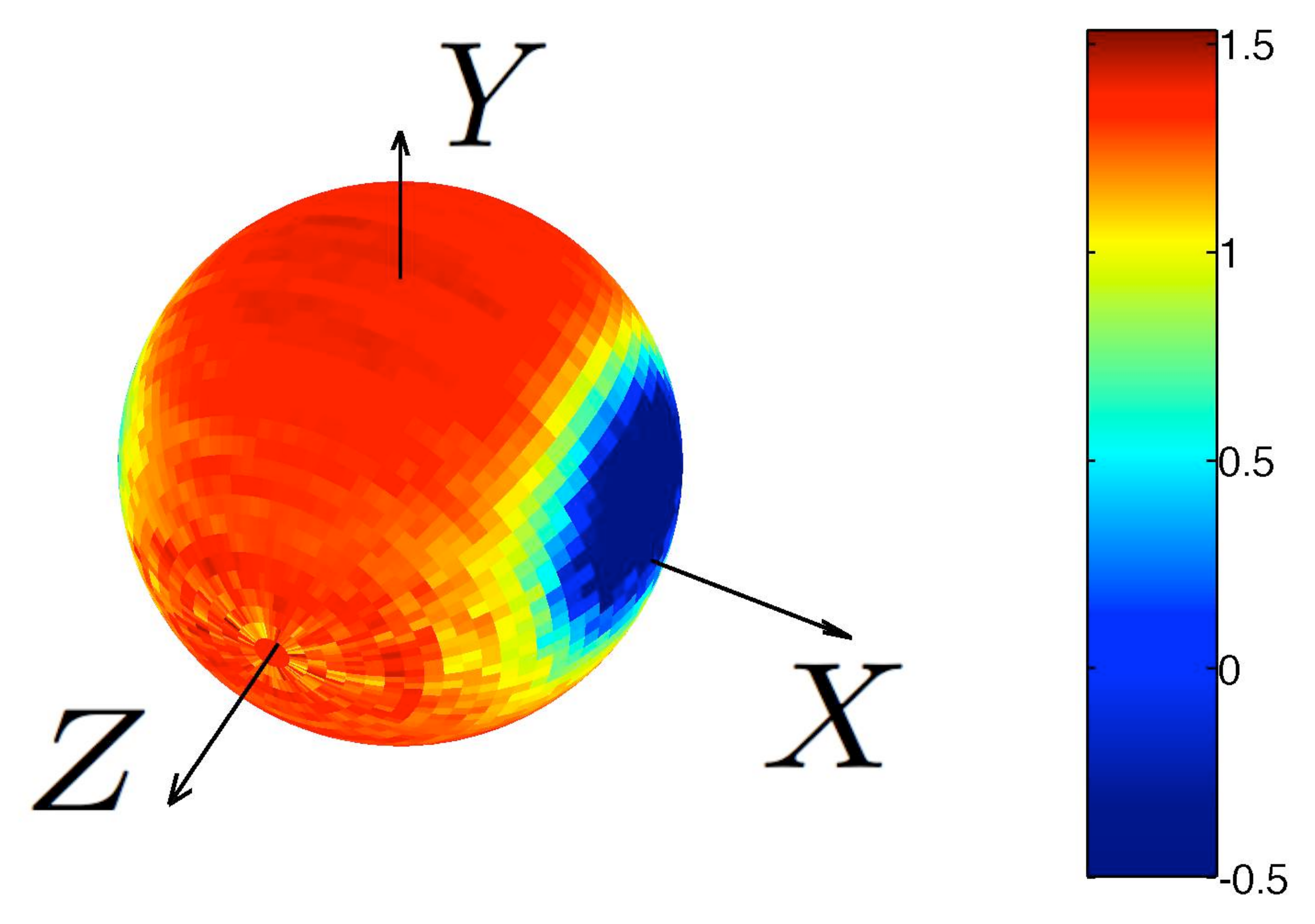}}
\subfigure[$Re = 0.6$, $\phi = 0.15$]{\includegraphics[totalheight=0.12\textheight,]{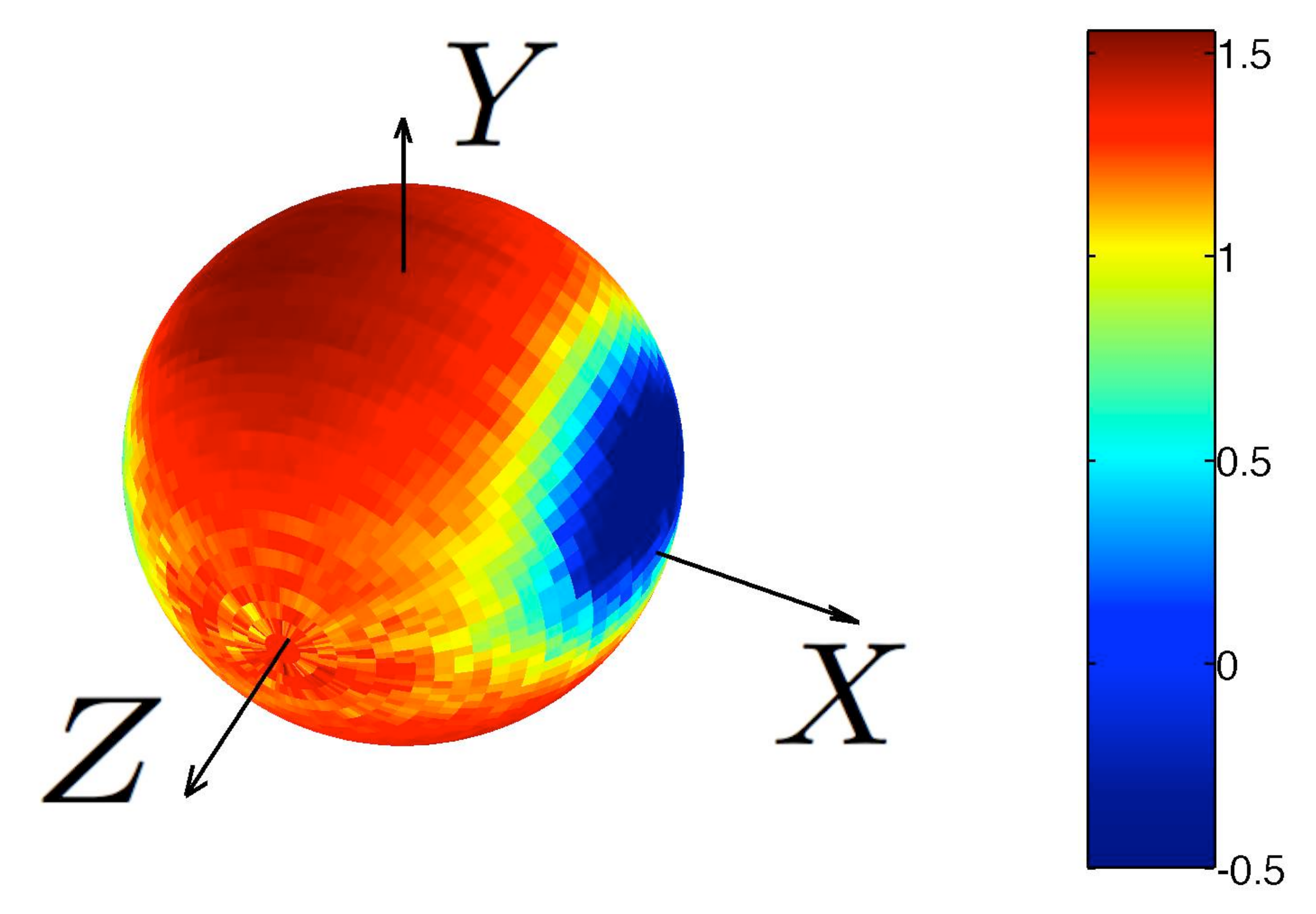}}
\subfigure[$Re = 1.2$, $\phi = 0.15$]{\includegraphics[totalheight=0.12\textheight,]{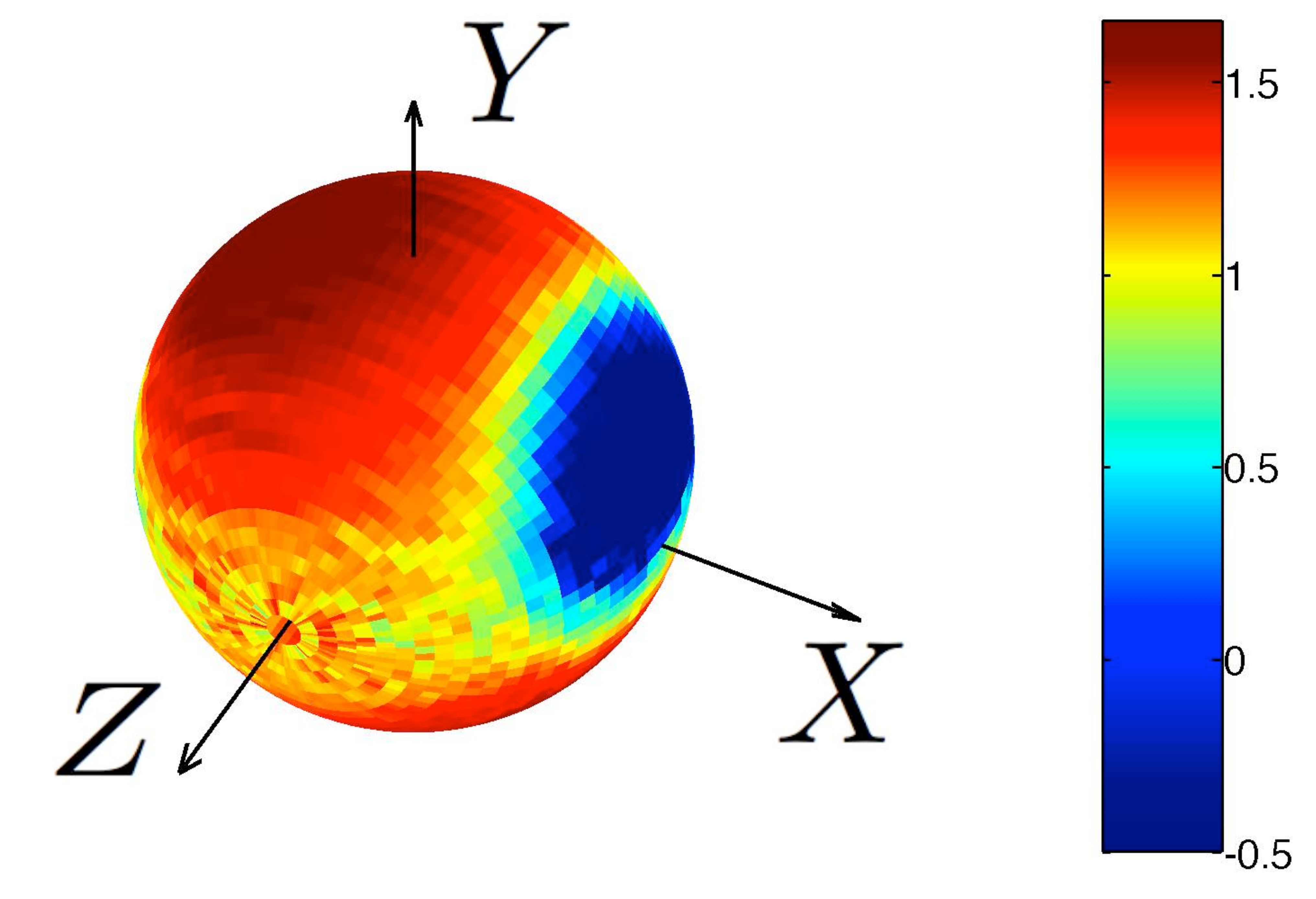}}
\subfigure[$\phi = 0.1$, $Re = 0.6$]{\includegraphics[totalheight=0.12\textheight,]{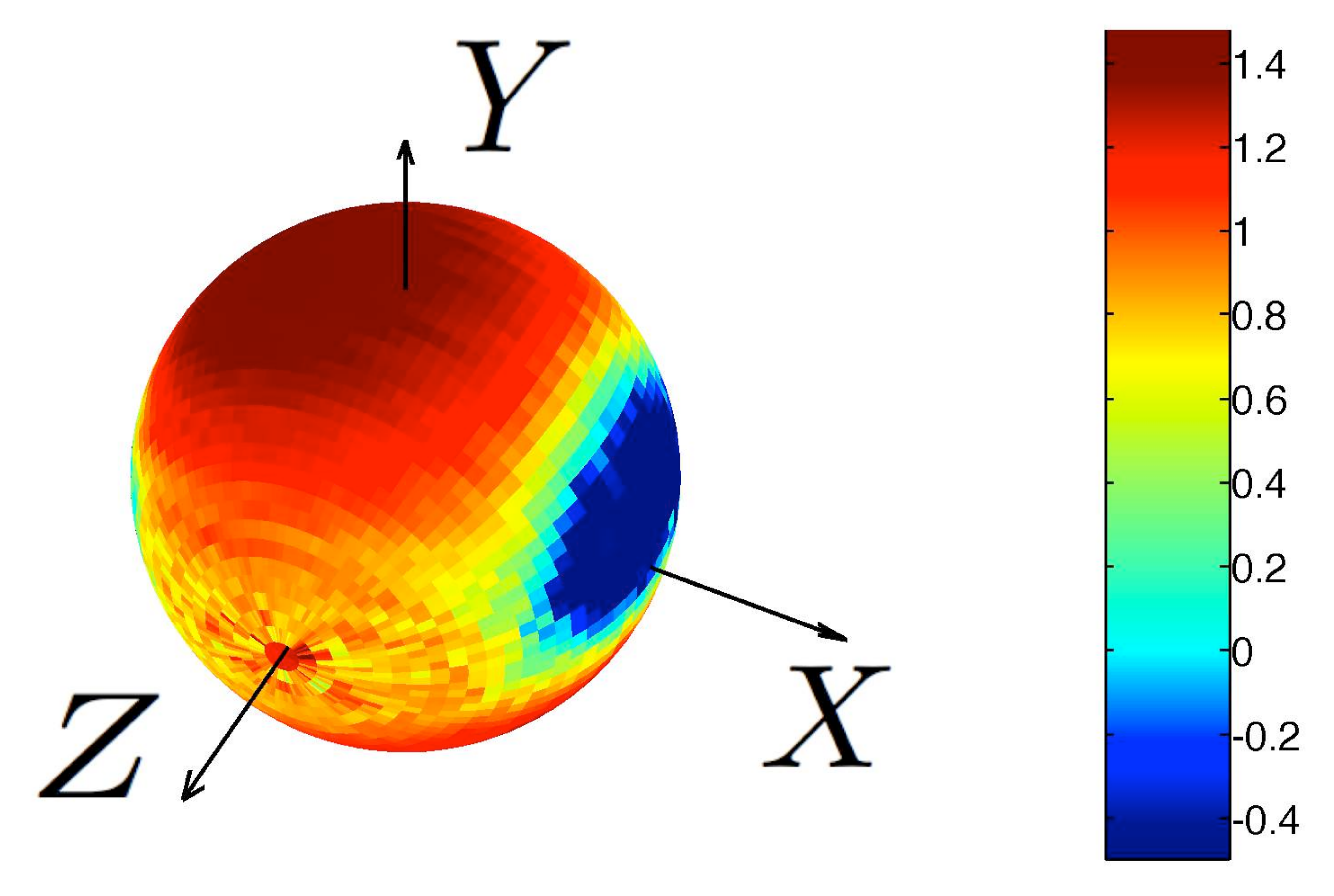}}
\subfigure[$\phi = 0.2$, $Re = 0.6$]{\includegraphics[totalheight=0.12\textheight,]{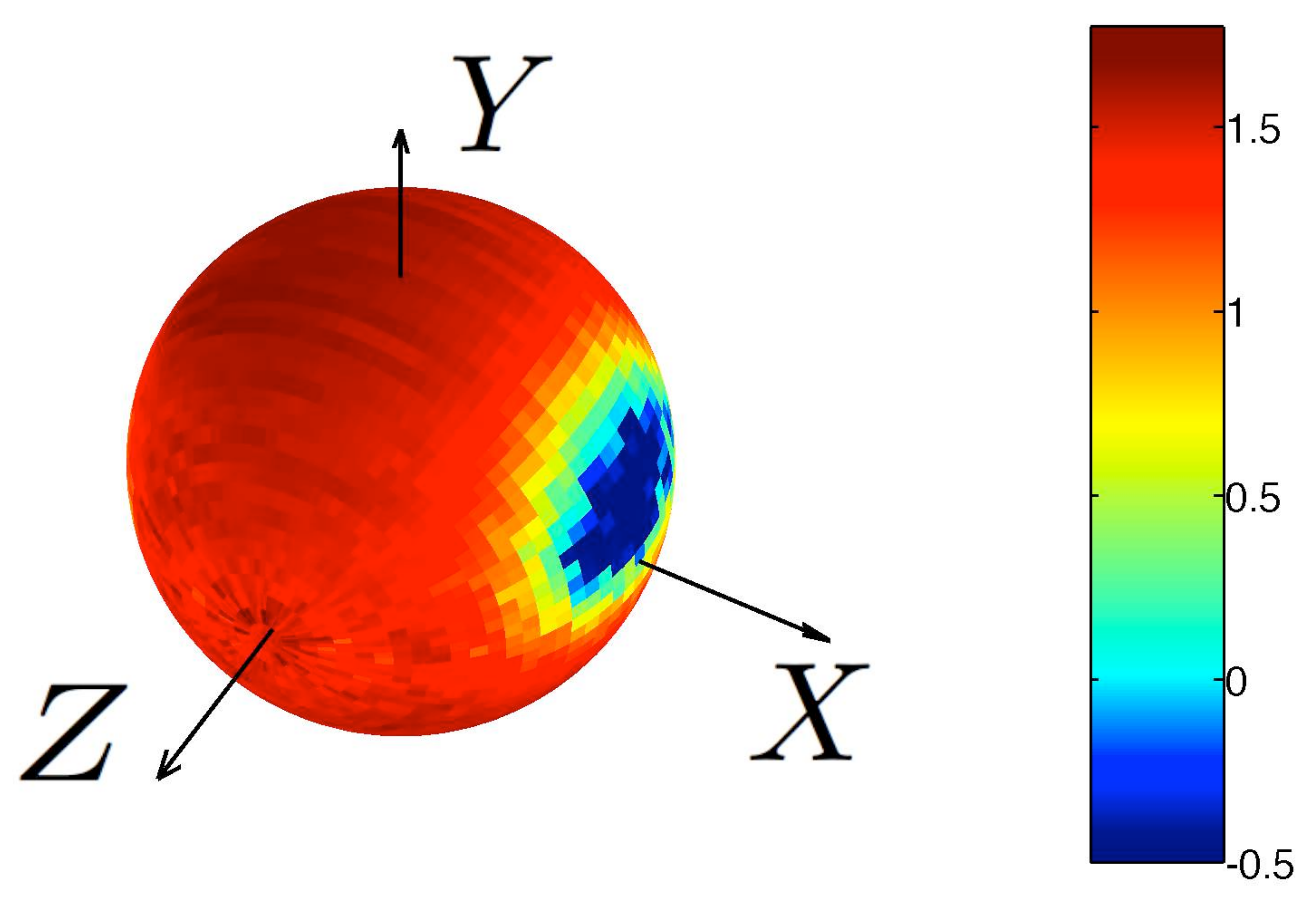}}
\subfigure[$\phi = 0.3$, $Re = 0.6$]{\includegraphics[totalheight=0.12\textheight,]{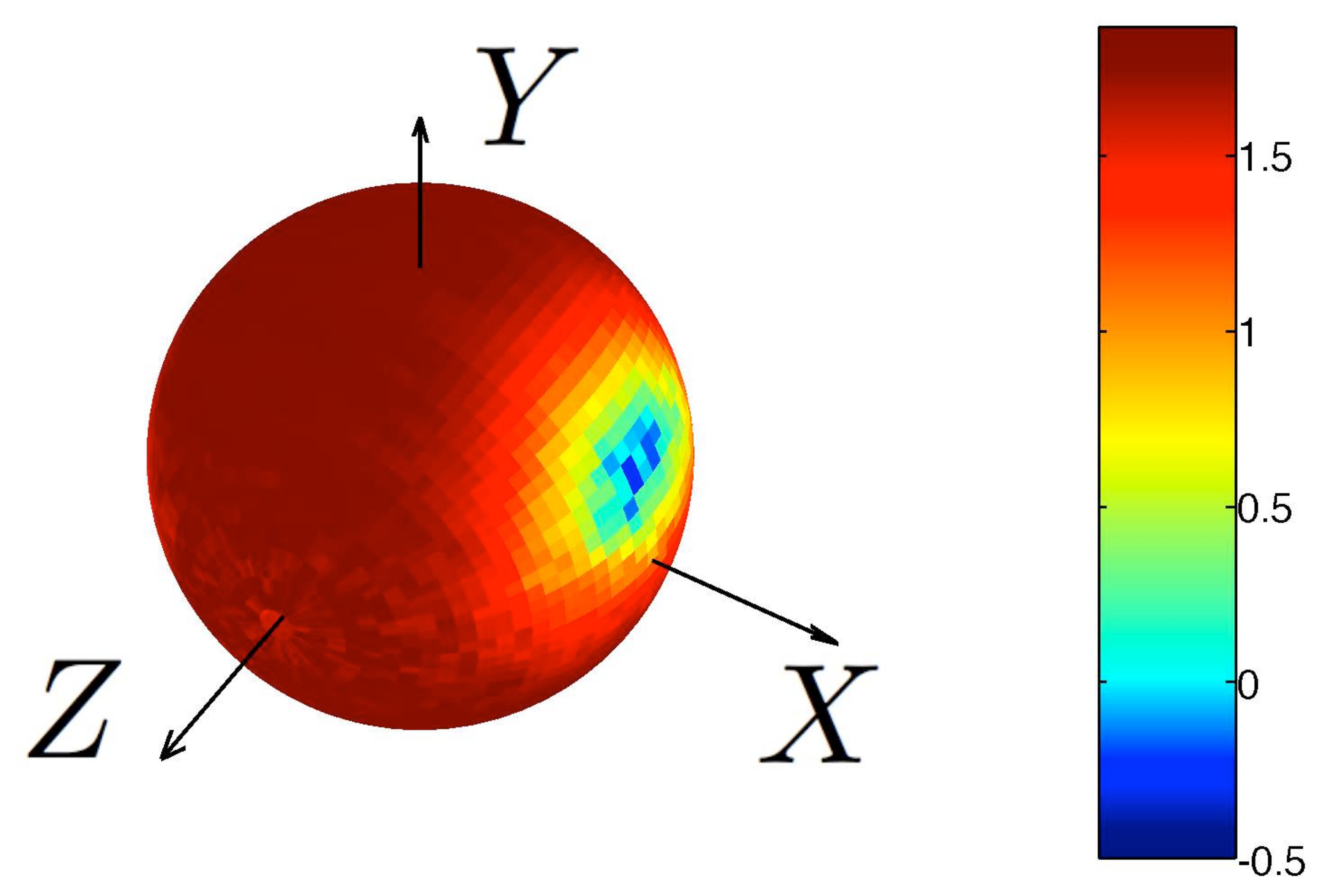}}
\caption{Three dimensional form of contact pair distribution:  (\emph{a}) - (\emph{c}) exhibit the effect of $Re$ for suspensions at $\phi = 0.15$, where increasing $Re$ leads to the decrease of the pair distribution function around the vorticity axis; (\emph{d})-(\emph{f}) illustrate the effect of volume fraction at $Re = 0.6$, showing the pair correlation increases around vorticity axis at higher $\phi$. }
 \label{fig:G-3D-Re}
\end{figure}

\begin{figure}
\centering
\subfigure[]{\includegraphics[totalheight=0.25\textheight,]{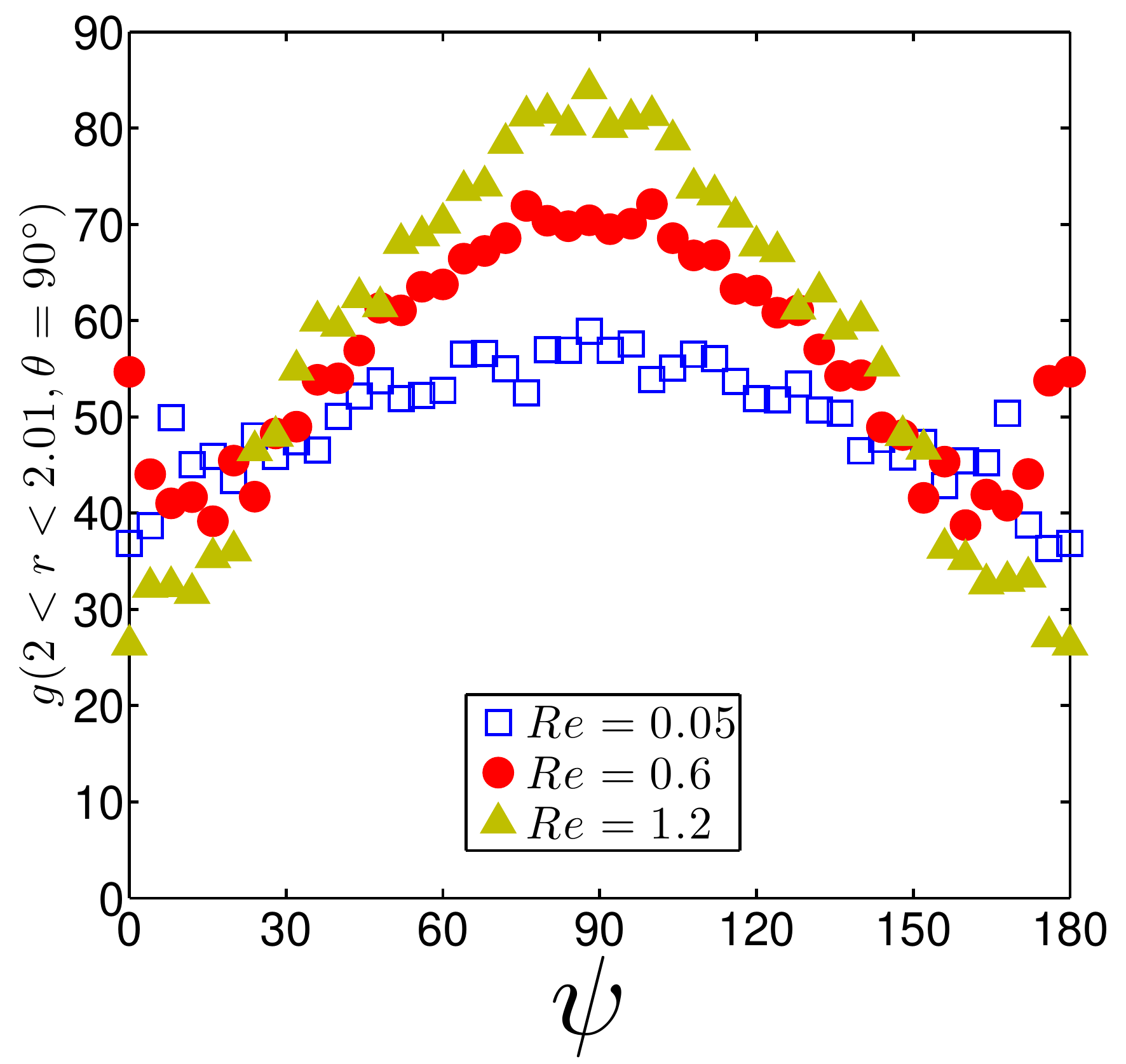}}
\subfigure[]{\includegraphics[totalheight=0.25\textheight,]{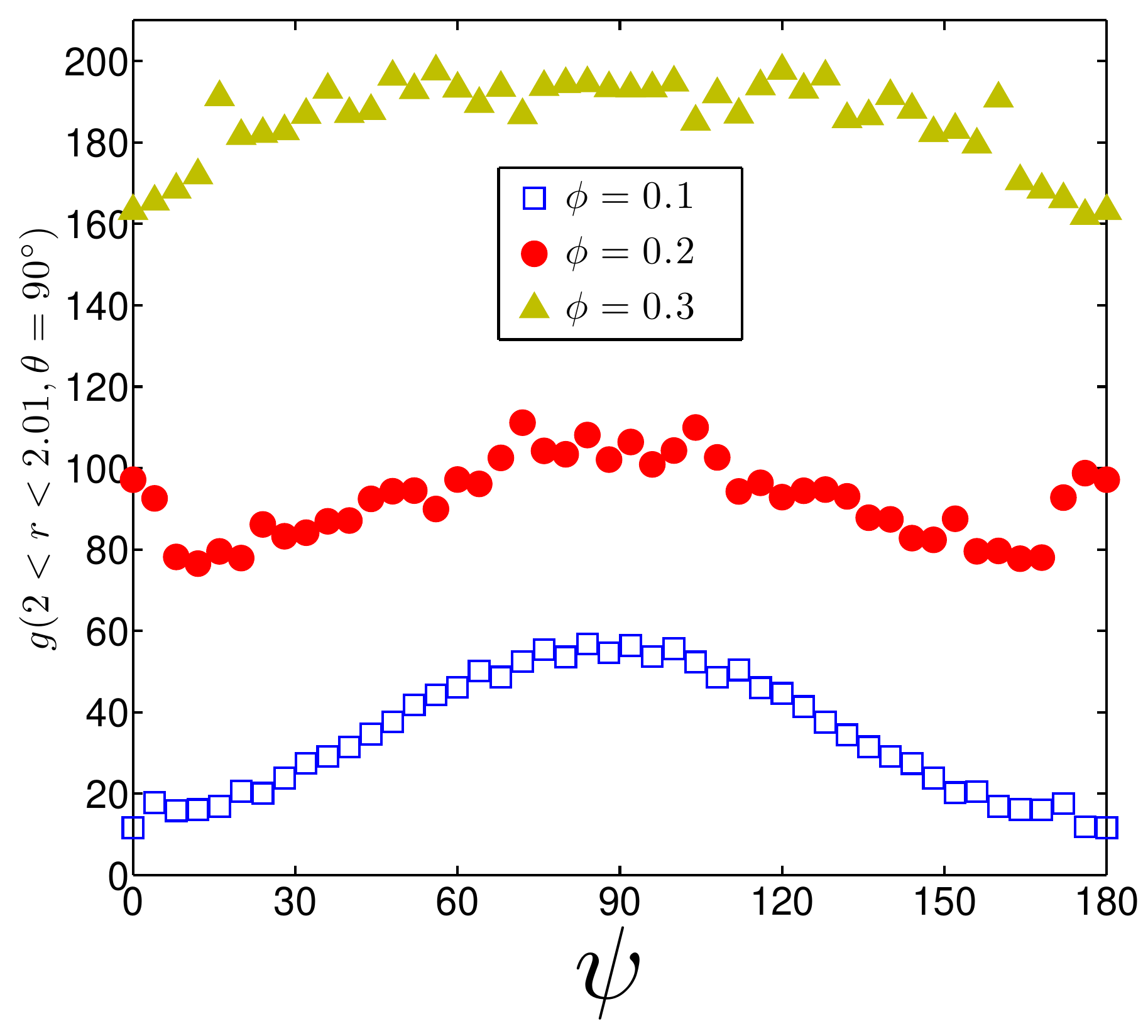}}
\caption{The pair correlation on the vorticity-velocity gradient plane ($g(2 < \boldsymbol{r} < 2.01, \psi, \theta = 90^\circ)$) (\emph{a}) the effect of $Re$ at $\phi = 0.15$ and (\emph{b}) the effect of $\phi$ at $Re = 0.6$.}
 \label{fig:GV}
\end{figure}

Figure ~\ref{fig:G-3D-Re} illustrates $g(2<\boldsymbol{r}<2.01)$ (note that here a slightly larger radial range is taken) in a three-dimensional surface plot, where we can observe the effect of $Re$. At all conditions, a wake is seen in the extensional zone. Figure ~\ref{fig:G-3D-Re} (\emph{a})-(\emph{c}) show that increasing $Re$ reduces the pair correlation around the vorticity axis, with the value on the shear plane thus more pronounced. On the other hand, figure ~\ref{fig:G-3D-Re} (\emph{d}) - (\emph{f}) show that with increasing $\phi$, $g(\boldsymbol{r})$ becomes more uniformly distributed with the angle away from the shear plane. We show this more quantitatively by computing $g(2 < \boldsymbol{r} < 2.01, \psi)$ at $\theta = 90^\circ$, which is on the vorticity-velocity gradient plane.  Figure ~\ref{fig:GV} (\emph{a})  exhibits the effect of $Re$ on $g(2 < \boldsymbol{r} < 2.01, \psi, \theta = 90^\circ)$:  with increasing $Re$, the accumulation of $g(\boldsymbol{r})$ on the plane of shear increases. In figure  ~\ref{fig:GV} (\emph{b}) we see an increase of near-contact $g$ at larger $\phi$, and also observe clearly that larger volume fraction tends to homogenize $g$ in the $\psi$ direction.  \newline 

We turn next to the stress generated by the suspension under shear flow, seeking to relate features of the microstructure to the rheology of the mixture.
  
\section{Suspension stress: rheology}
The stress developed in a sheared suspension at finite particle Reynolds number is considered for volume fractions of $\phi = 0.1$ to $\phi = 0.35$.  We begin with a consideration of the stresslet contribution to the rheology, as this is the dominant Stokes flow contribution and it is useful to first consider how inertia causes it to vary.  Considering the stress mechanisms which are absent at $Re=0$, we find the acceleration stress negligible up to $Re = 5$ while fluctuational transport of momentum, or Reynolds stress, is non-negligible at $Re = O(1)$ and is discussed following the stresslet contribution.

\subsection{Stresslet}
We begin by considering the generation of a hydrodynamic stress as captured by the symmetric first moment of surface tractions, termed the \emph{stresslet}, and described in (\ref{eq:accel}). The stresslet on each particle can be calculated by summing the moment of hydrodynamic force on the boundary nodes. Considering the frame invariance of (\ref{eq:stressletf}), the stresslet of each particle can be calculated as $\emph{sym}(\int_{A_p} \boldsymbol{r}\boldsymbol{\sigma.n} dA_p) \rightarrow \sum_{\boldsymbol{r}_b} \frac{1}{2}(\boldsymbol{f}_b (\boldsymbol{r}_b - \boldsymbol{R}) + (\boldsymbol{r}_b - \boldsymbol{R}) \boldsymbol{f}_b)$. For a very dilute suspension at $Re = 0$, normal stress differences are zero and the stresslet results in the Einstein viscosity contribution, $\eta(\phi) = \mu(1 + \frac{5\phi}{2})$. Inertia  results in normal stress differences even in the dilute limit (Lin {\em et al.} 1970; Mikulencak \& Morris 2004; Vivek Raja, Subramanian \& Koch 2010; Subramanian \emph{et al.} 2011).  Normal stress components of the stresslet are denoted $S_{11}$, $S_{22}$ and $S_{33}$, where $1$, $2$ and $3$ denote flow, velocity gradient and vorticity directions, repectively.  In this section we investigate the role of the stresslet on the first ($S_{11} - S_{22}$) and second ($S_{22} - S_{33}$) normal stress differences, the particle pressure [$\frac{1}{3}S_{ii} = \frac{1}{3}(S_{11} + S_{22} + S_{33}) $] and the viscosity ($1 +  S_{12}$). \newline

\begin{figure}
\centering
\subfigure[$N_1$]{\includegraphics[totalheight=0.2\textheight,]{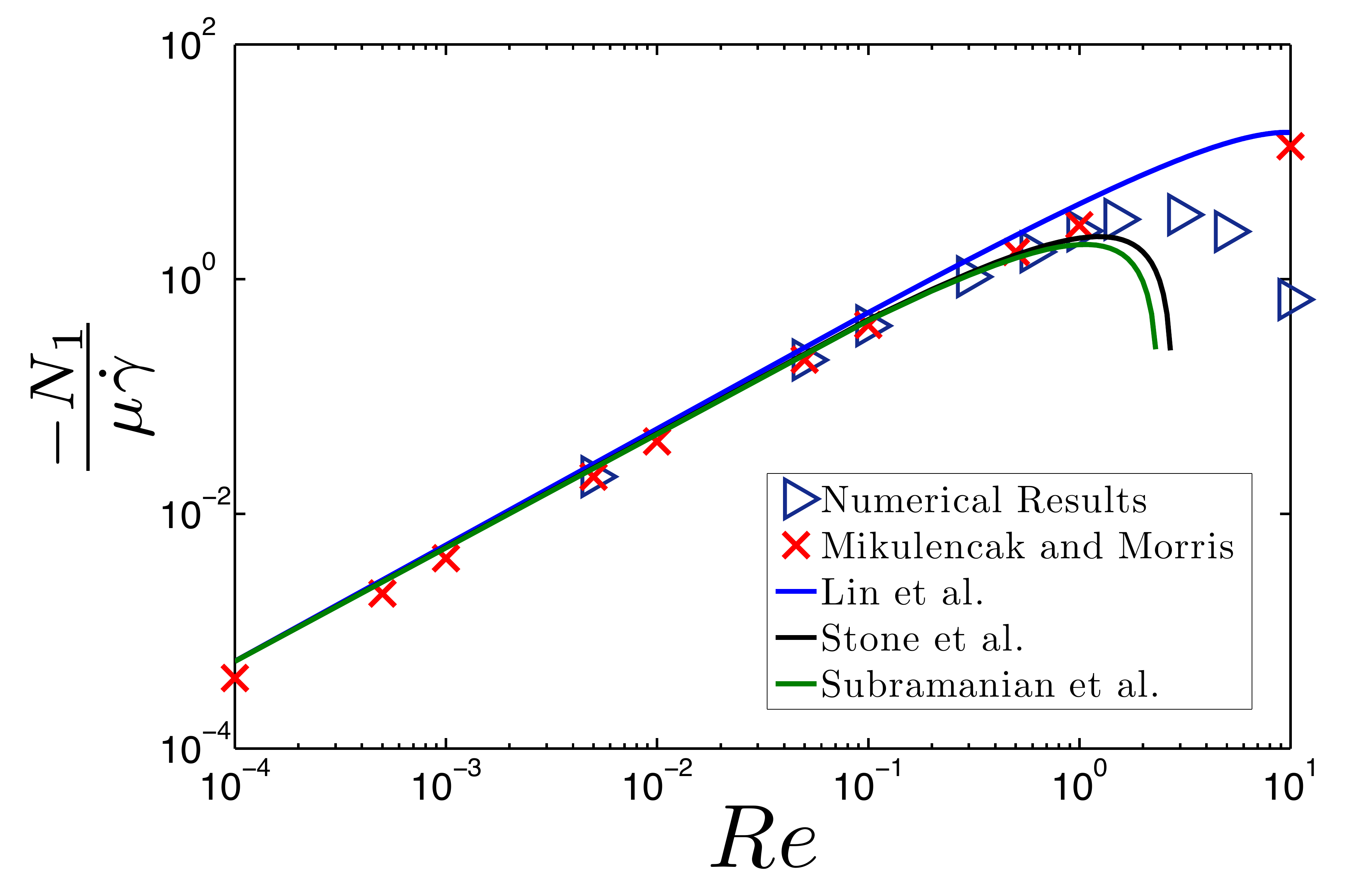}}
\subfigure[$N_2$]{\includegraphics[totalheight=0.2\textheight,]{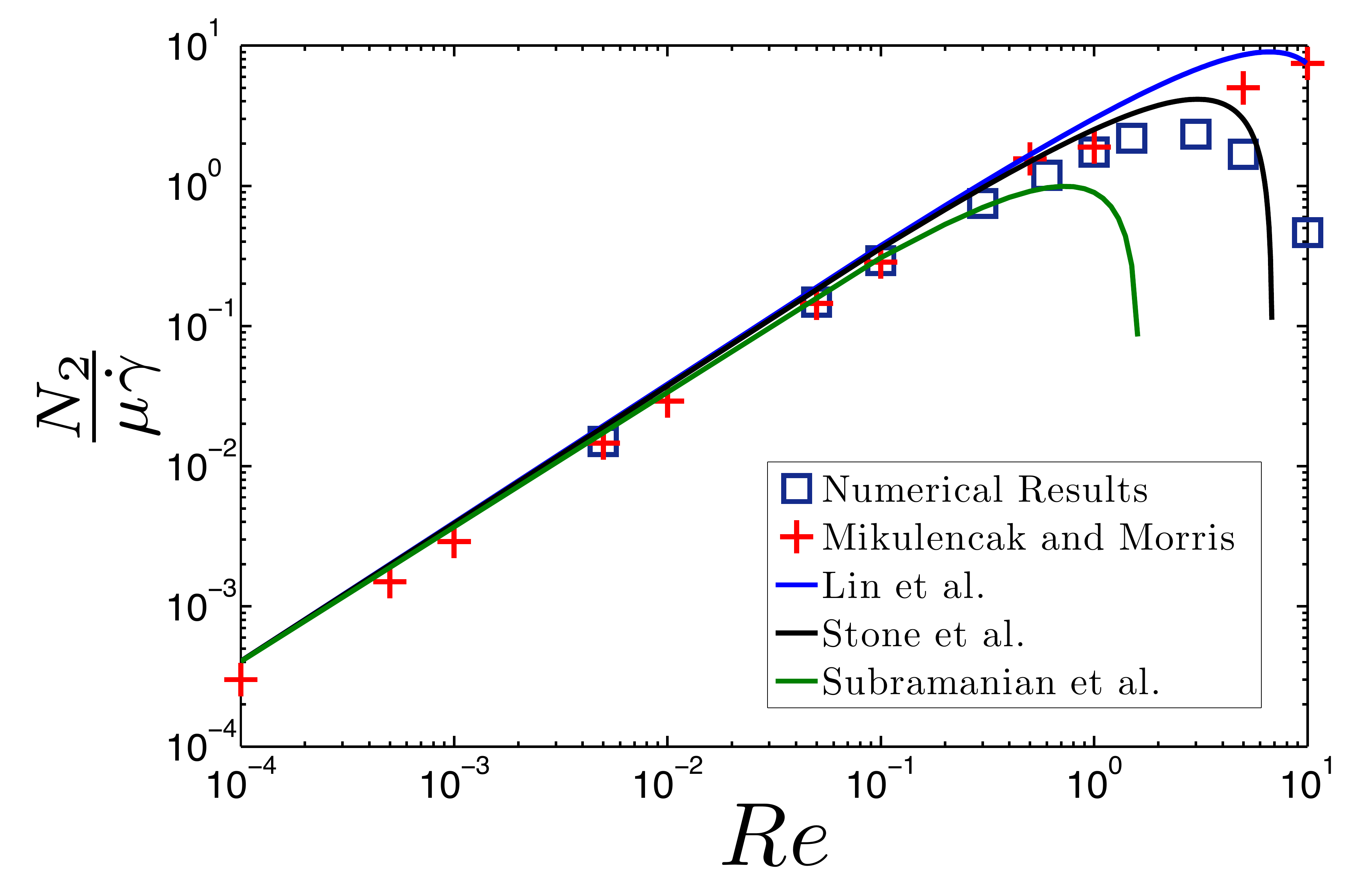}}
\subfigure[$\Pi$]{\includegraphics[totalheight=0.2\textheight,]{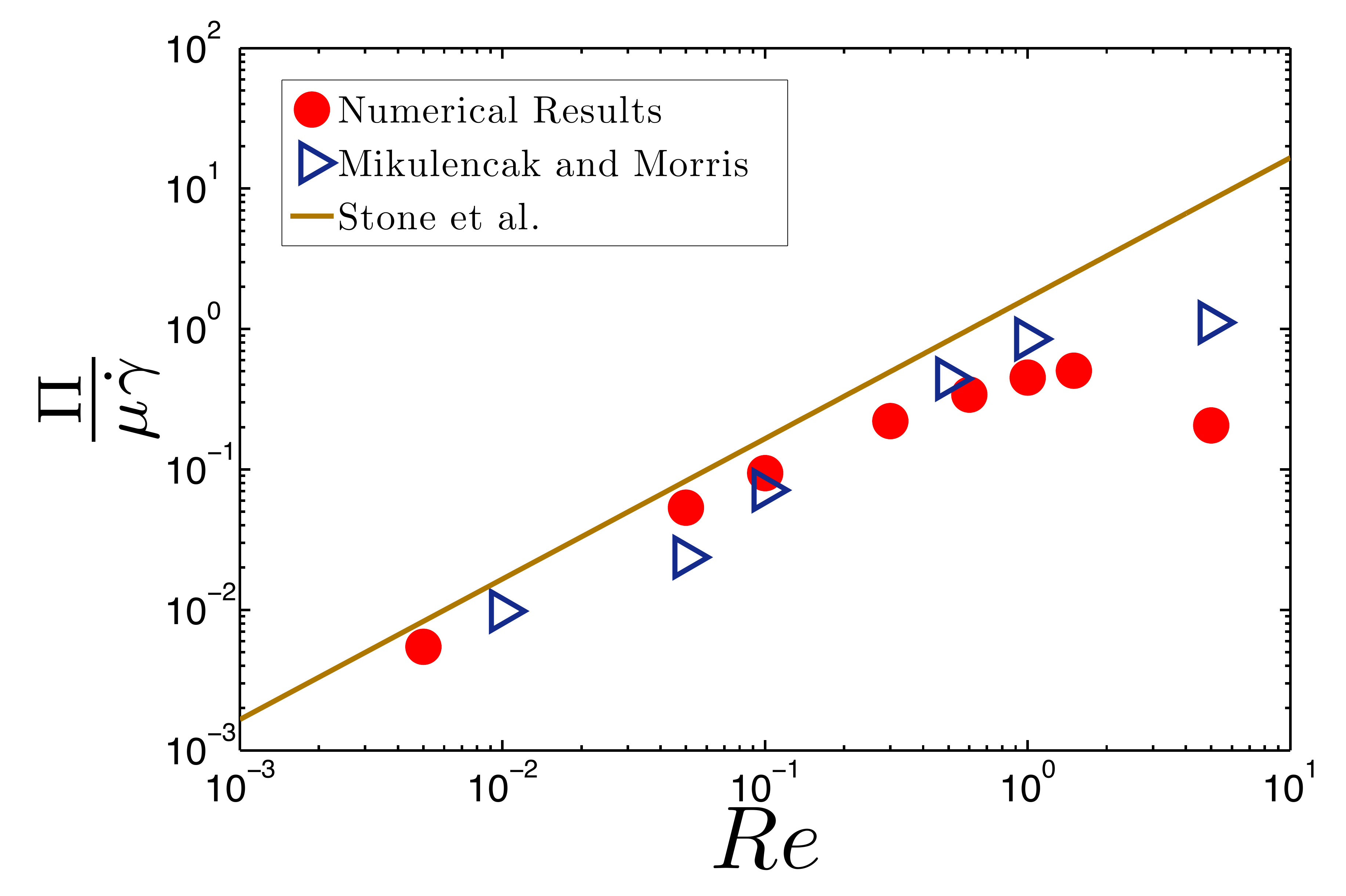}}
\subfigure[$S_{12}$]{\includegraphics[totalheight=0.2\textheight,]{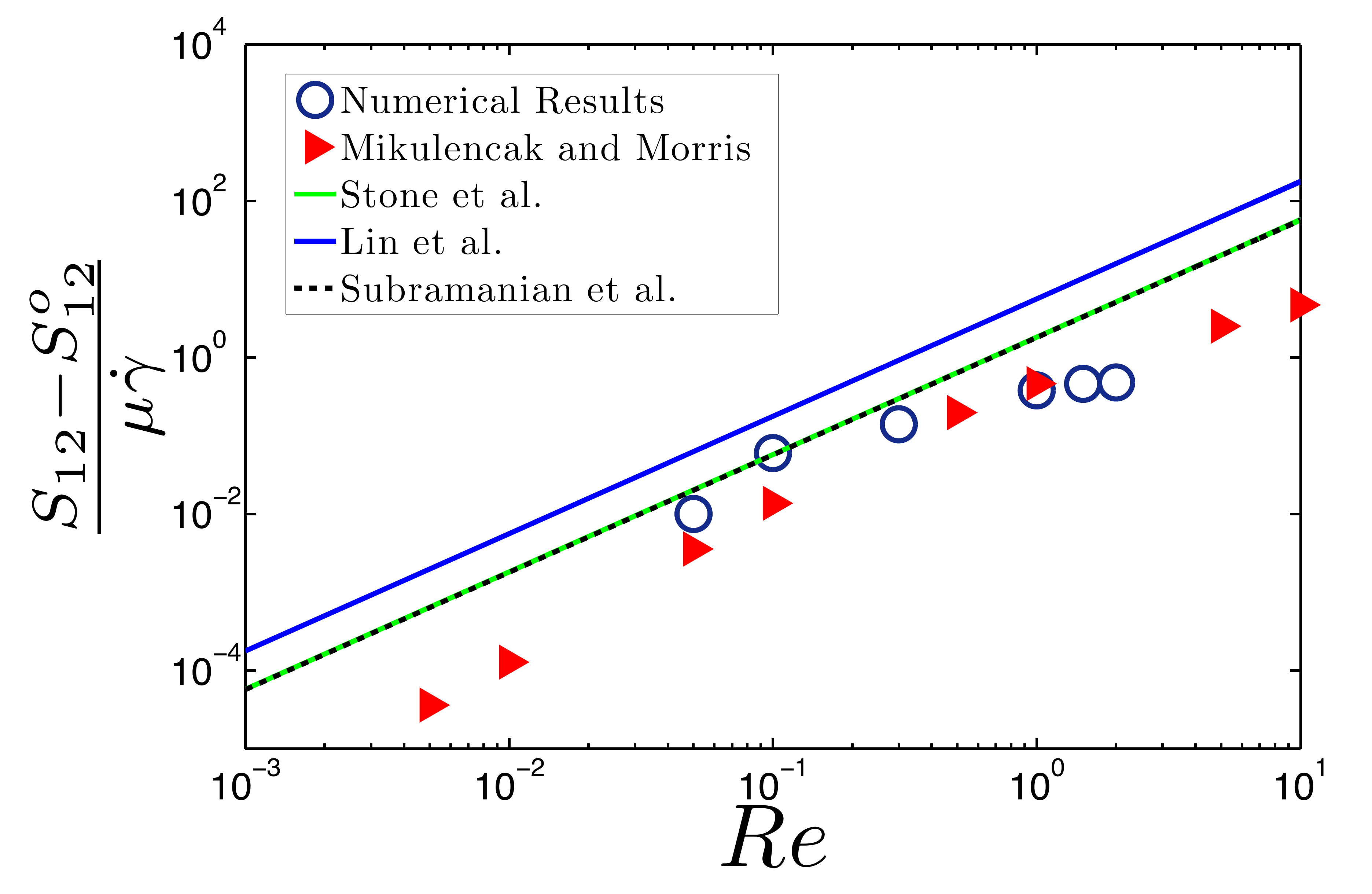}}
\caption{Results of the present numerical calculation scheme for the (\emph{a}) first normal stress difference, (\emph{b}) second normal stress difference, (\emph{c}) particle pressure and (\emph{d}) shear component of the stresslet tensor of a single particle, compared with theoretical predictions of Lin \emph{et al} (1970), Stone \emph{et al.} (2001), Subramanian \emph{et al.} (2011) and numerical calculations of Mikulencak \& Morris (2004)}.
 \label{fig:Single-S}
\end{figure}

In order to validate our calculation scheme, we compare our results with finite element based numerical calculations of Mikulencak \& Morris (2004) and analytical solutions of Lin \emph{et al.} (1970), Stone, Brady \& Lovalenti  (2001) (see also the reproduced results in Mikulencak \& Morris 2004 or Vivek Raja \emph{et al.} 2010) and Subramanian \emph{et al.} (2011) for the stresslet of a single particle in simple-shear flow. The latter reference presents $O(Re^{3/2})$ corrections to the previous theoretical predictions. As we observe in figure~\ref{fig:Single-S} (\emph{a}) and (\emph{b}), our $N_1$ and $N_2$ results are in very good agreement with previous numerical calculations and analytical solutions for $Re \leqslant 1$. For larger $Re$, while calculations of Mikulencak \& Morris show a negligible reduction of normal stress differences with increasing $Re$, our results begin to drop; thus our results are similar to theoretical predictions of Stone \emph{et al.} and Subramanian \emph{et al.}, which predict a drop of $N_1$ and $N_2$ at this range of $Re$. There is close agreement between our numerical calculations of $N_2$ and the analytical solution of Stone \emph{et al}. It is observed in figure~\ref{fig:Single-S} (\emph{c}) that the numerical calculations and theoretical predictions of the particle pressure agree at $Re < 0.1$ but there is a deviation from theoretical predictions at larger $Re$, while the normal stress difference predictions agree with theory to $Re\approx 1$. For larger $Re$, our calculations are close to the results of Mikulencak \& Morris except for high $Re$, where we observe a more pronounced drop of particle pressure. \newline

To study the variation of the shear stress contribution of a single particle, $S_{12}$, in more detail, we present the difference between $S_{12}$ at each $Re$ and $S_{12}^0$, defined as the value at $Re= 0.005$, with the result normalized by $\mu\dot{\gamma}$ to yield the increment in viscosity.  All theories against which comparison is made predict the added viscosity is proportional to $Re^{\frac{3}{2}}$. However, $S_{12} - S_{12}^0$ appears to be proportional to $Re^{\frac{3}{2}}$ for $Re < 0.1$ and to  $Re^{1}$ beyond $Re = 0.1$.  Mikulencak \& Morris and Yeo \& Maxey (2013) observed similar scalings using finite elements and force-coupling calculations, respectively. \newline 

\begin{figure}
\centering
\subfigure[$N_1$ versus $\phi$ at various $Re$]{\includegraphics[totalheight=0.21\textheight,]{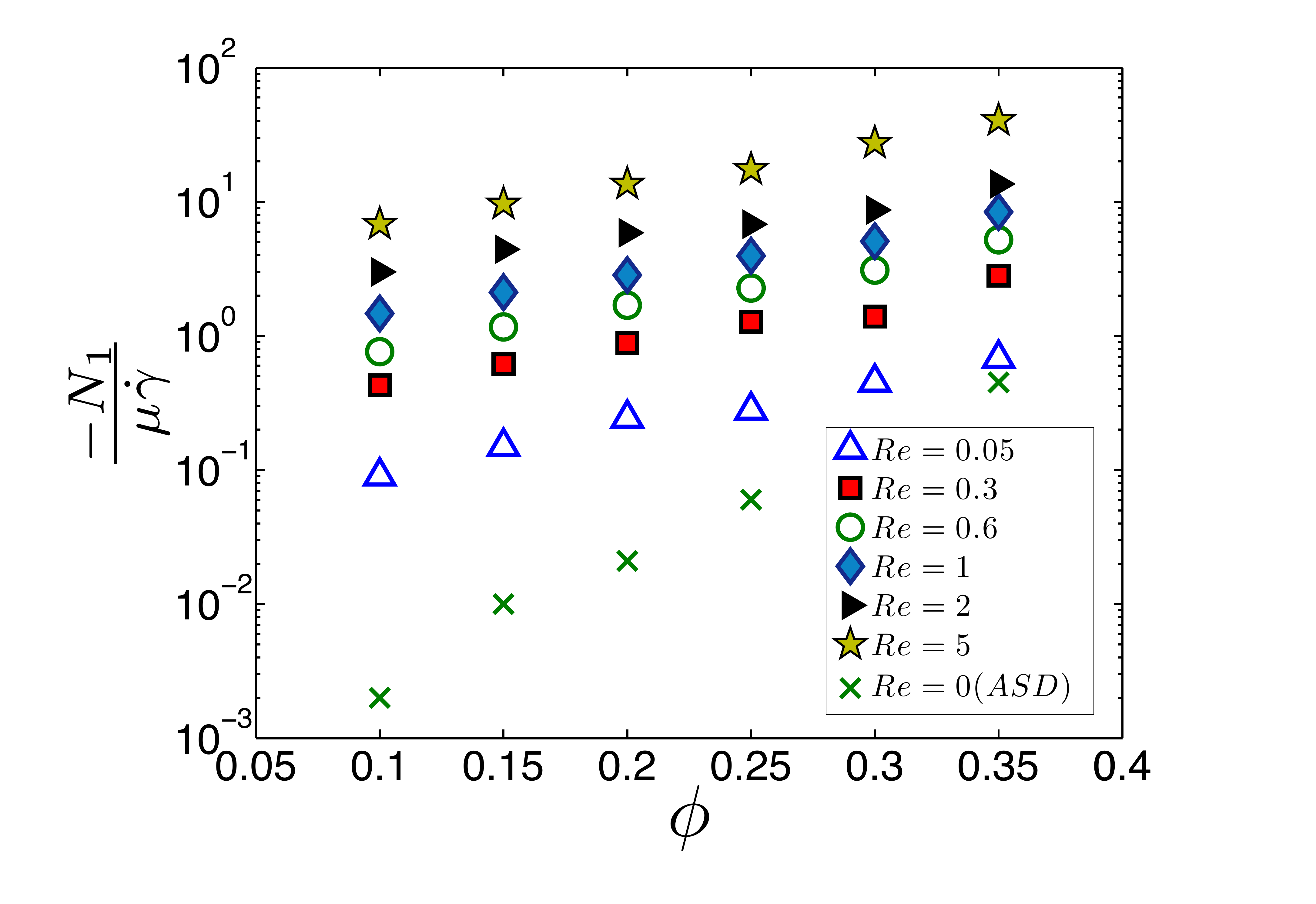}}
\subfigure[$N_2$ versus $\phi$ at various $Re$]{\includegraphics[totalheight=0.21\textheight,]{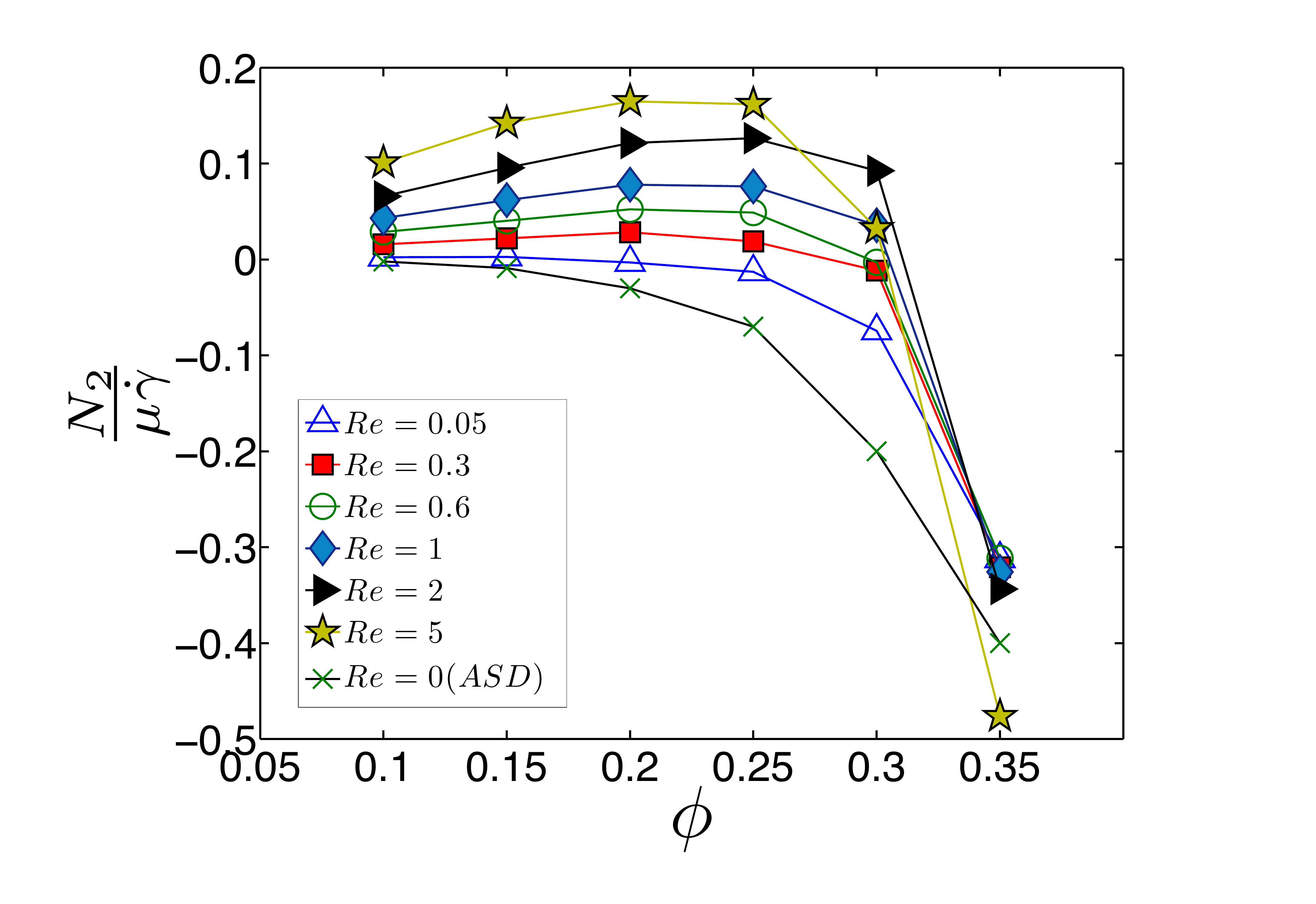}}
\subfigure[$\Pi$ versus $\phi$ at various $Re$]{\includegraphics[totalheight=0.21\textheight,]{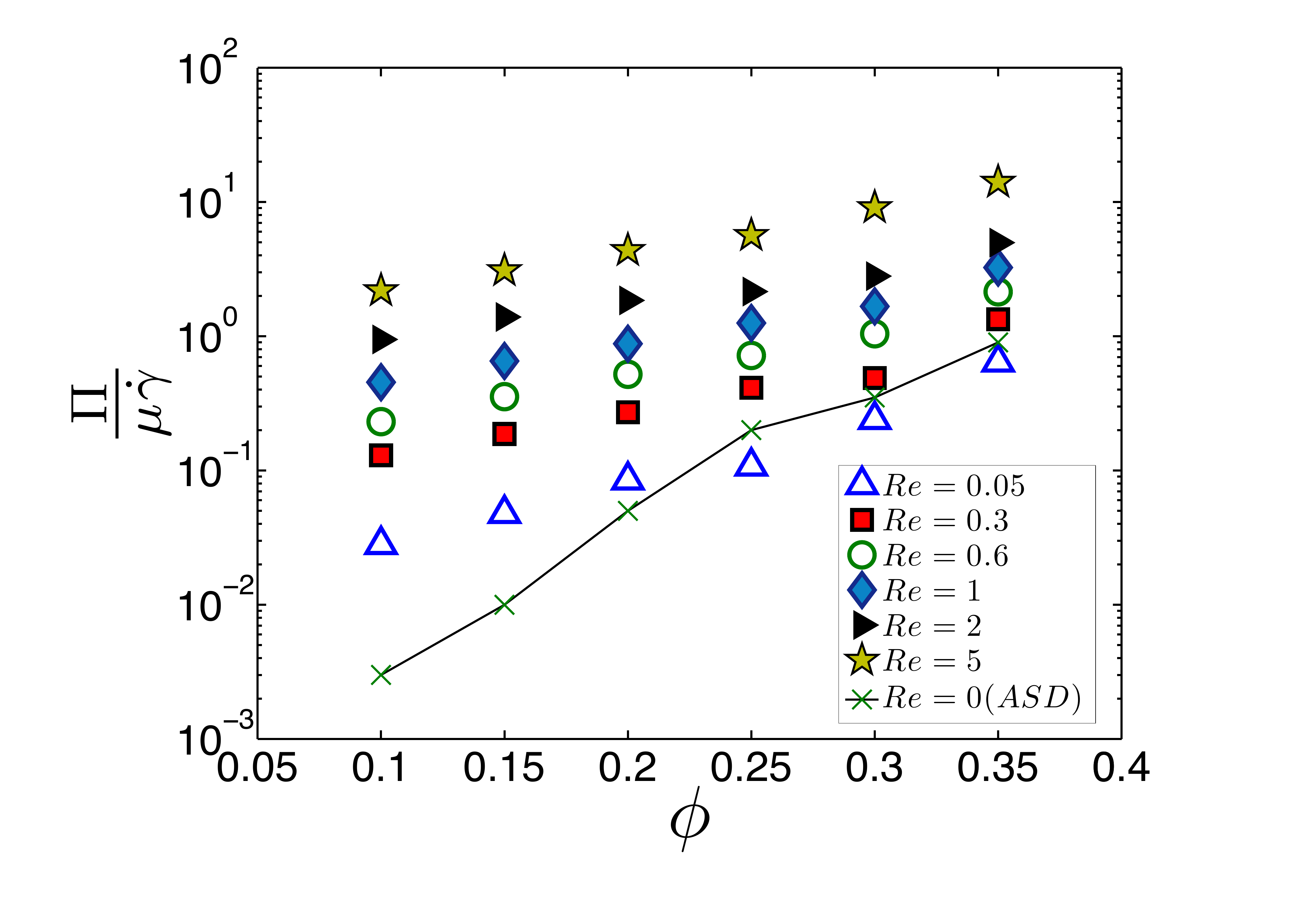}}
\subfigure[$\mu_r$ versus $\phi$ at various $Re$]{\includegraphics[totalheight=0.21\textheight,]{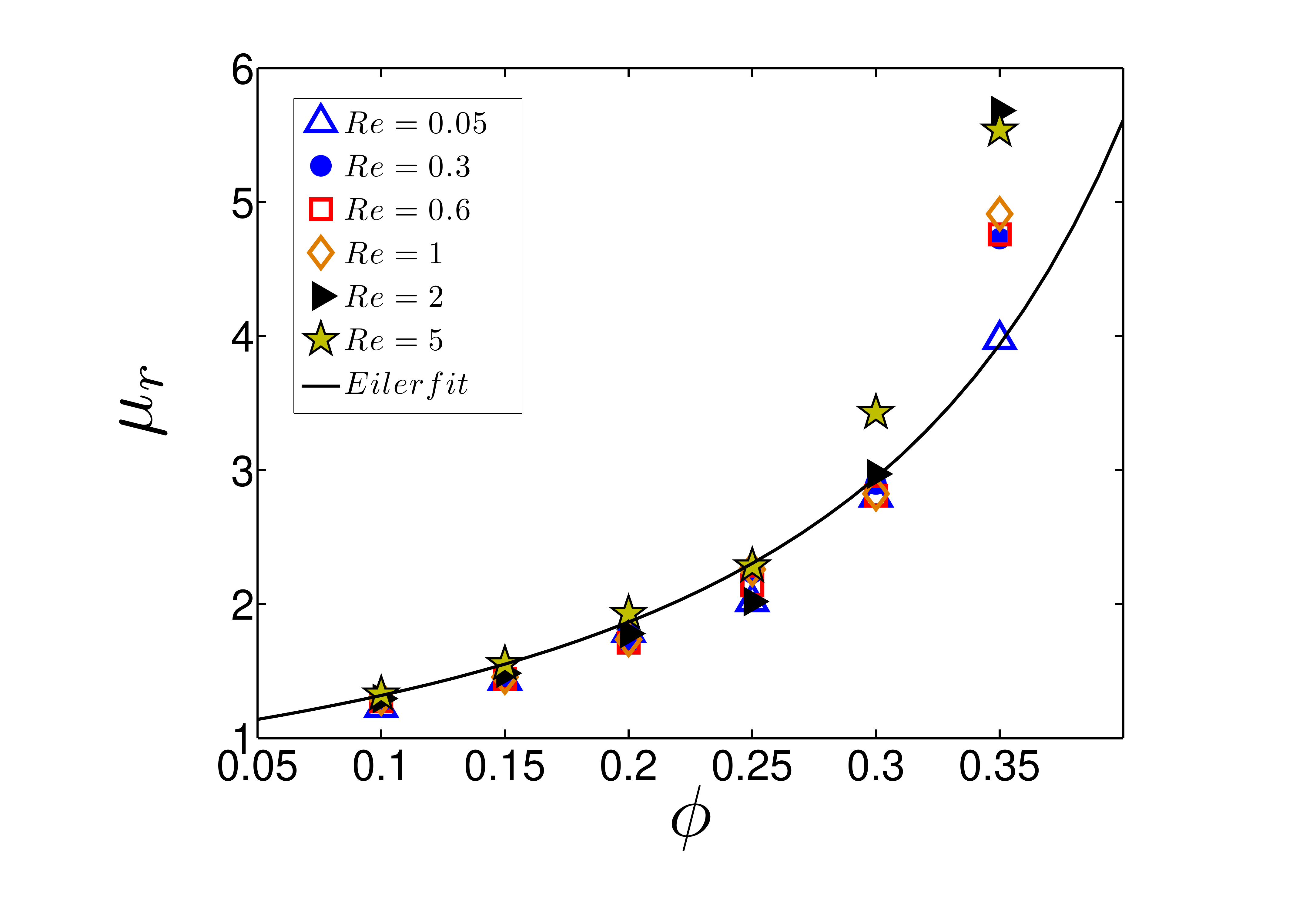}}
\caption{(\emph{a}) First normal stress difference (\emph{b}) second normal stress difference and (\emph{c}) particle pressure and (\emph{d}) viscosity generated by the stresslet contribution in suspensions at various $\phi$ and $Re$. }
 \label{fig:Normal}
\end{figure}

\begin{figure}
\centering
\subfigure[$\phi = 0.15$]{\includegraphics[totalheight=0.21\textheight,]{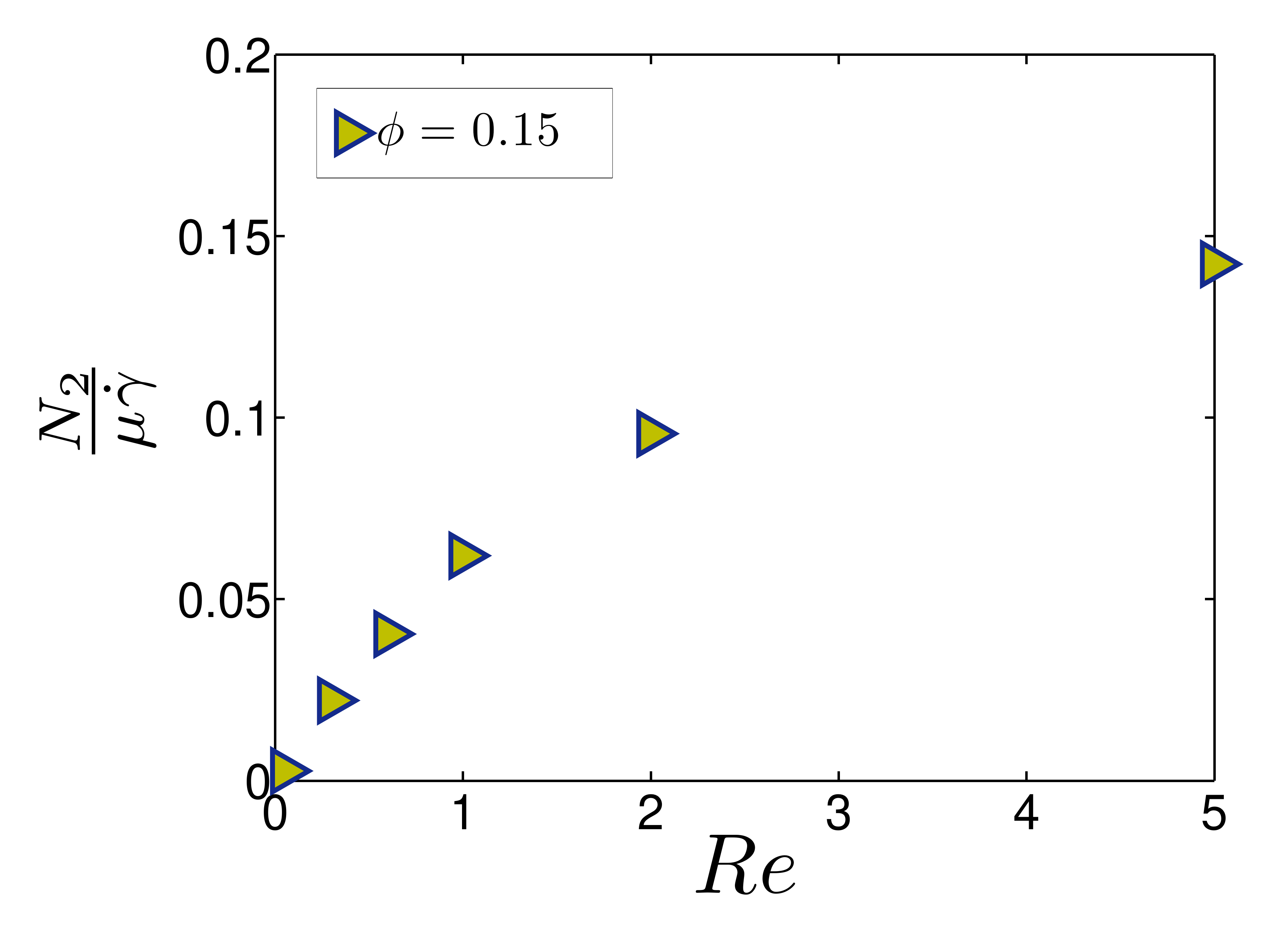}}
\subfigure[$\phi = 0.35$]{\includegraphics[totalheight=0.21\textheight,]{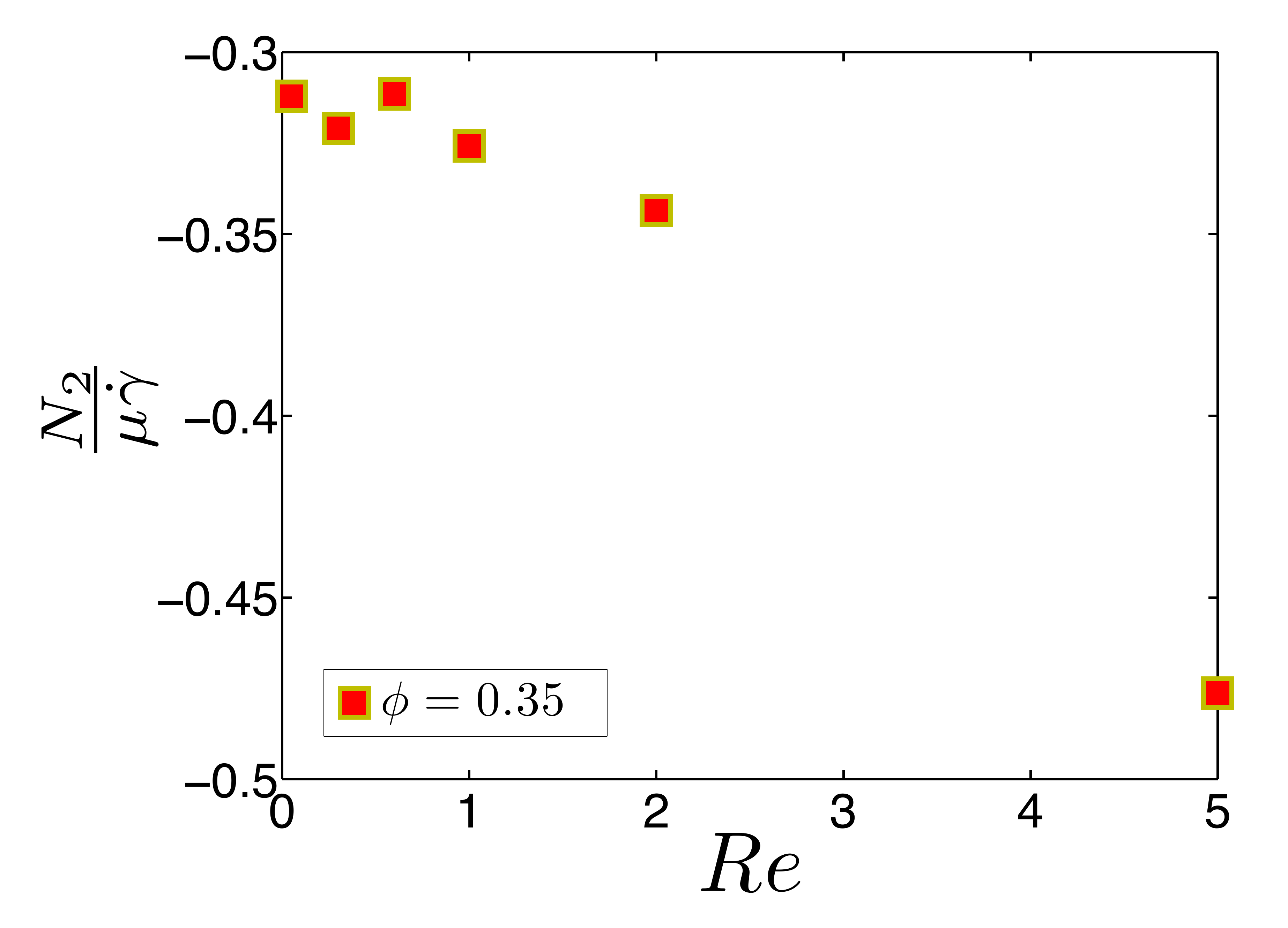}}
\caption{$N_2$ versus $Re$ at (\emph{a}) $\phi = 0.15$ and (\emph{b}) $\phi = 0.35$. }
 \label{fig:NIN}
\end{figure}

The bulk suspension rheology at finite $\phi$ generally shows features observed for single-body, results just outlined for small $\phi$, with qualitative changes arising at larger $\phi$.  We consider the bulk suspension rheology here.   In figure ~\ref{fig:Normal} we report the first and second normal stress differences, the particle pressure and the viscosity of suspensions for various $\phi$ and $Re$. In KM08, a preliminary evaluation of viscometric functions has been presented. We have calculated the viscometric functions with more enhanced statistics and compare our findings with the KM08 data. Similar to their calculations, $N_{1}$ is always negative and its magnitude increases with increasing $\phi$ and $Re$. At $\phi = 0.35$ and $Re = 0.05$, the results become close to $Re = 0$ results obtained by Accelerated Stokesian Dynamics (ASD) simulations of Sierou \& Brady (2004). For $N_2$, we find positive values for $0.1 < \phi < 0.2$, which increase in magnitude at larger $Re$, consistent with dilute theory and calculation which predict a positive $N_2$ at finite $Re$. With further increase of $\phi$ a deviation from dilute theory is observed and a decline of $N_2$ is seen at all $Re$. The volume fraction beyond which $N_{2}$ starts to decrease depends on the $Re$:  at $Re = 2$,  $N_2$ decreases at $\phi = 0.25$, while at $Re = 0.05$ the decline begins at $\phi = 0.15$. We observe for $Re = 5$, $N_2$ begins to decline at $\phi = 0.2$ and takes on the largest magnitude negative value at $\phi = 0.35$. We show the variation of $N_2$ with $Re$ for  $\phi = 0.15$ and $0.35$ in figures  ~\ref{fig:NIN} (\emph{a}) and (\emph{b}). We observe that in general, the magnitude of $N_2$ increases with $Re$: at $\phi = 0.15$, $N_2>0$ increases with $Re$, while at $\phi = 0.35$, $N_2<0$ but the larger $Re$ values have the largest magnitude. The transition of $N_2$ from positive to negative values in an inertial suspension is indicative of the alteration of suspension rheology from low $\phi$ where fluid mechanical inertia is dominant, to large $\phi$ where the excluded volume plays the dominant role. However, at large $\phi$, inertia is seen to amplify the effect of excluded volume.

In KM08, the patterns for variation of $N_1$ and $N_2$ with $Re$ and $\phi$ are similar except that our $N_1$ values are smaller at each $\phi$. In KM08 the transition to negative $N_2$ with $Re$ occurs at a smaller $\phi $. It should be mentioned here that for calculation of stress the effect of the walls and formation of the particle layers close to the walls should be considered. We observed a negligible difference in stresslets by excluding the particle layer close to the walls. However, the excluded volume due to the particle-wall interactions should be considered for calculation of the volume fraction. Therefore, the actual $\phi$ is $0.03 - 0.05$ larger than the value which is obtained by computing the ratio of the volume occupied by particles and the total volume. This point was not considered in the calculation of the volume fraction in KM08. By making this correction and increasing the volume fractions of KM08 by about $0.05$, the numbers reported in KM08 are quite close to our results. Yeo \& Maxey (2013) have recently reported the normal stress differences of suspensions at finite inertia employing a Force Coupling Method. In their calculations, the particle stress is the sum of surface traction stresses and hard sphere potential forces, which are applied on particles when the center-to-center distance is less than $2.004a$. Although the patterns in our calculations for $N_2$ are similar to their findings, we find larger $|N_1|$. \newline

We demonstrate the particle pressure of the suspensions in figure~\ref{fig:Normal} (\emph{c}) where the particle pressure increases with increasing $\phi$ and $Re$, consistent with KM08 calculations. Figure ~\ref{fig:Normal} (\emph{d}) exhibits the viscosity together with the Eiler's fit given by \newline

\begin{equation}
\mu_r = (1 + \frac{\frac{1}{2}\left[\mu\right]\phi}{1-\frac{\phi}{\phi_m}})^{2},
\end{equation}
where we have chosen $[\mu] = 2.5$ and $\phi_m = 0.63$, similar to Yeo \& Maxey (2013). In KM08, $[\mu] = 3$ and $\phi_m = 0.58$ were chosen as fitting parameters. The simulated viscosity is in good agreement with the empirical relationship, and shows only a weak $Re$ dependence for the largest $\phi$. We have also computed the viscosity by calculating the wall shear stress and obtained similar results. The magnitude of $\mu_r$ is in close agreement with Yeo \& Maxey (2013).  \newline

\begin{figure}
\centering
\subfigure[$S_{11}$ versus $\phi$ at various $Re$]{\includegraphics[totalheight=0.19\textheight,]{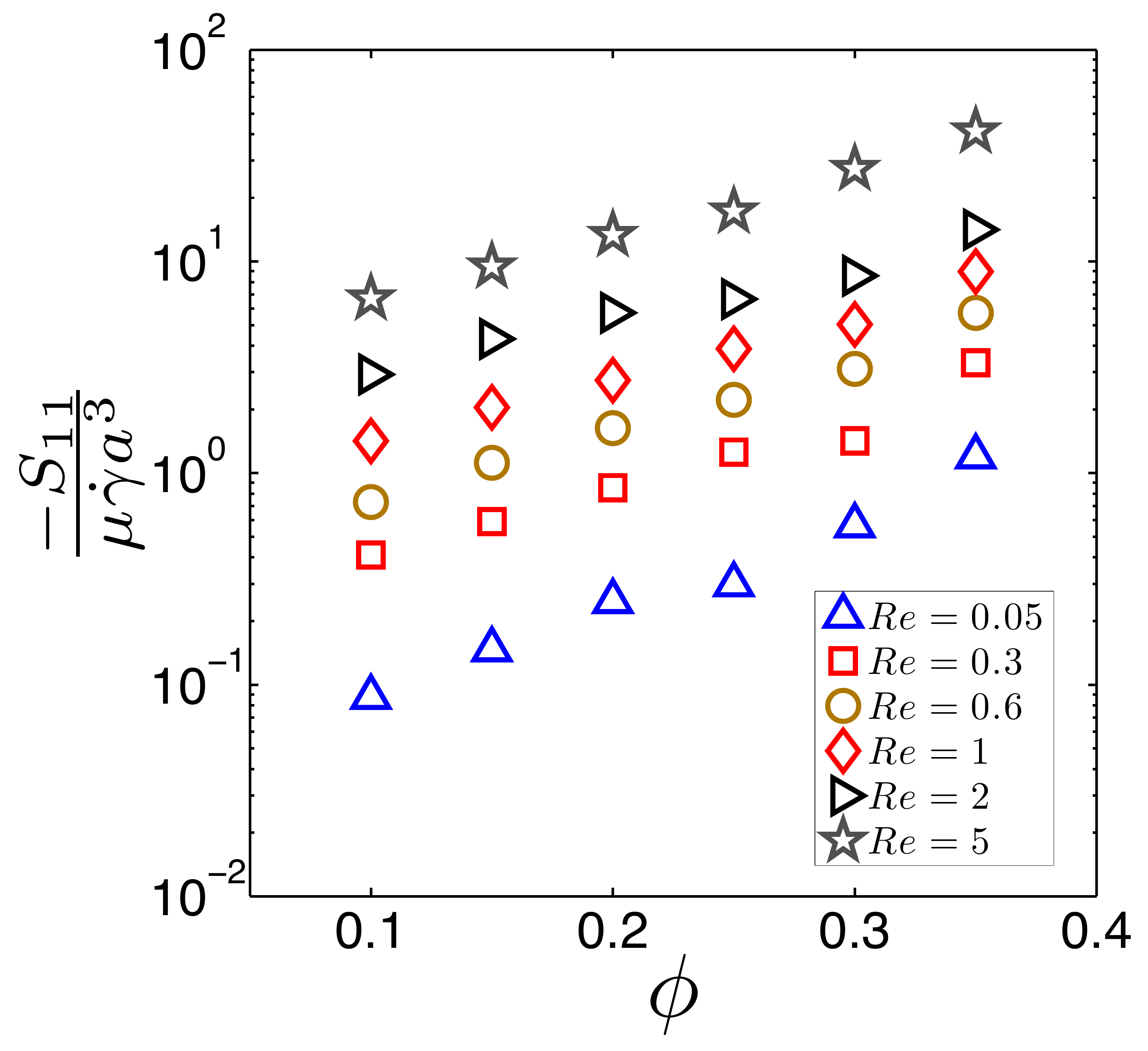}}
\subfigure[$S_{22}$ versus $\phi$ at various $Re$]{\includegraphics[totalheight=0.19\textheight,]{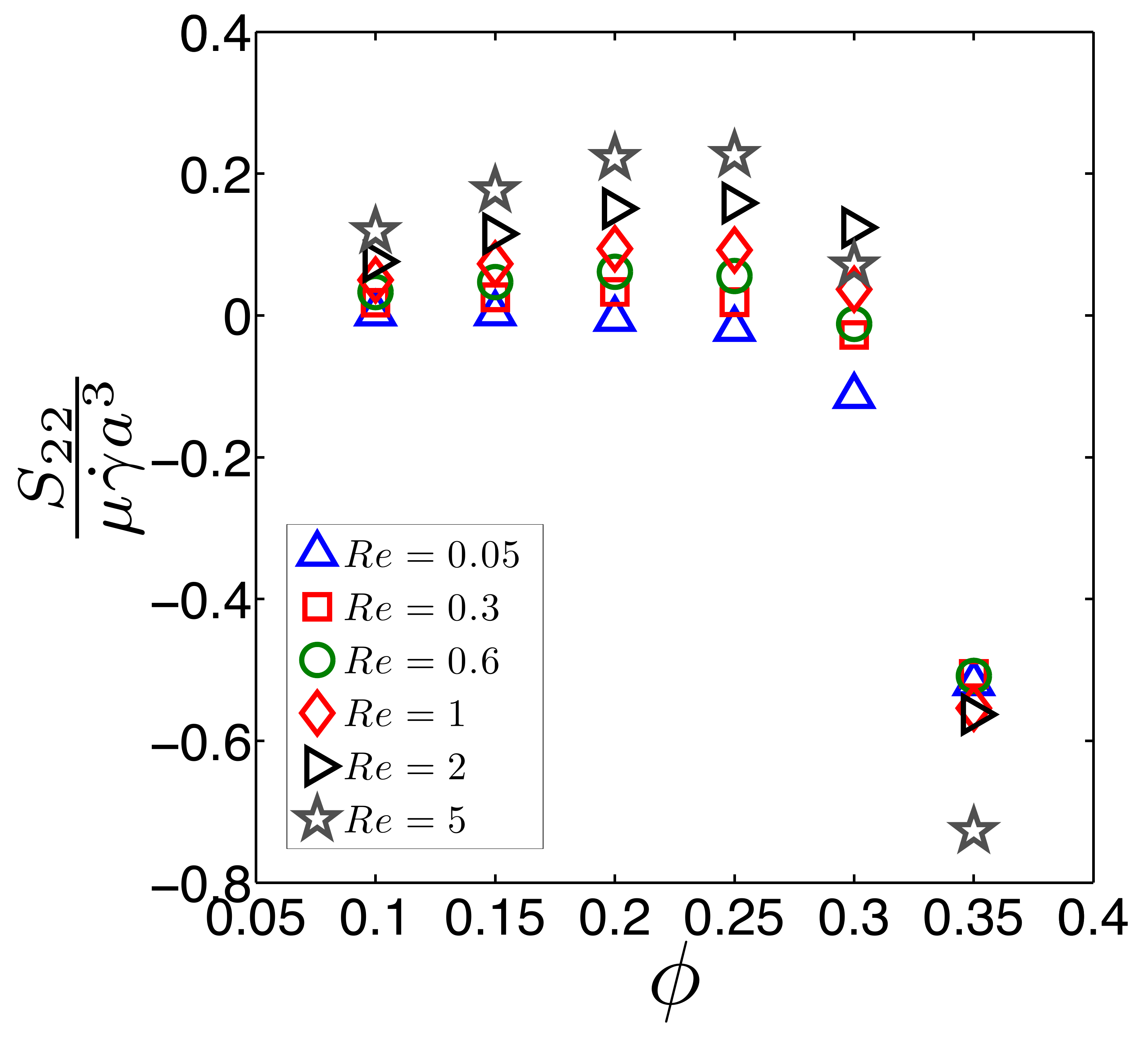}}
\subfigure[$S_{33}$ versus $\phi$ at various $Re$]{\includegraphics[totalheight=0.19\textheight,]{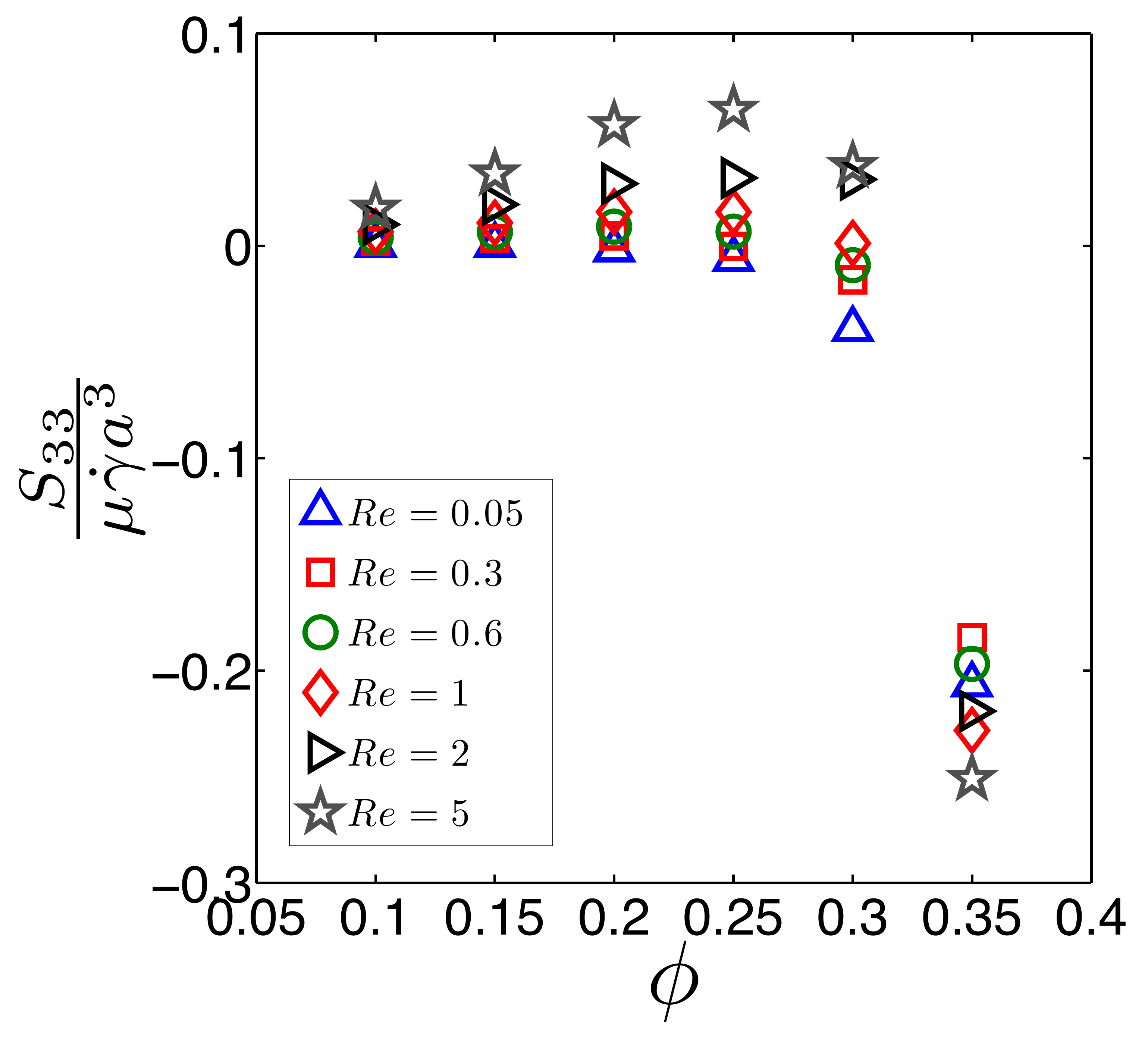}}
\caption{Average value of normal components of the stresslet tensor in (\emph{a}) $x$, (\emph{b}) $y$ direction and (\emph{c}) $z$ directions.}
 \label{fig:S-Comp}
\end{figure}

Figure ~\ref{fig:S-Comp} illustrates the individual normal components of the average stresslet, i.e. $S_{ii}$ for $i =1,2$ or 3, at different $Re$ and $\phi$. Here we choose to present the stress information in terms of the mean value of the individual particle quantity contributing to the bulk stress, i.e. the stresslet, as we will in the following consider how the surface tractions leading to the stresslet. 
For suspensions at finite inertia, $S_{11}$ is negative and its magnitude increases with $Re$ and $\phi$.  We find that $S_{22}$ and $S_{33}$ are positive for solid fractions up to $ \phi = 0.2$ or 0.25 (this result depends on $Re$) followed by a decline toward negative values at larger $\phi$. For suspensions at $Re = 0.05$, the decrease begins around $\phi = 0.15$. $S_{22}$ and $S_{33}$ of the suspensions at $Re = 5$ rapidly decrease at $\phi = 0.25$. The magnitude of $S_{22}$ and $S_{33}$ of the suspensions is the largest. Based on the magnitude of normal components,  $S_{11}$ is the dominant contribution to $N_1$ and the particle pressure. The negative $S_{11}$ is indicative of \emph{compressive} stress along the flow direction, an effect which we find to be quite pronounced at larger $Re$ and $\phi$. \newline

\begin{figure}
\centering
\subfigure[$S_{11}$  at $Re = 0.05$]{\includegraphics[totalheight=0.135\textheight,]{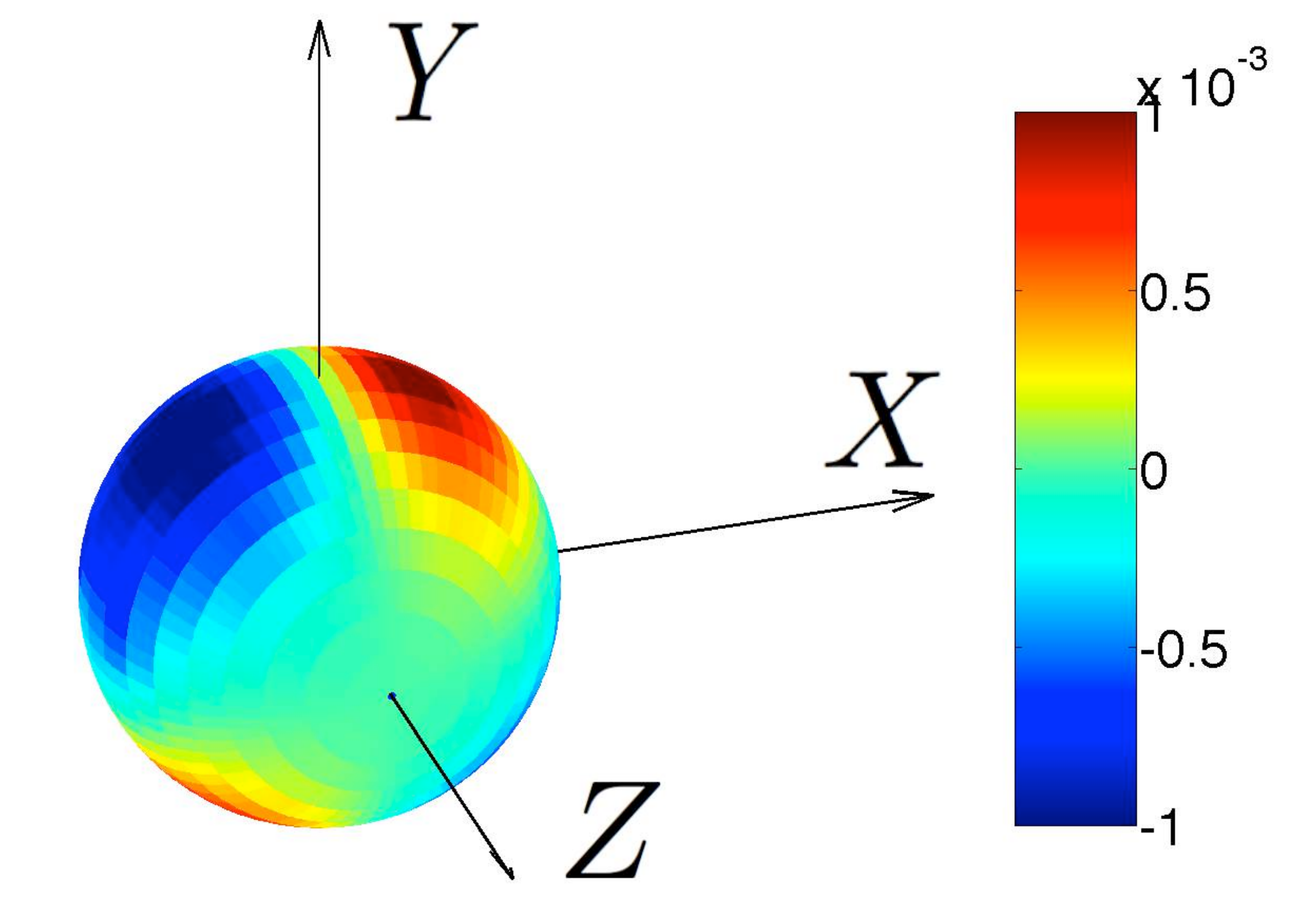}}
\subfigure[$S_{11}$  at $Re = 0.1$]{\includegraphics[totalheight=0.135\textheight,]{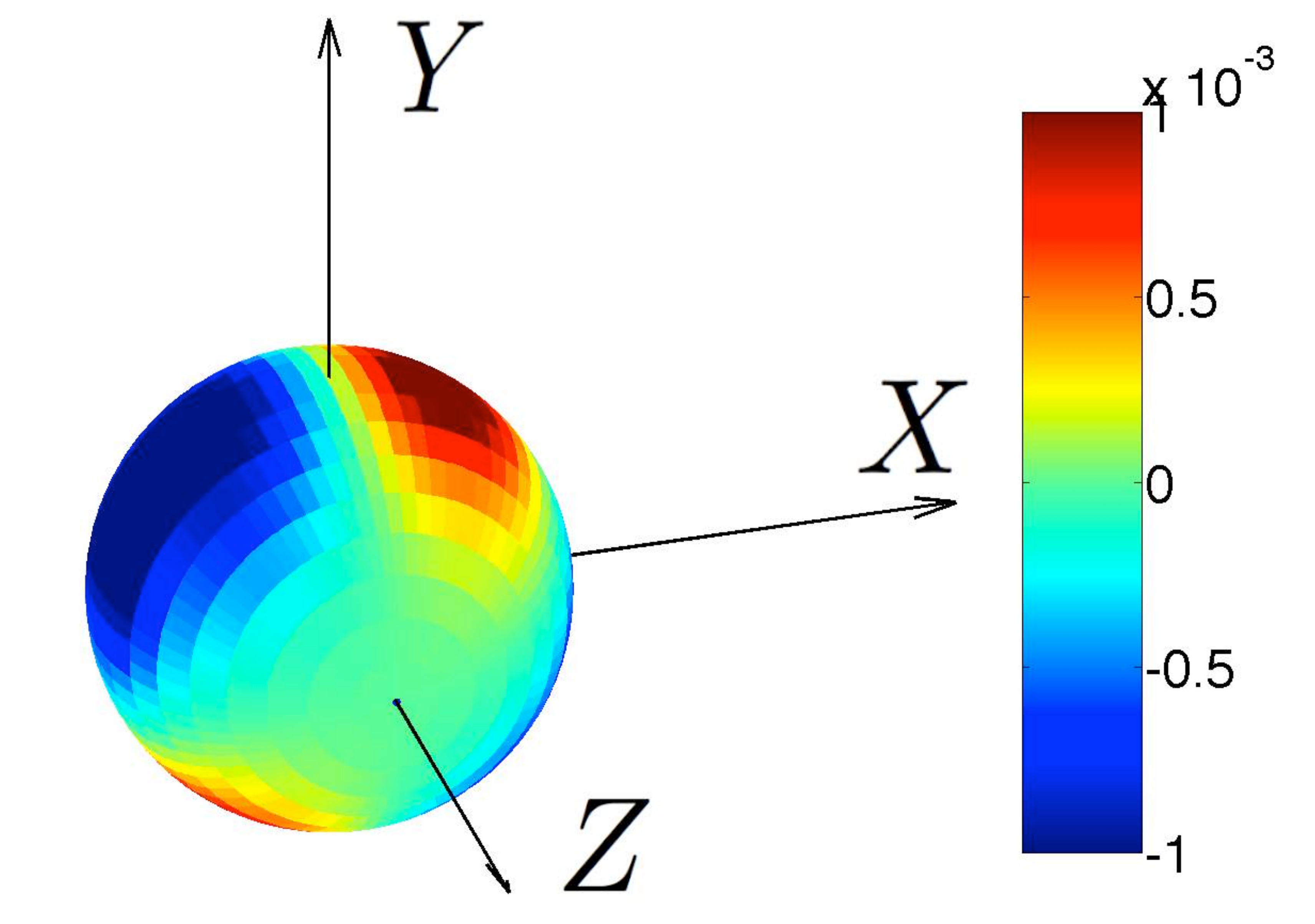}}   % Re = 0.1 NOT 0.2
\subfigure[$S_{11}$  at $Re = 0.6$]{\includegraphics[totalheight=0.135\textheight,]{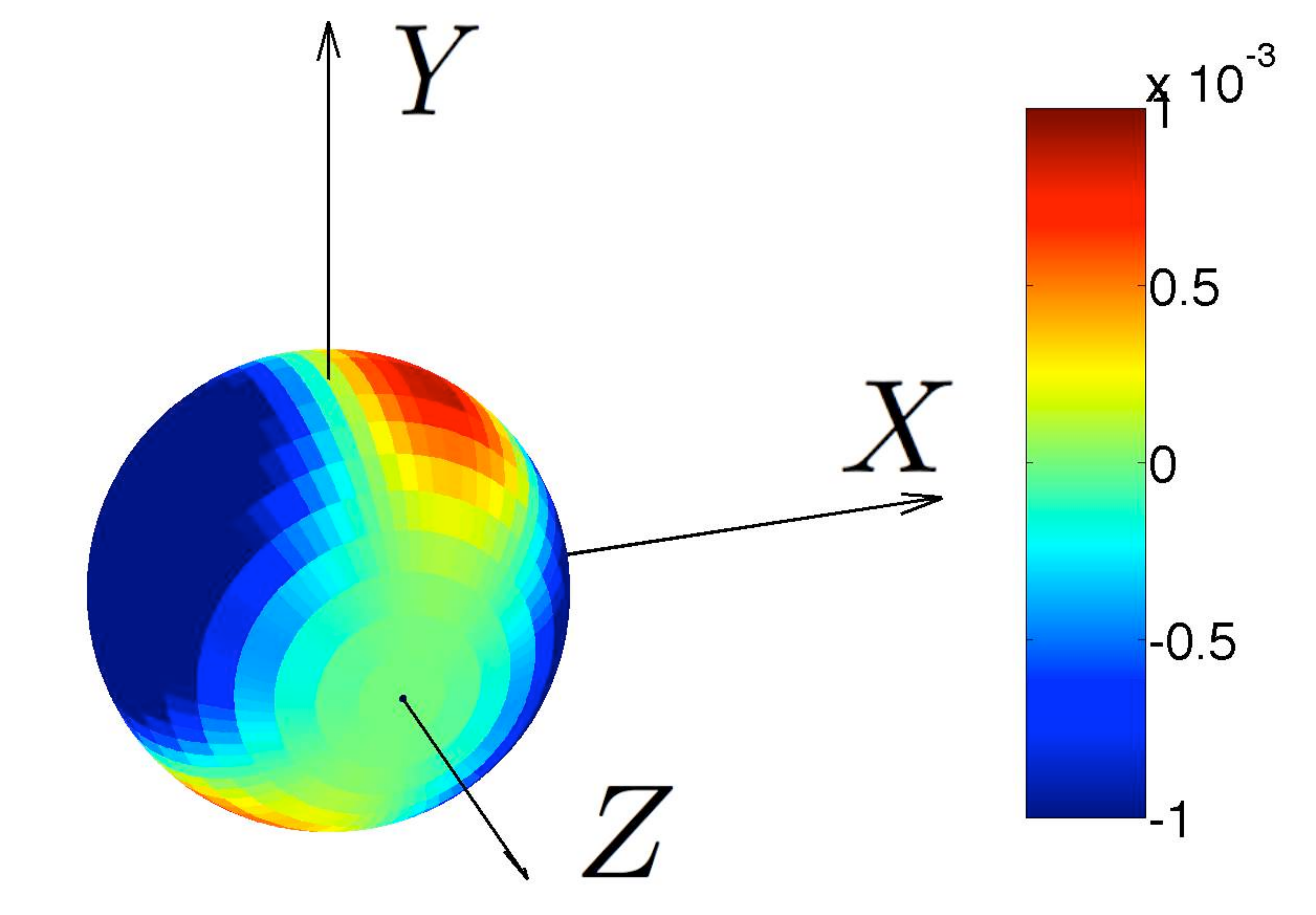}}
\subfigure[$S_{22}$  at $Re = 0.05$]{\includegraphics[totalheight=0.14\textheight,]{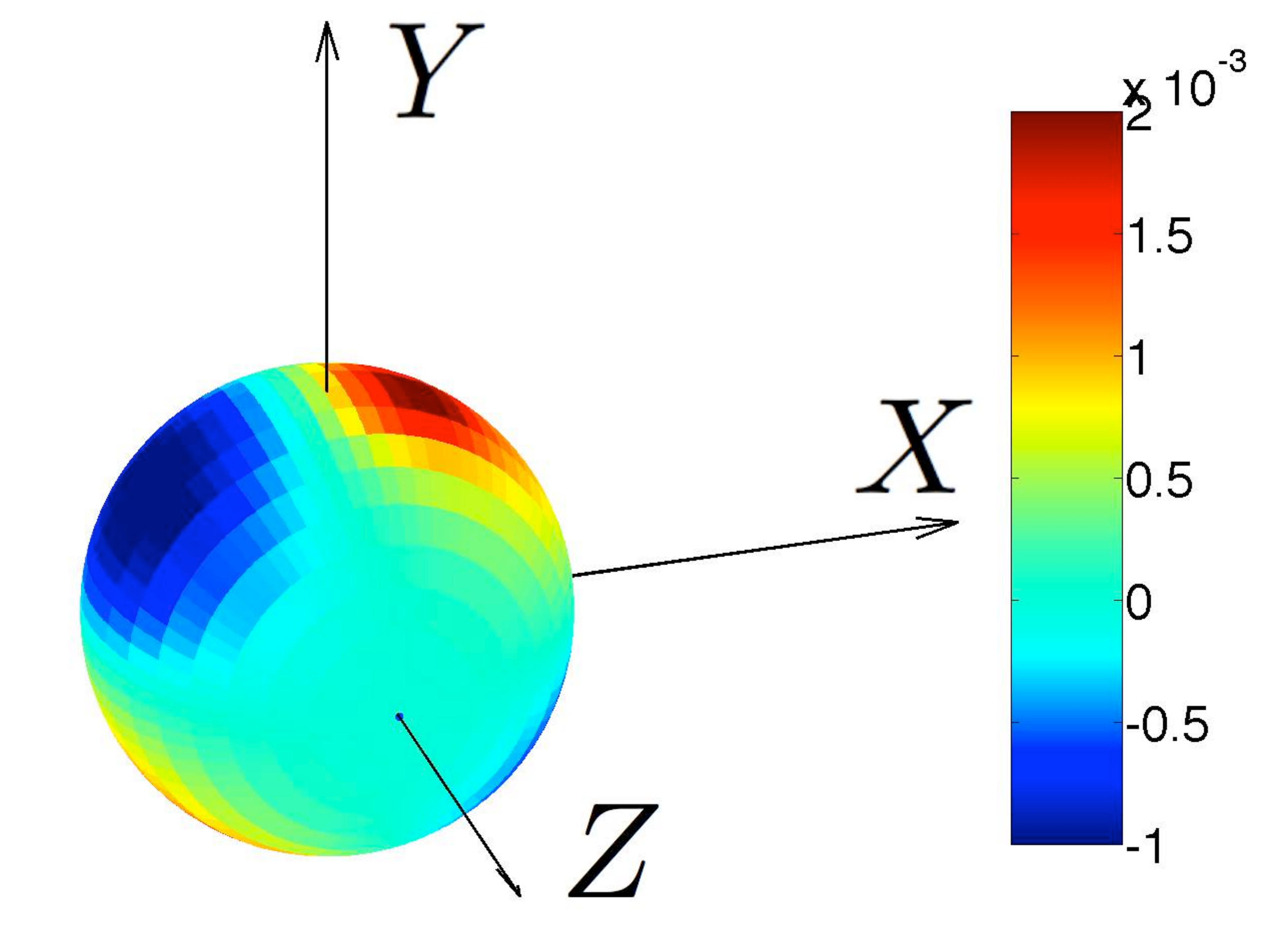}}
\subfigure[$S_{22}$  at $Re = 0.1$]{\includegraphics[totalheight=0.14\textheight,]{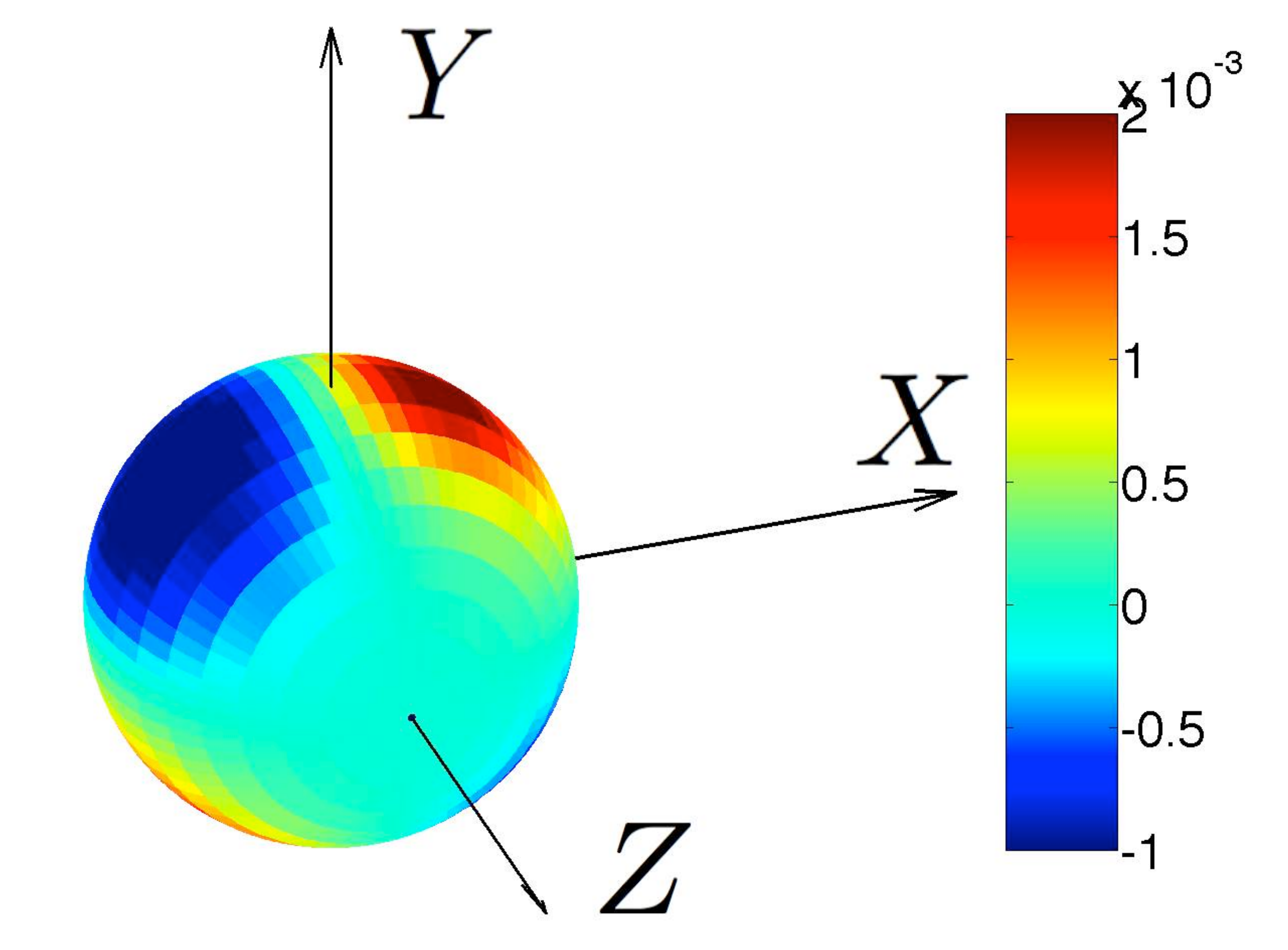}} % Re = 0.1 NOT 0.2
\subfigure[$S_{22}$  at $Re = 0.6$]{\includegraphics[totalheight=0.14\textheight,]{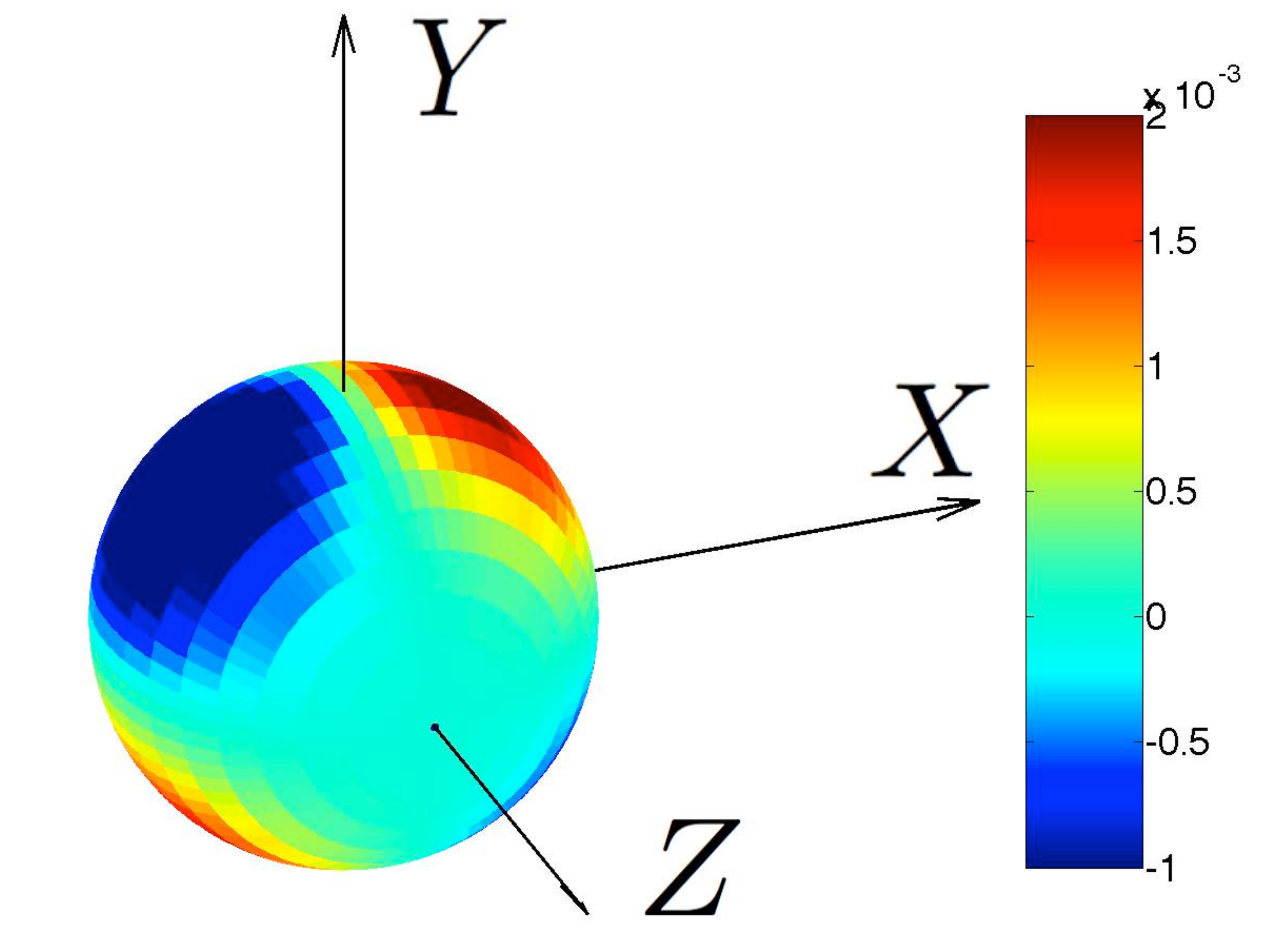}}
\subfigure[$S_{33}$  at $Re = 0.05$]{\includegraphics[totalheight=0.14\textheight,]{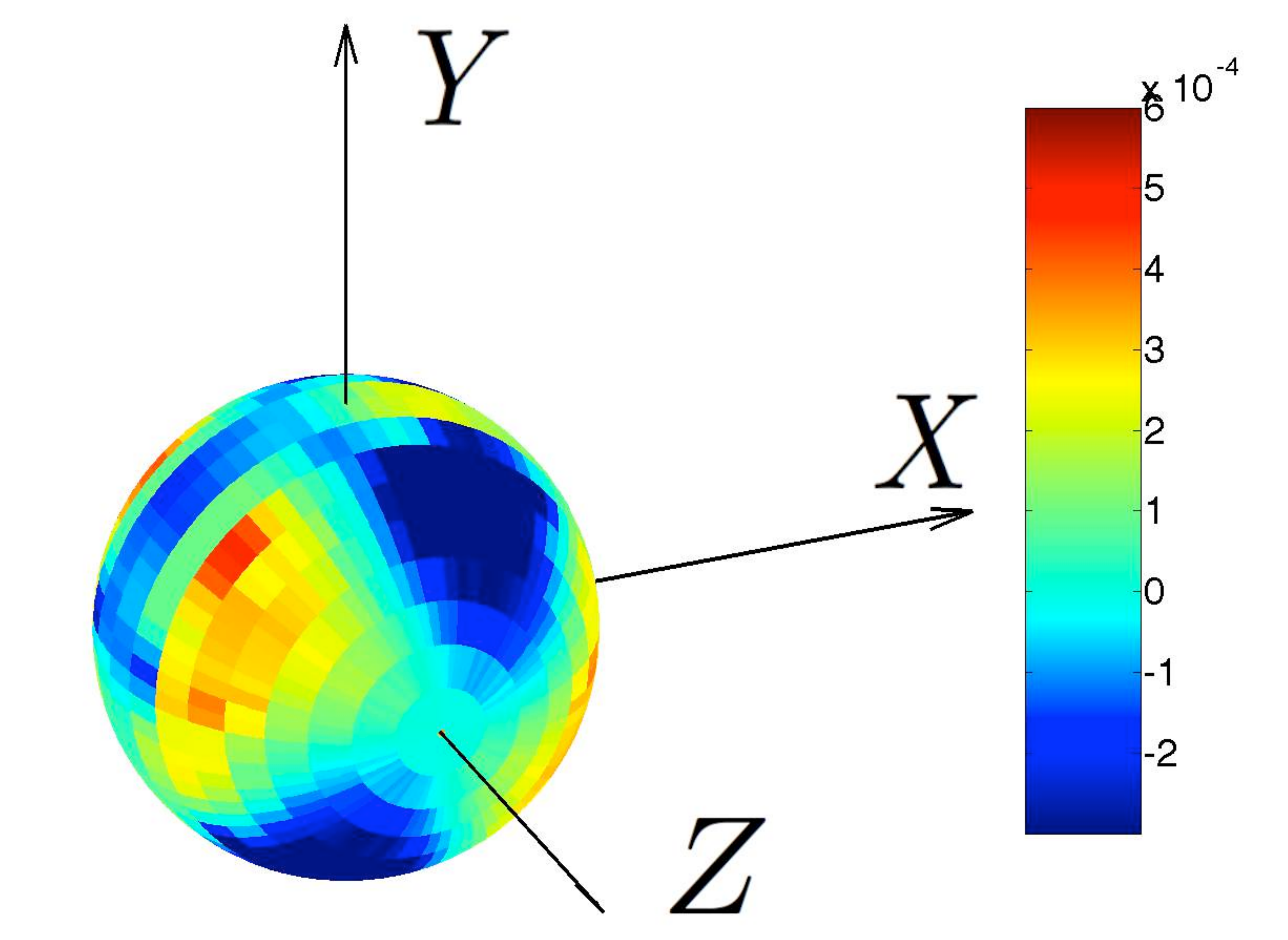}}
\subfigure[$S_{33}$  at $Re = 0.1$]{\includegraphics[totalheight=0.14\textheight,]{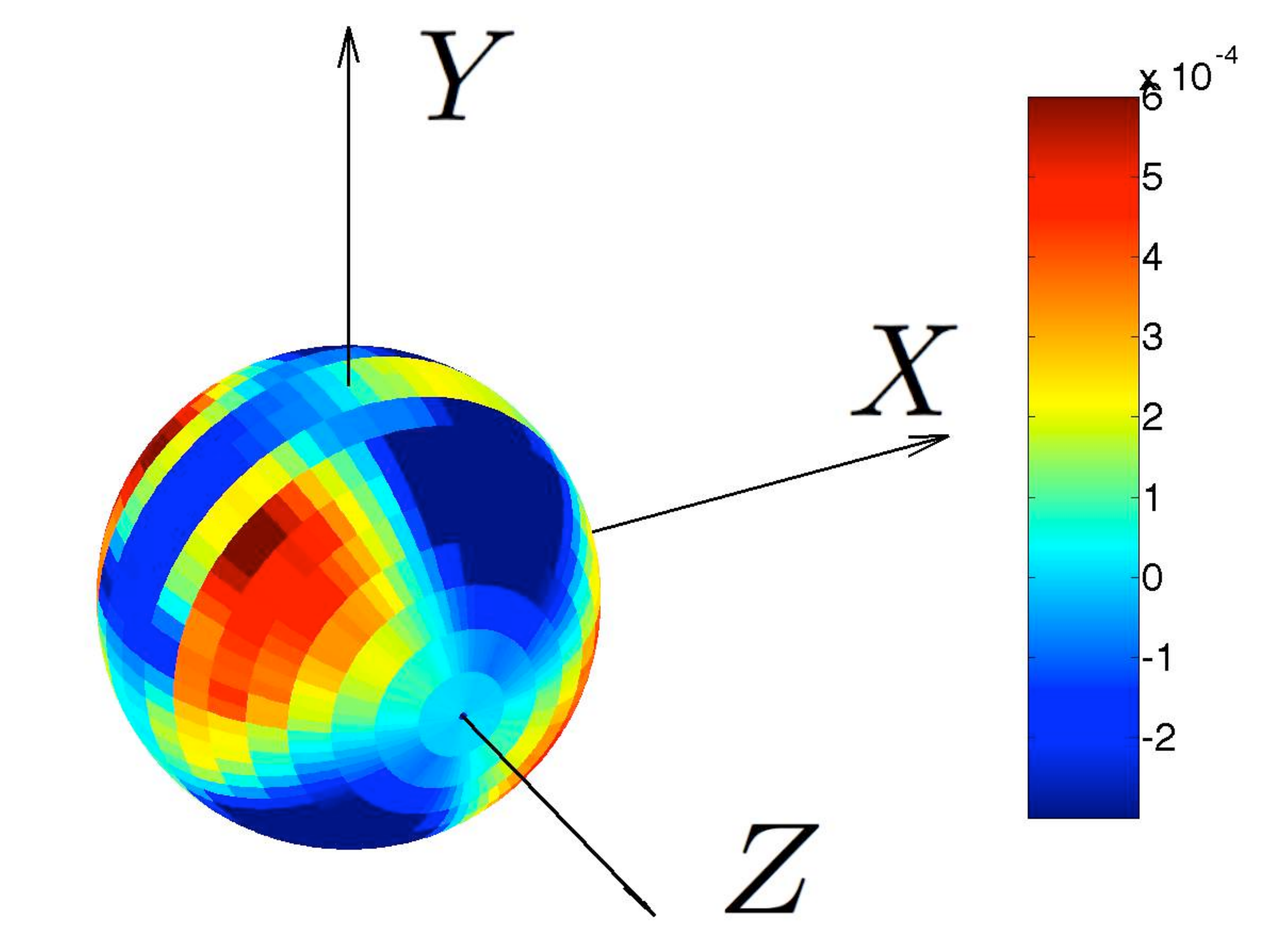}} % Re = 0.1 NOT 0.2
\subfigure[$S_{33}$  at $Re = 0.6$]{\includegraphics[totalheight=0.14\textheight,]{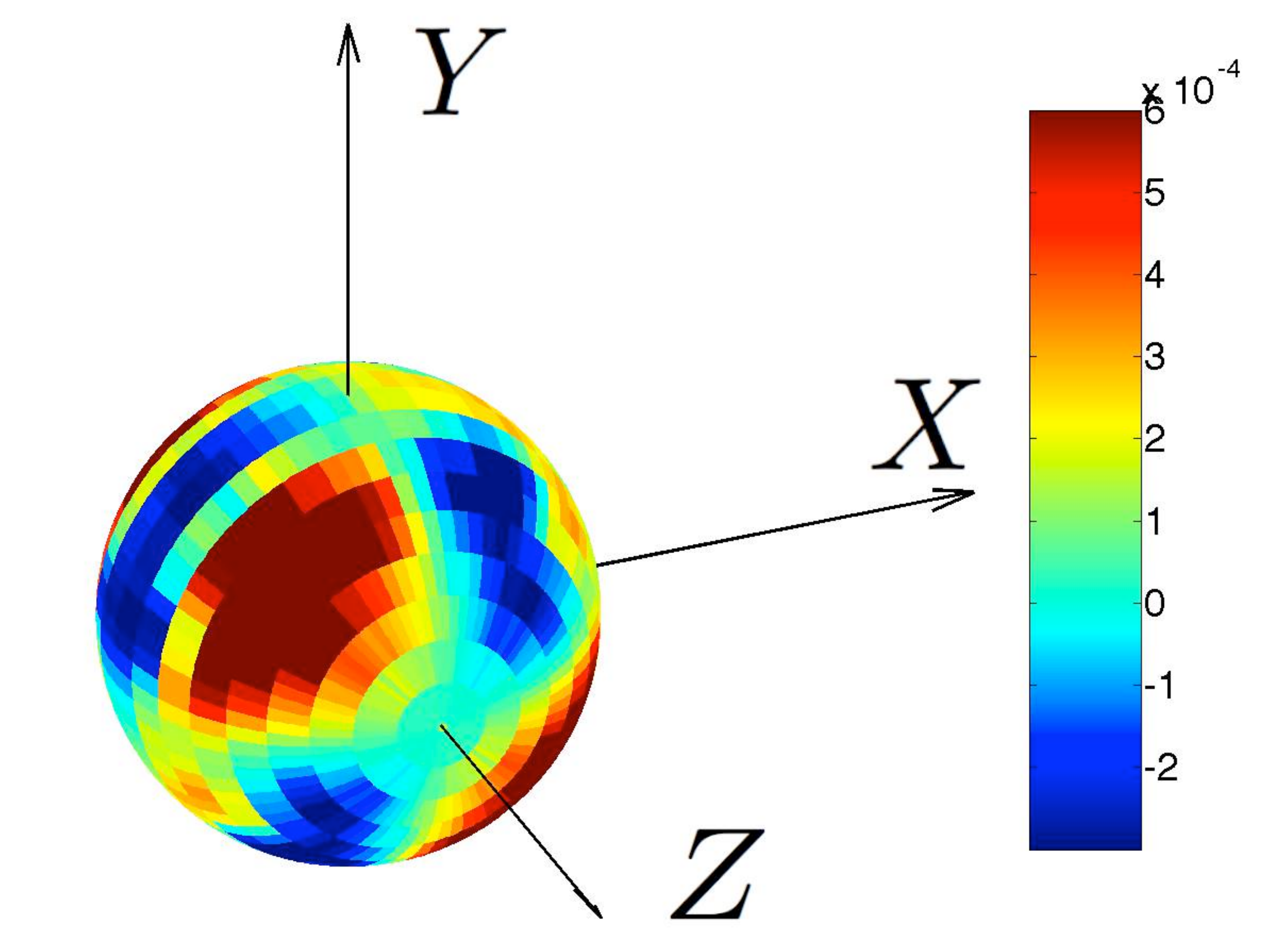}}
\caption{The effect of $Re$ on the distribution of the normal components of the stresslet on the particle surface. The volume fraction of all distributions is $0.15$. (\emph{a}) $S_{11}$ at $Re = 0.05$ (\emph{b}) $S_{11}$ at $Re = 0.1$ (\emph{c}) $S_{11}$ at $Re = 0.6$ (\emph{d}) $S_{22}$ at $Re = 0.05$ (\emph{e}) $S_{22}$ at $Re = 0.1$ (\emph{f}) $S_{22}$ at $Re = 0.6$ (\emph{g}) $S_{33}$ at $Re = 0.05$ (\emph{h}) $S_{33}$ at $Re = 0.1$ (\emph{i}) $S_{33}$ at $Re = 0.6$. }
 \label{fig:S-Dist}
\end{figure}

We consider the details of the contribution to the stresslet integrand, i.e. the local moment of the surface traction as a function of the position on the particle surface, in order to relate the flow field to the normal stress differences. The stresslet distribution on the particle is obtained by computing the traction moment at each boundary node on the surface of each particle and averaging these values. In figure ~\ref{fig:S-Dist} we show the effect of inertia on the surface distribution of stresslet for a suspension at $\phi = 0.15$.  The distribution of $S_{11}$ at $Re = 0.05$ is displayed in figure  ~\ref{fig:S-Dist} (\emph{a}). The distribution has the expected form for a near-Stokes flow with antisymmetry: negative in compression and positive in extension.  In figure  ~\ref{fig:S-Dist}(\emph{a}) - (\emph{c}), the magnitude of positive $S_{11}$ in the extensional zone decreases with increasing $Re$ and a compressive stress builds up around the flow direction in both compressional and extensional regions. \newline

The surface distribution of $S_{22}$ and its variation with $Re$ is displayed in figure  ~\ref{fig:S-Dist} (\emph{d}) - (\emph{f}). At small inertia ($Re = 0.05$), the distribution is, as described for $S_{11}$, roughly antisymmetric. By comparing figure ~\ref{fig:S-Dist} (\emph{d}) with (\emph{a}) we see that at $Re = 0.05$, the magnitudes of $S_{11}$ and $S_{22}$ distributions are similar.  Figure ~\ref{fig:S-Dist} (\emph{f}) shows the increase of the magnitude of $S_{22}$ results in a more pronounced anti-symmetry at larger $Re$. The distribution of $S_{33}$ is shown in figure ~\ref{fig:S-Dist} (\emph{g})-(\emph{i}). $S_{33}$ is symmetric with respect to the shear plane. Close to the plane of shear, $S_{33}$ changes sign; the basis for this behavior is unclear.  \newline      

\begin{figure}
\centering
\subfigure[]{\includegraphics[totalheight=0.26\textheight,]{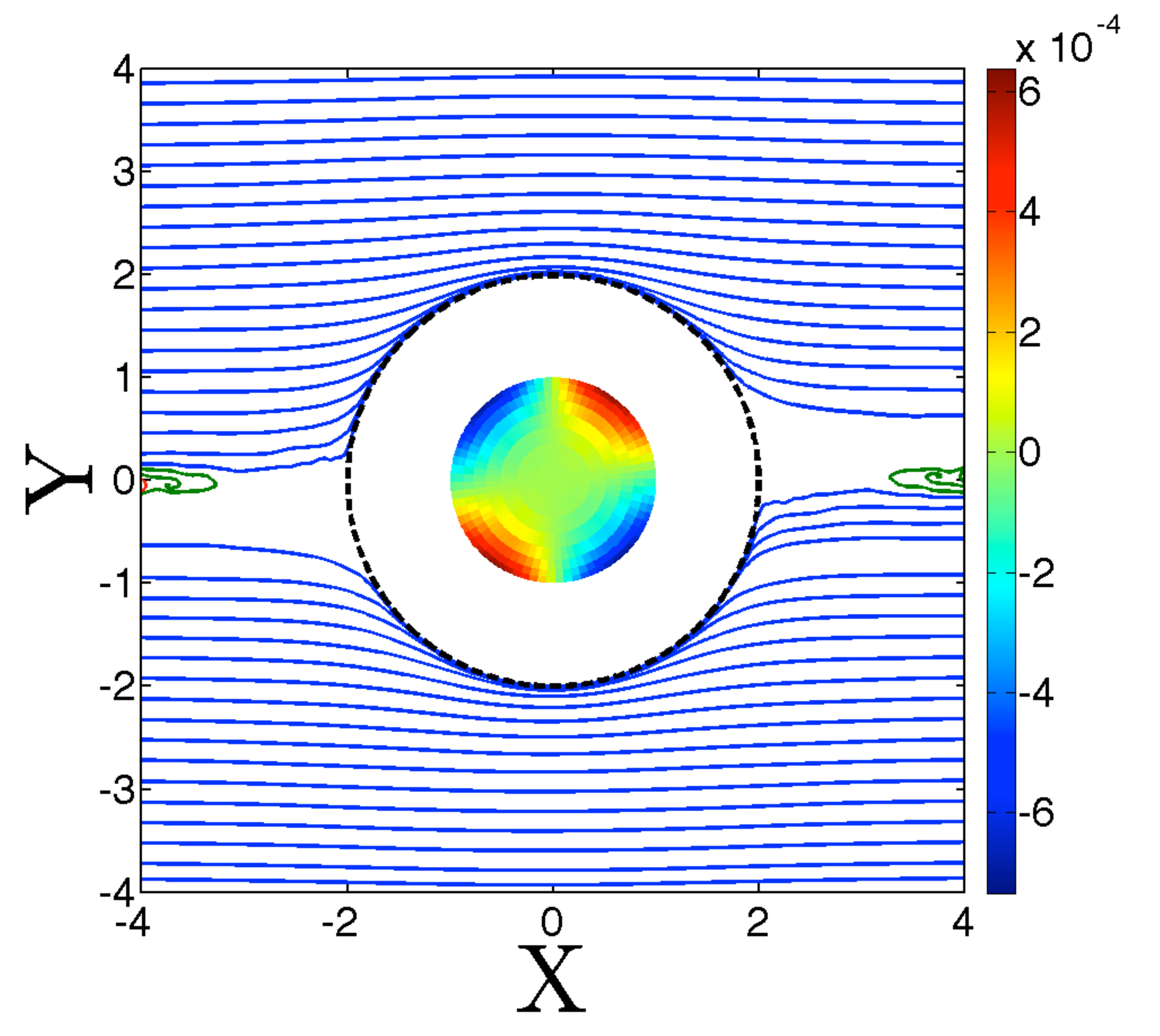}}
\subfigure[]{\includegraphics[totalheight=0.26\textheight,]{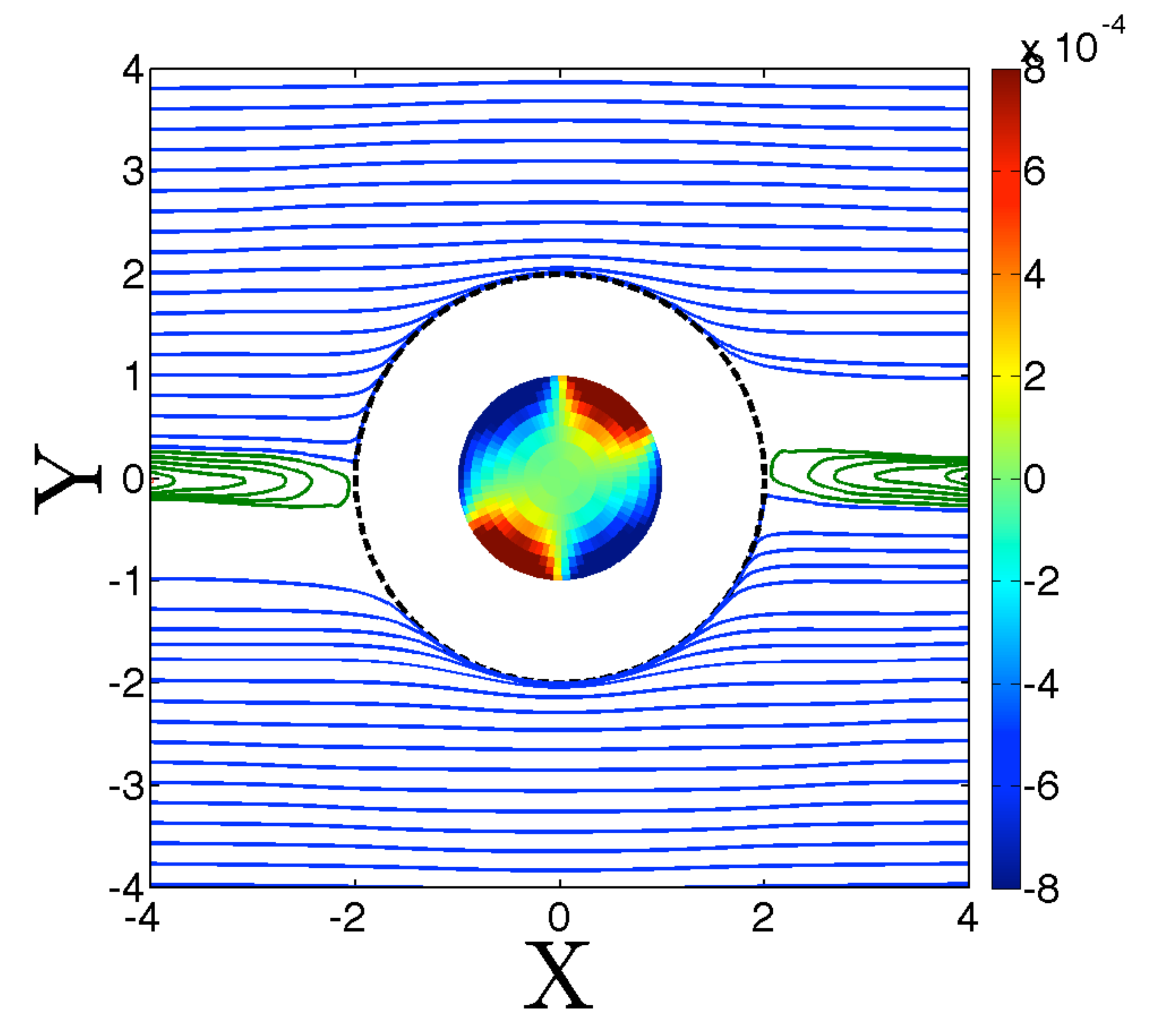}}
\subfigure[]{\includegraphics[totalheight=0.26\textheight,]{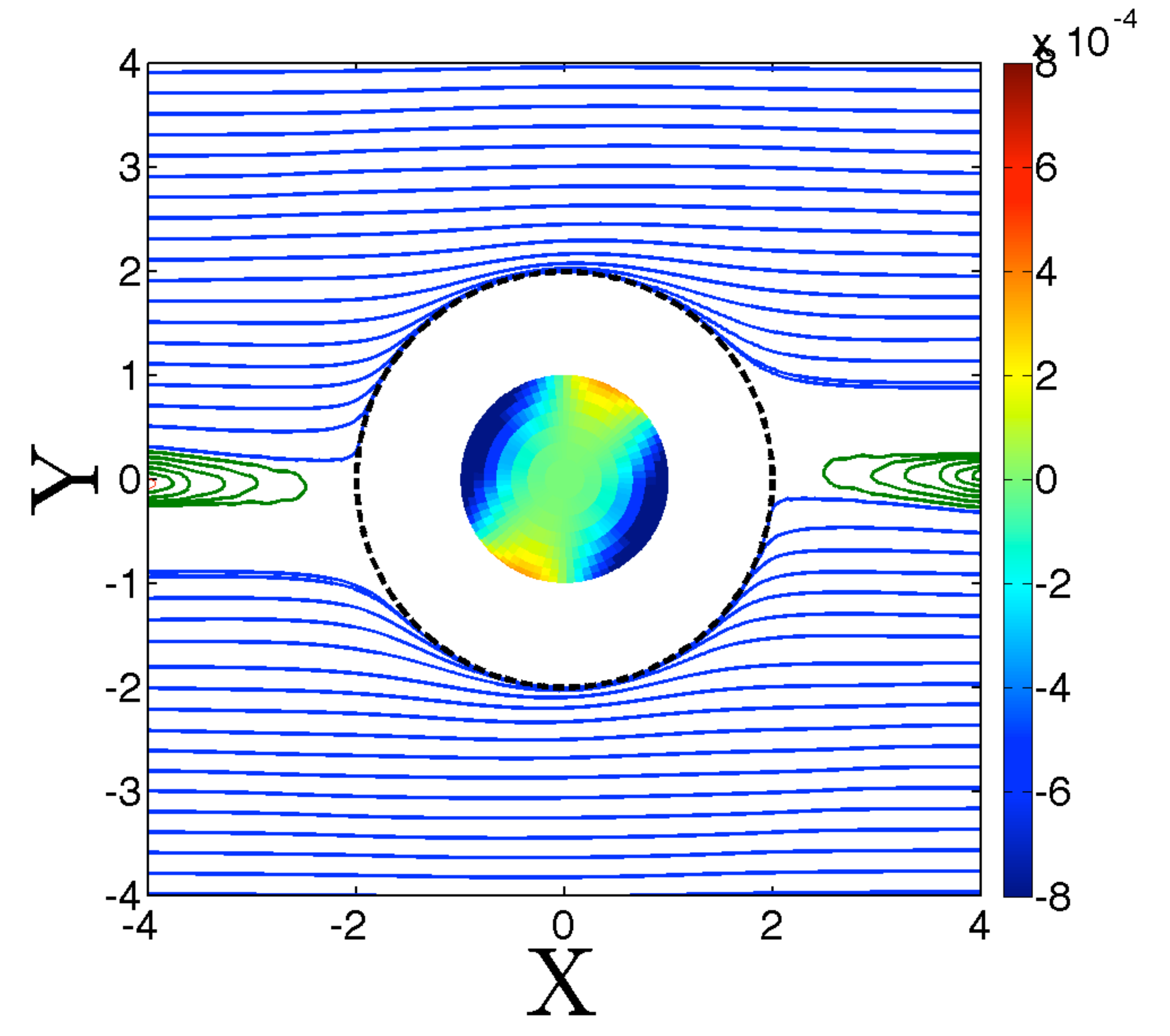}}
\subfigure[]{\includegraphics[totalheight=0.26\textheight,]{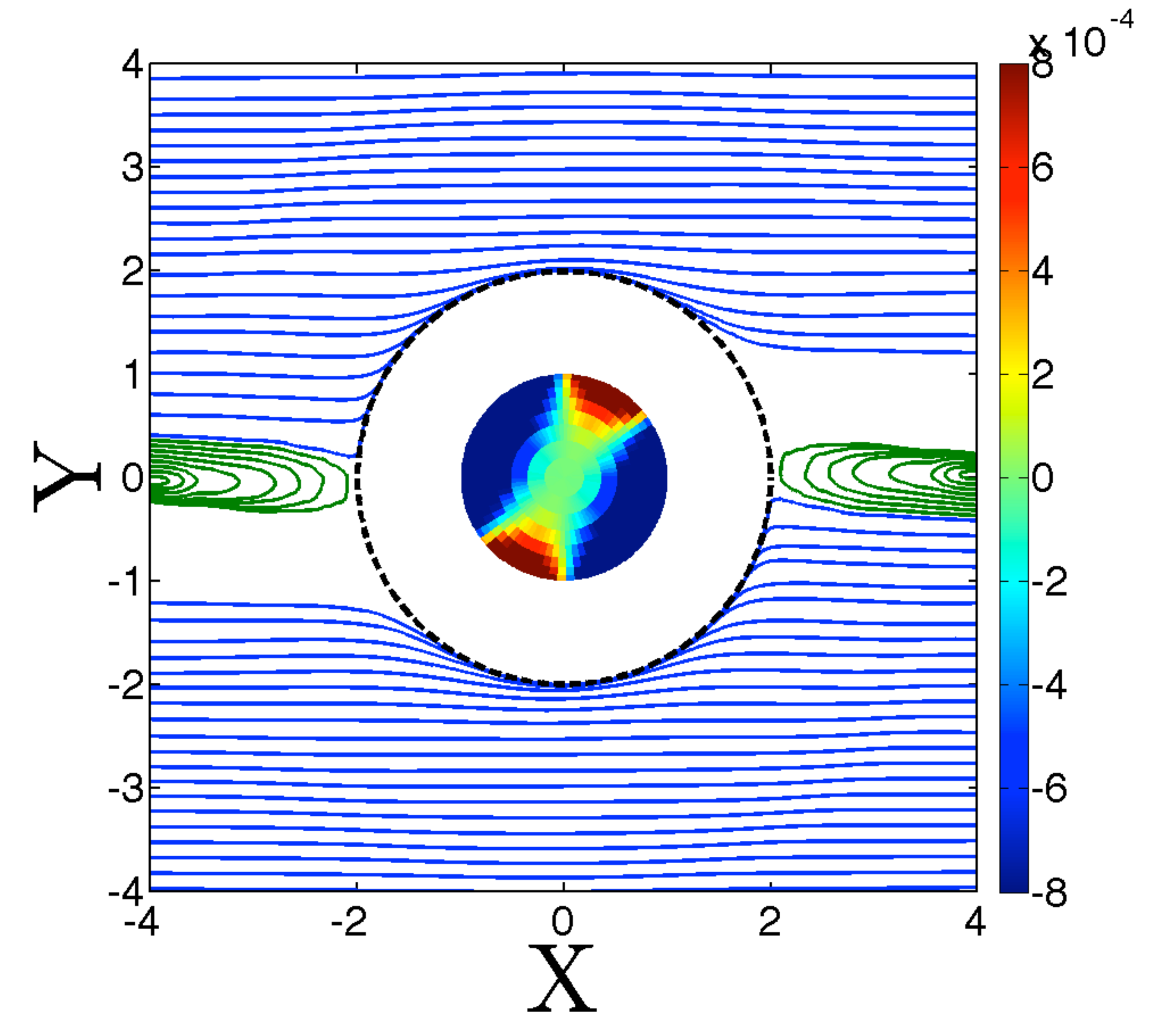}}
\caption{The portrait of the shear plane average pair trajectories along with the surface distribution of $S_{11}$ for suspensions at (\emph{a}) $Re = 0.05$ and $\phi = 0.1$ (\emph{b}) $Re = 0.05$ and $\phi = 0.24$ (\emph{c}) $Re = 0.6$ and $\phi = 0.1$ and (\emph{d}) $Re = 0.6$ and $\phi = 0.24$. The boundary of the excluded volume is indicated by dashed line. Pair trajectory space is symmetric with respect to the origin.}
 \label{fig:REV}
\end{figure}

The stresslet distributions are determined by the flow field at each condition of $Re$ and $\phi$. The relative motion of a pair of particles can be partially explained by the isolated pair trajectories. The increasing asymmetry of the pair trajectories with increasing $Re$ results in the asymmetric stresslet distributions.  This loss of symmetry directly impacts the normal stress differences. It has been observed by studying isolated pairs that in finite inertia three types of trajectories emerge (Kulkarni \& Morris 2008a). Open trajectories cover the majority of the pair space and form when two particles approach and pass each other. The open trajectories are fore-aft asymmetric and there is a positive offset in the gradient direction for neutrally buoyant particles.  The reversing trajectories emerge when two particles with a small separation in the gradient direction approach one another. In this situation, they reverse their path and separate moving in the opposite direction. Another prominent feature of trajectory space at finite particle-scale inertia is formation of spiraling trajectories, appearing both in the shear plane and off-plane. Two particles with a very small separation in the shear plane spiral around each other before leaving the close interaction region. Off-plane spirals along the vorticity axis carry pairs toward one another, i.e. they approach the plane of shear. For off-plane spirals, the distance of the trajectory from the vorticity axis increases as the pair approach along this axis. The topology of the pair trajectories is a replica of the streamline space. A detailed description of the streamline and isolated pair trajectory structure at finite inertia has been presented in Subramanian \& Koch (2006 a,b) and Kulkarni \& Morris (2008a). In figure  ~\ref{fig:REV} we display the average pair trajectories, sampled from many-body simulations, on the shear plane along with the $S_{11}$ distribution on the particle surface at $\phi = 0.1$ and $0.25$ for $Re = 0.05$ and $0.6$. It is outside the scope of the present paper to discuss the average pair trajectories in finite inertia and the influence of $\phi$ and $Re$ on the pair-particle dynamics. The topology of the pair trajectories on the shear plane are rather utilized to explain the stresslet data. \newline

The average trajectory structure on the shear plane consists of open and reversing trajectories. We have not identified average in-plane spiraling trajectories at finite $\phi$.  These in-plane spirals either cease to exist at the volume fractions studied ($\phi\ge 0.1$) or are compressed into a very small zone near contact that can not be captured by our calculations. We see that in the compressional zone, particles are pushed toward each other by open trajectories and thus hydrodynamic interaction generates a compressive stress on the particle surface around the compressional axis. In the extensional zone, the hydrodynamic stress resists the pair separation, resulting in a tensile stress: we observe a positive $S_{11}$ strip in the extensional region. By comparing figures ~\ref{fig:REV} (\emph{a}) and (\emph{c}) we see that with increasing $Re$, the separation point of open trajectories shifts to larger $\theta$ which results in a decrease of the size of the positive $S_{11}$ strip in the extensional region.  Additionally, the size of the reversing trajectory zone grows with increasing $Re$ and $\phi$, a feature which appears to be associated with the generation of large compressive stress along the flow direction on both sides of the reference particle. \newline

\begin{figure}
\centering
\subfigure[$S_{11}$  at $\phi = 0.15$]{\includegraphics[totalheight=0.16\textheight,]{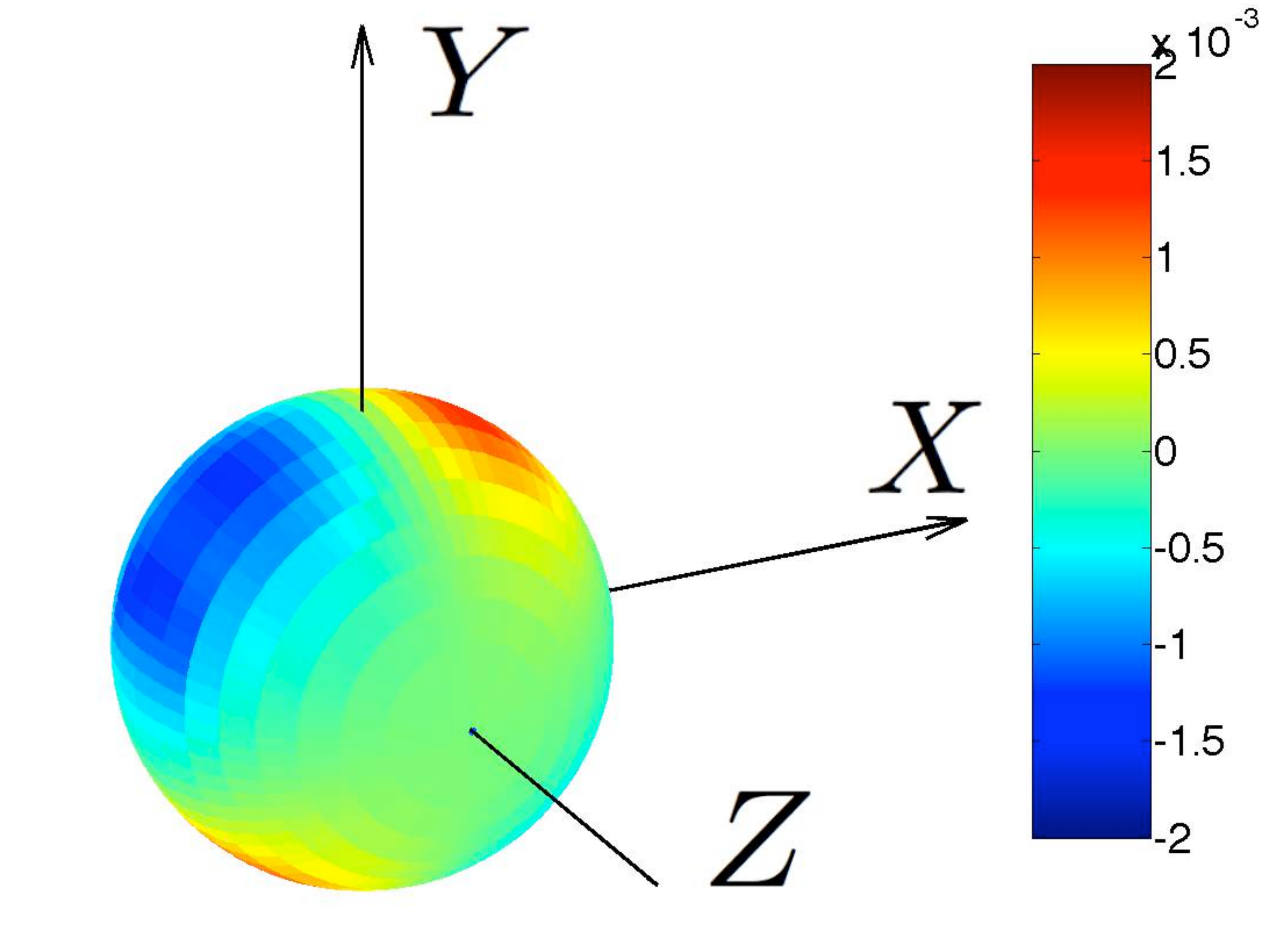}}
\subfigure[$S_{11}$  at $\phi = 0.25$]{\includegraphics[totalheight=0.16\textheight,]{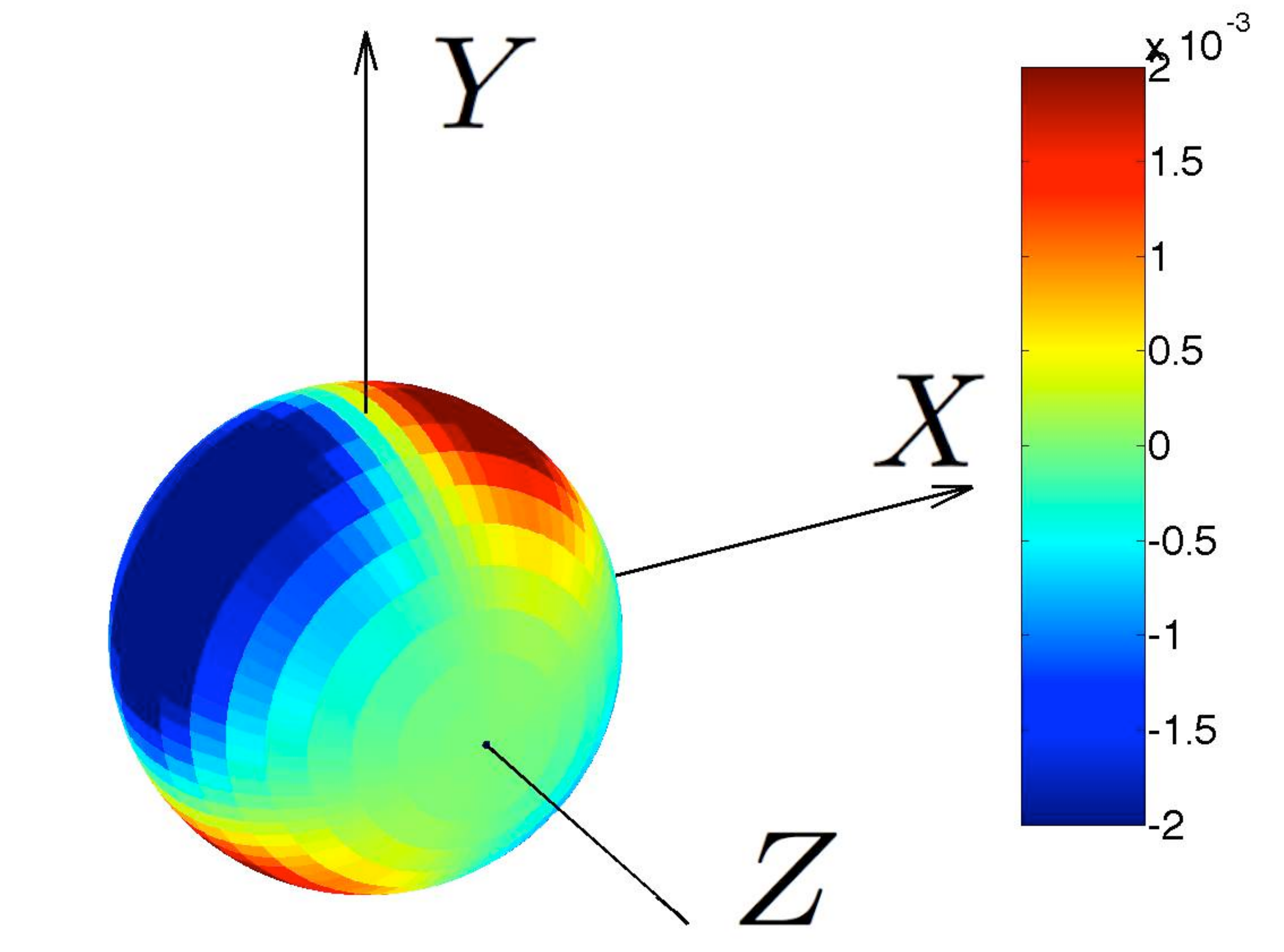}}
\subfigure[$S_{22}$  at $\phi = 0.15$]{\includegraphics[totalheight=0.16\textheight,]{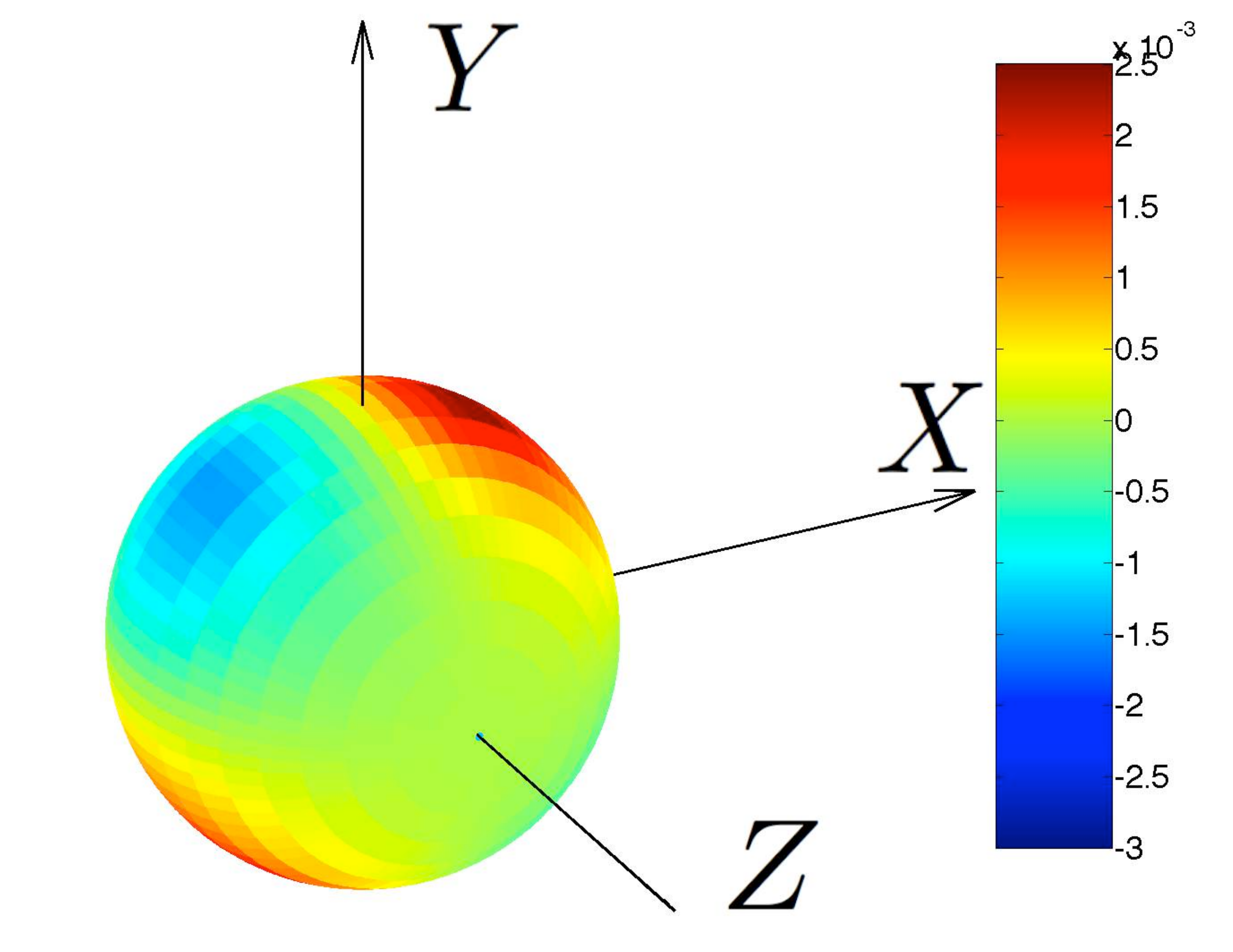}}
\subfigure[$S_{22}$  at $\phi = 0.25$]{\includegraphics[totalheight=0.16\textheight,]{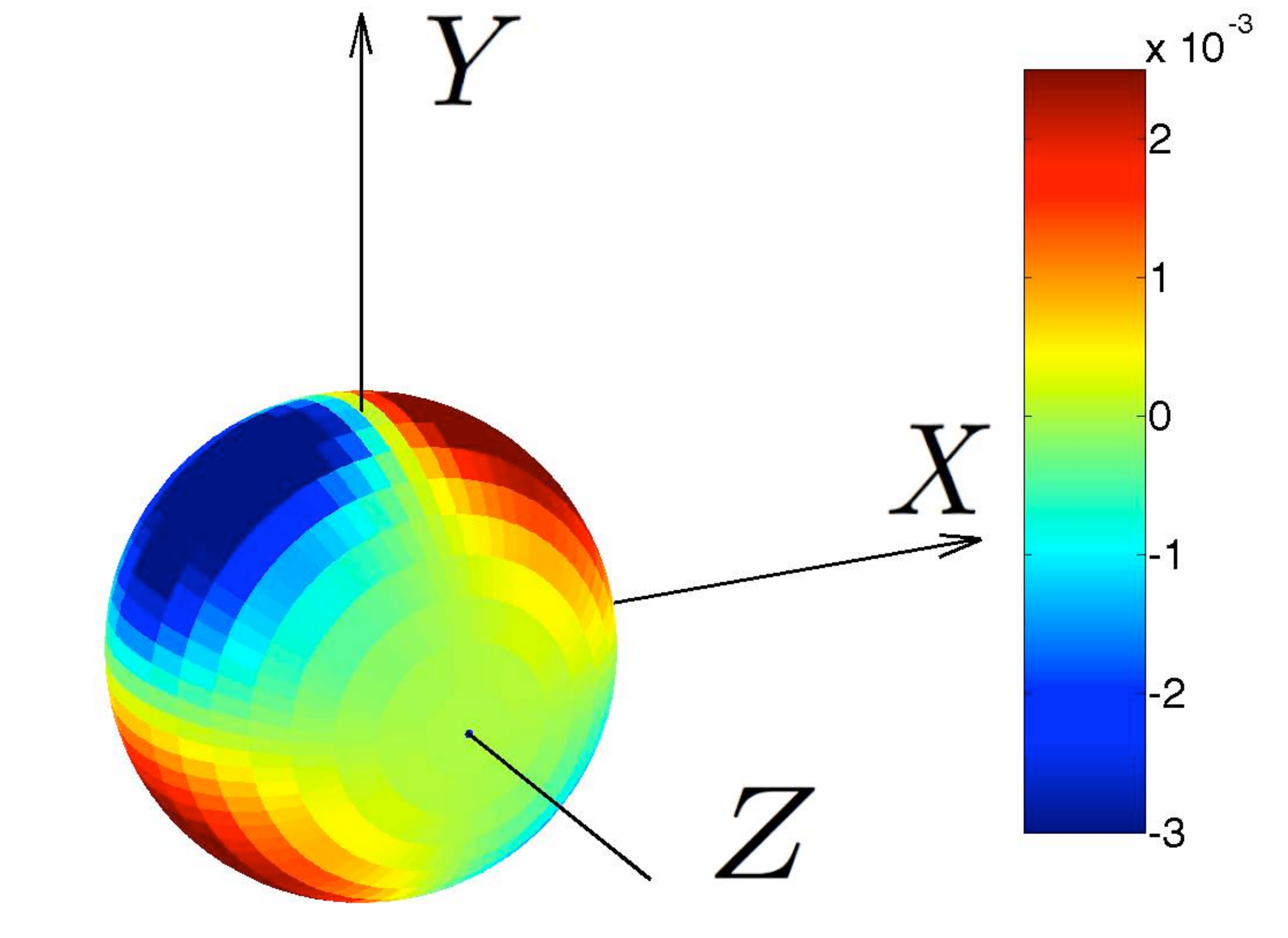}}
\subfigure[$S_{33}$  at $\phi = 0.15$]{\includegraphics[totalheight=0.16\textheight,]{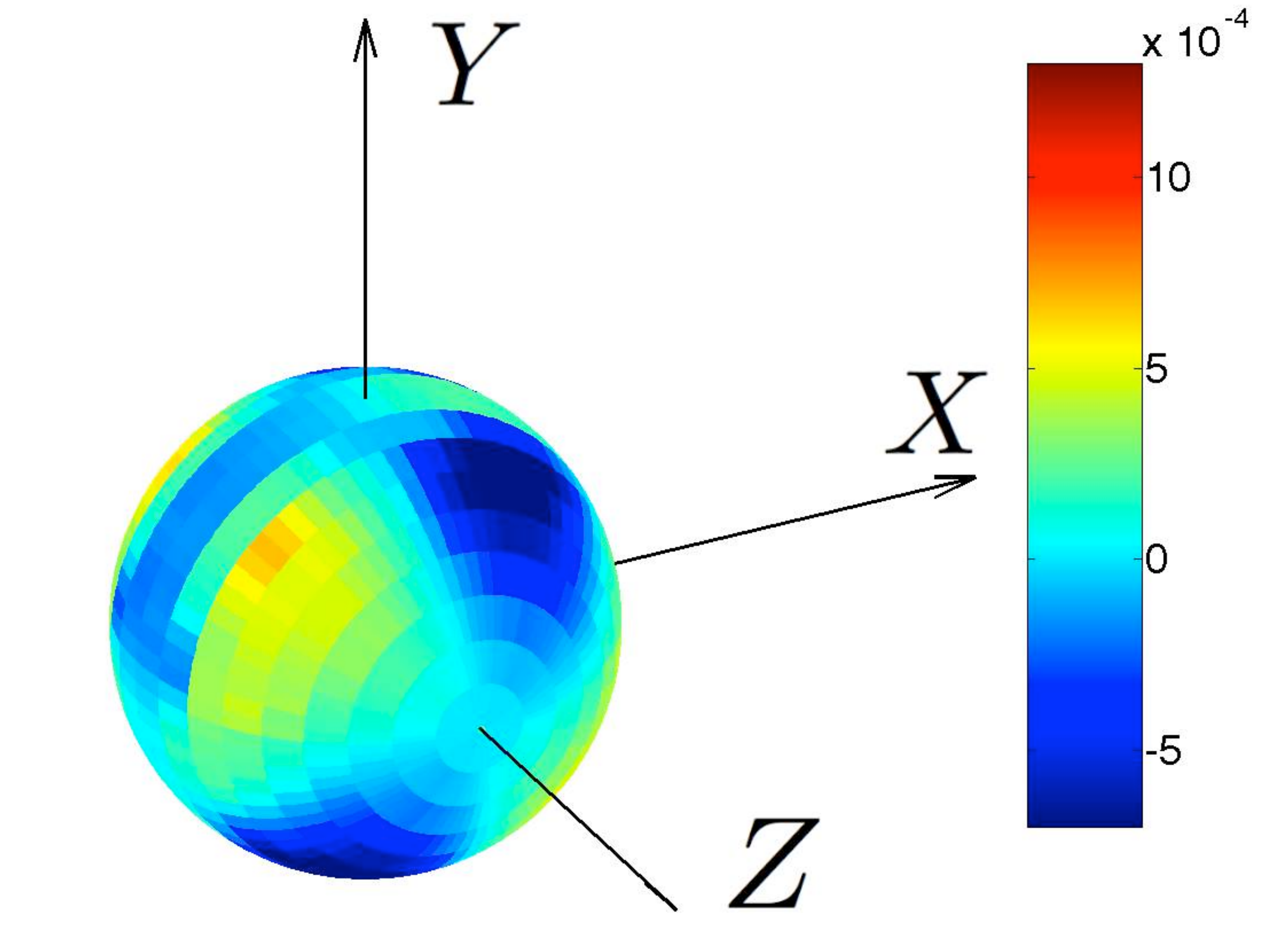}}
\subfigure[$S_{33}$  at $\phi = 0.25$]{\includegraphics[totalheight=0.16\textheight,]{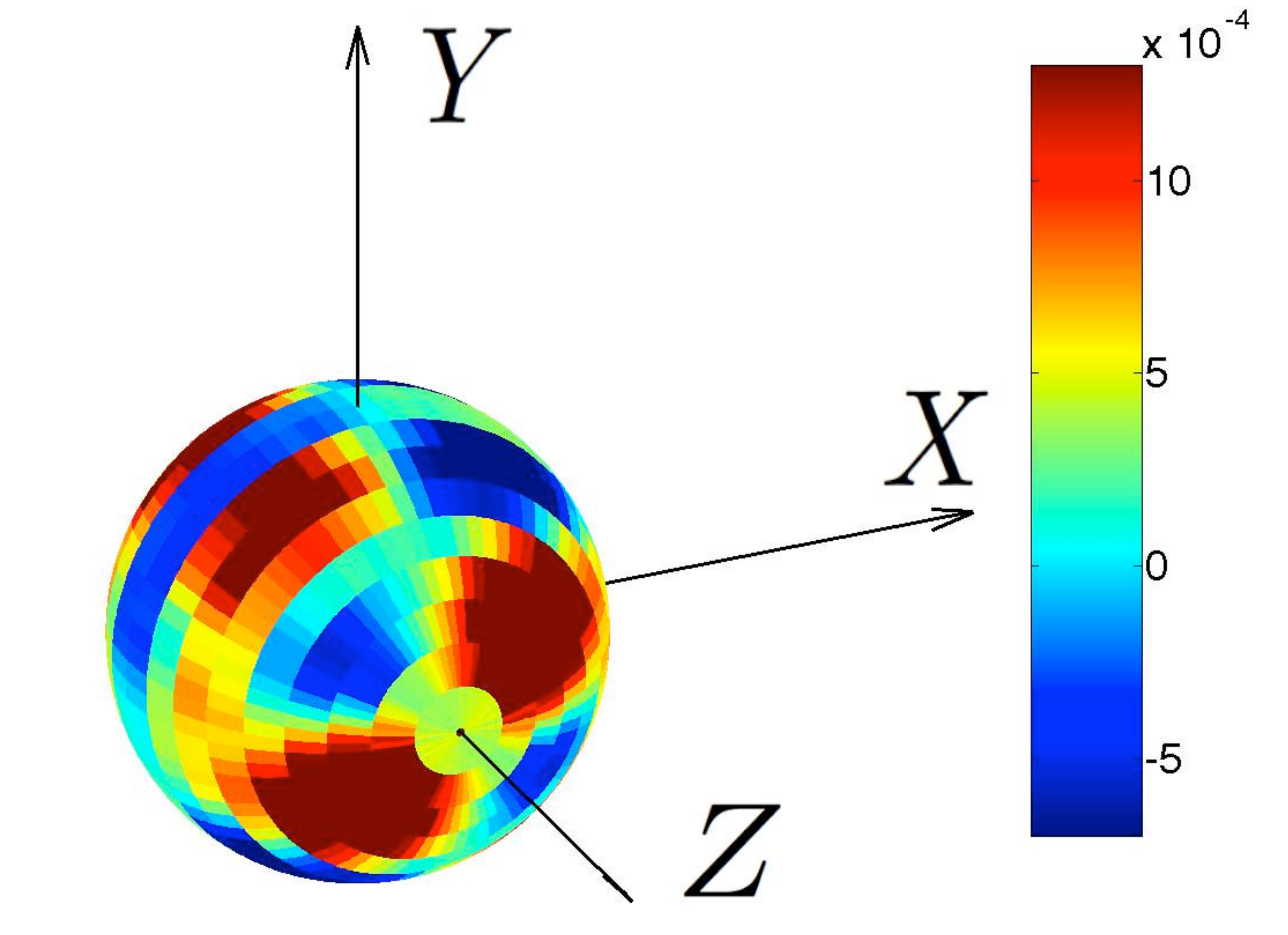}}
\caption{The influence of volume fraction on the distribution of the normal stress components on the particle surface. In all distributions $Re$ is $0.1$.(\emph{a}) $S_{11}$ at $\phi = 0.15$ (\emph{b}) $S_{11}$ at $\phi = 0.25$ (\emph{c}) $S_{22}$ at $\phi = 0.15$ (\emph{d}) $S_{22}$ at $\phi = 0.25$ (\emph{e}) $S_{33}$ at $\phi = 0.15$ (\emph{f}) $S_{33}$ at $\phi = 0.25$. }
 \label{fig:PHI-S-Dis}
\end{figure}

In addition to the increase of the size of the reversing zone, increasing $\phi$ results in larger magnitudes of $S_{11}$ and $S_{22}$ on the particle surface; this is apparent in figure ~\ref{fig:PHI-S-Dis} for suspensions at $Re = 0.1$ and $\phi = 0.15$ and $0.25$. For the distribution of $S_{33}$ at larger $\phi$, regions of negative values in compression and positive in extension form close to the vorticity axis.

\subsection{Reynolds stress}
Particle interactions generate velocity fluctuations of both the fluid and particles. The momentum transfer associated with velocity fluctuations is captured by Reynolds stress, given by $\frac{1}{V} \int_V \rho\boldsymbol{u^{\prime}u^{\prime}}dV$.  Because the Reynolds stress scales with density, its influence is seen only at finite inertia. In this section we report the Reynolds stress determined in our simulations. \newline

\begin{figure}
\centering
\subfigure[fluid Reynolds stress]{\includegraphics[totalheight=0.22\textheight,]{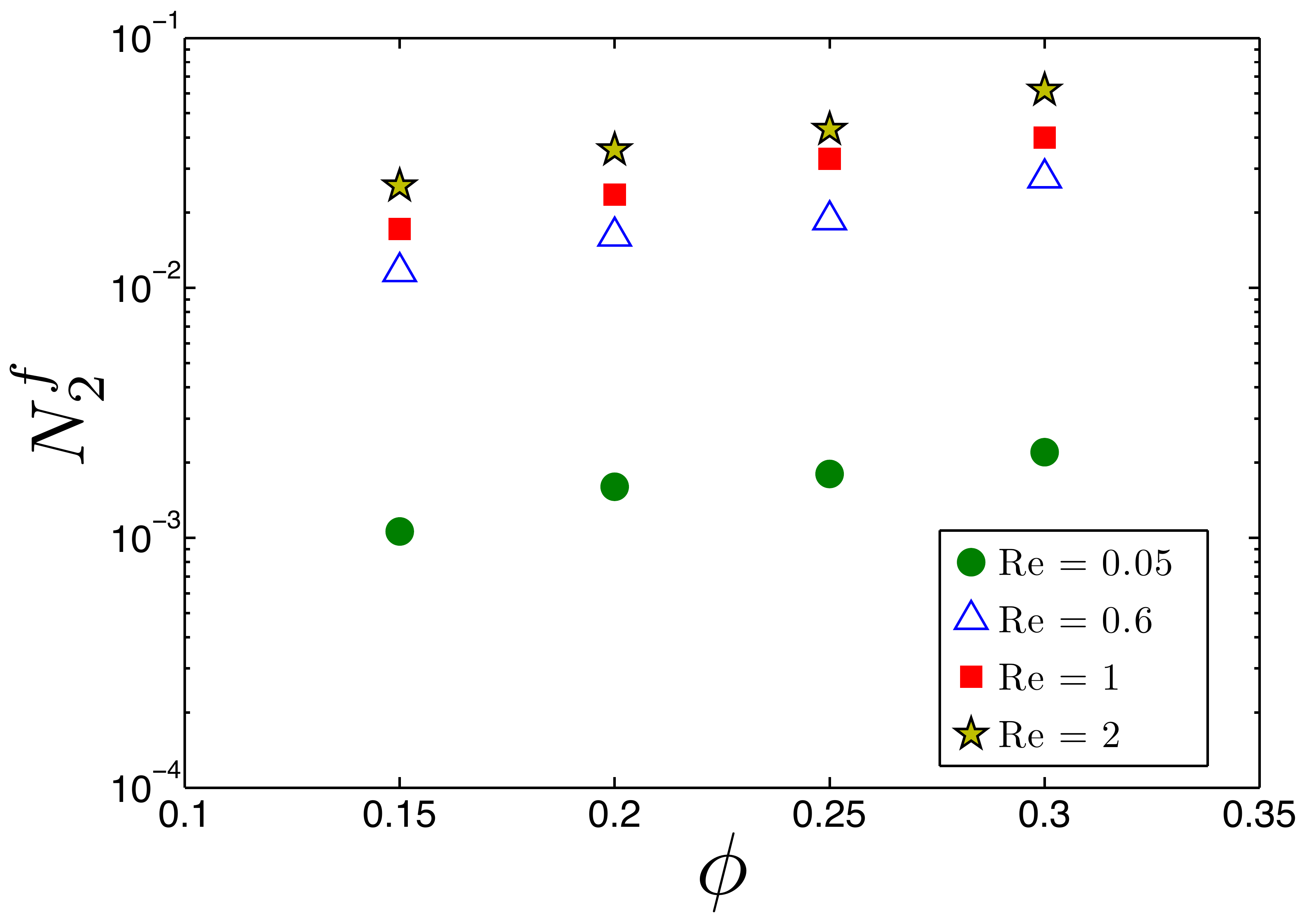}}
\subfigure[particle Reynolds stress]{\includegraphics[totalheight=0.22\textheight,]{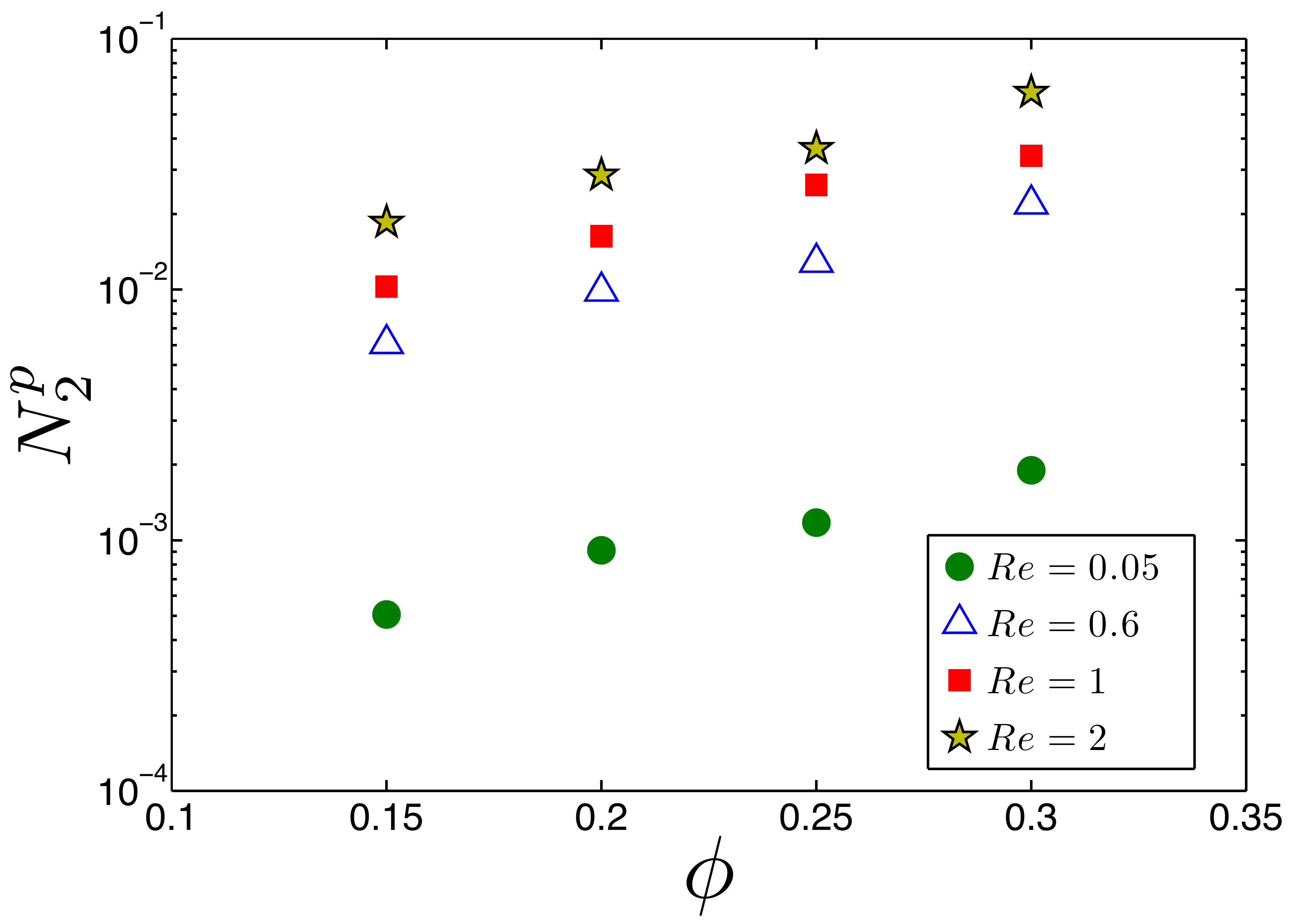}}
\caption{The Reynolds stress $N_{2}$ associated with velocity fluctuations of the \emph{(a)} fluid and \emph{(b)} particles.}
 \label{fig:RSPF}
\end{figure}

The work of KM08 found that Reynolds stress, although increasing linearly with $Re$, have a negligible influence on the bulk suspension stress. Using a scaling argument, Yeo \& Maxey made a comparison between the magnitudes of the Reynolds stress and the stresslet and deduced that Reynolds stress contribution is negligible, but considered the fluid phase only. However, the fluctuations in both solid and liquid phase generate the Reynolds stress as seen in equation(~\ref{eq:stresslet}). The particle phase Reynolds stress is calculated by finding the difference between rigid body velocity of the solid node and the average velocity. The rigid body velocity is computed as $\boldsymbol{U}_i + \boldsymbol{\Omega}_i \times (\boldsymbol{r} - \boldsymbol{R})$; where $\boldsymbol{r} - \boldsymbol{R}$ is the distance between the solid node and the center of mass of the sphere. Figure ~\ref{fig:RSPF} displays Reynolds stress contribution to $N_{2}$ generated by fluid and particle phase. We observe that fluctuations in fluid and solid phase generate a comparable $N_{2}$. \newline

\begin{figure}
\centering
\subfigure[]{\includegraphics[totalheight=0.16\textheight,]{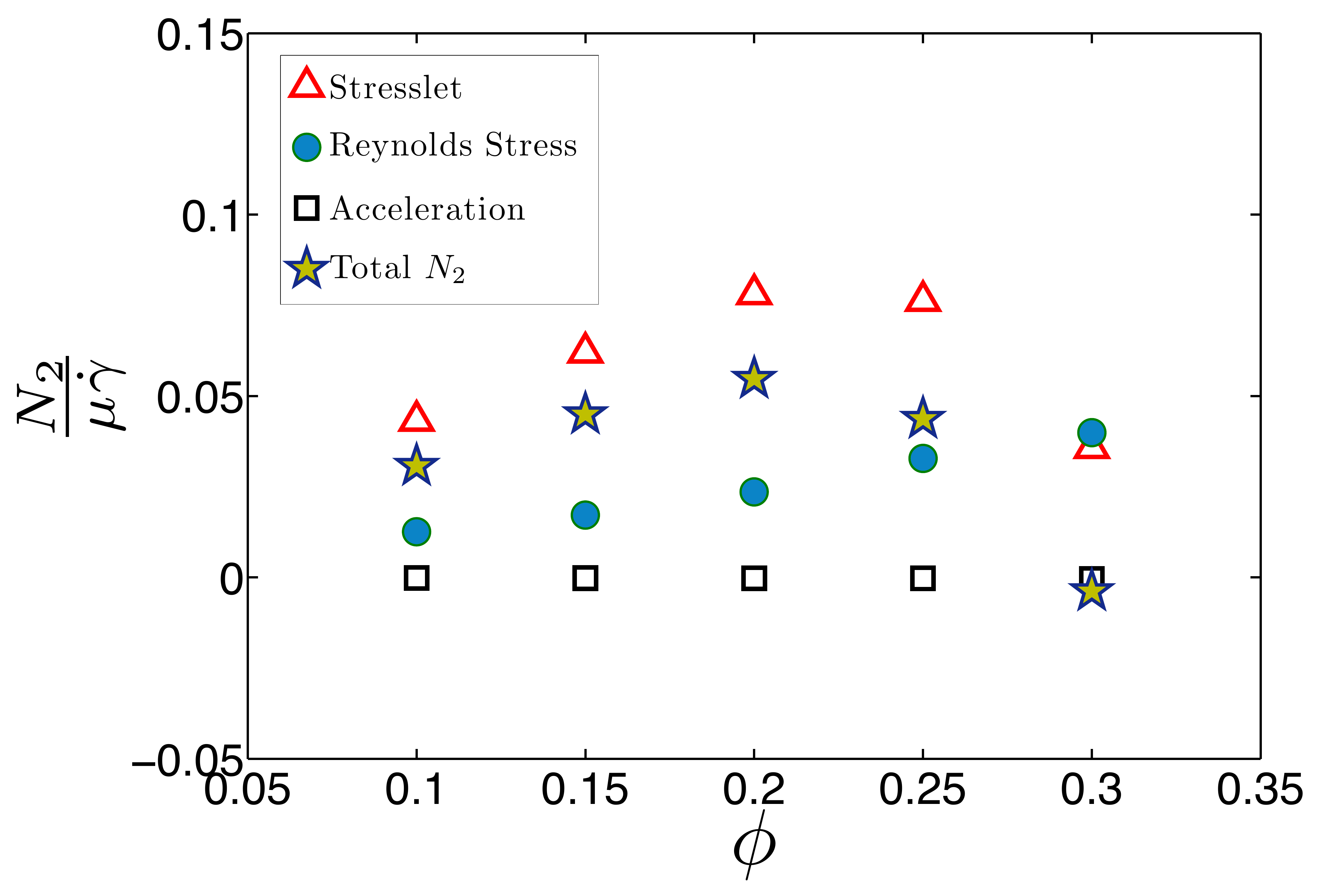}}
\subfigure[]{\includegraphics[totalheight=0.16\textheight,]{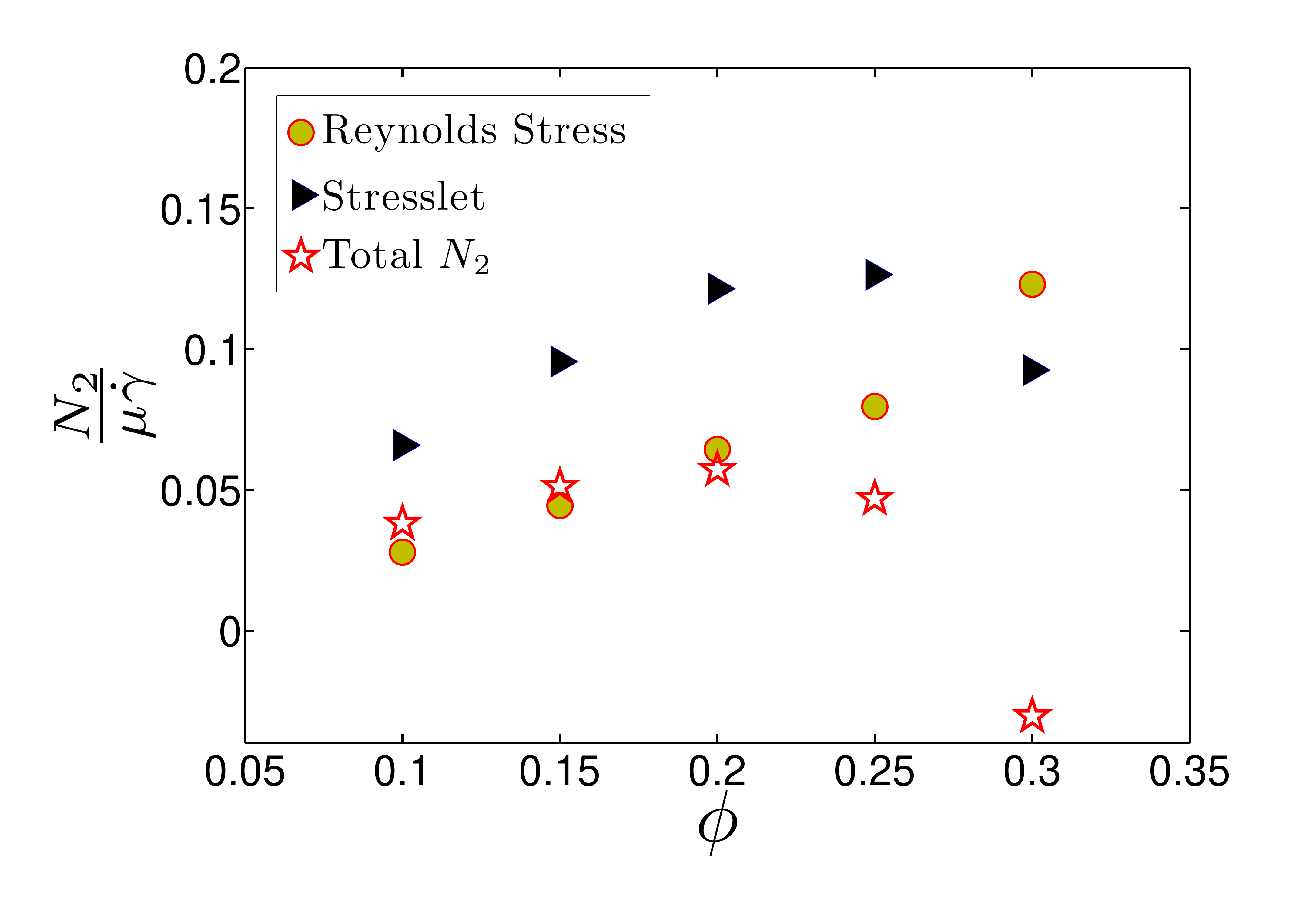}}
\subfigure[]{\includegraphics[totalheight=0.16\textheight,]{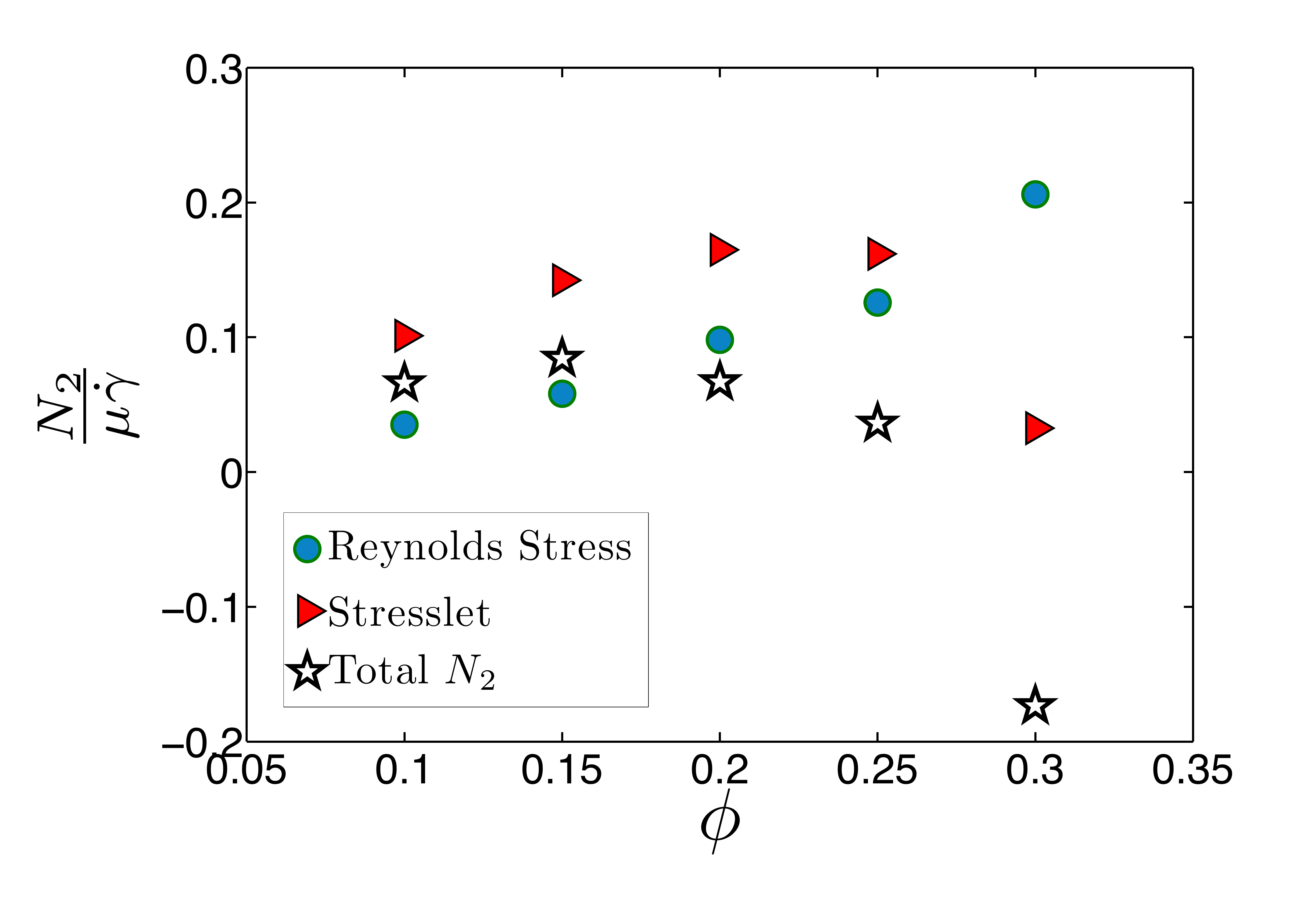}}
\caption{(\emph{a}) Comparison between the magnitude of $N_2$ generated by the acceleration stress, Reynolds stress and stresslet at $Re = 1$. Comparison between the Reynolds stress and stresslet contributions on the $N_2$ of the suspension at (\emph{b}) $Re = 2$ and (\emph{c}) $Re = 5$.}
 \label{fig:WS}
\end{figure}

We have found that Reynolds stress does not appreciably influence the first normal stress difference, as the large compressive $S_{11}$ (the stresslet contribution)  dominates $N_1$. However, Reynolds stress $N_2$ increases noticeably when $Re\sim O(1)$. Figure ~\ref{fig:WS} \emph{(a)} shows the contributions to $N_2$ from stresslet, acceleration stress, and Reynolds stress at $Re = 1$, along with the total $N_2$.  As noted, the acceleration stress is negligible. The magnitude of $N_2$ generated by the Reynolds stress is smaller than stresslet, but is non-negligible.  Figure  ~\ref{fig:WS} (\emph{b}) and (\emph{c}) shows $N_2$ contributions at $Re = 2$ and $5$, respectively, where we see significant contribution of Reynolds stress.  The $N_2$ contribution from the stresslet remains larger for $\phi < 0.25$ but falls below the Reynolds stress contribution at $\phi = 0.3$.  At $Re = 2$ and $5$, the contribution from Reynolds stress decreases the total $N_2$ toward negative values. \newline

%********************************** SECTION 5 ***********************************%
%************** HYDRODYNAMIC FORCE AND TORQUE DISTRIBUTIONS *****************%

\section{Statistics: Velocity Fluctuations; Hydrodynamic Forces and Torques}

\begin{figure}
\centering
\subfigure[]{\includegraphics[totalheight=0.165\textheight,]{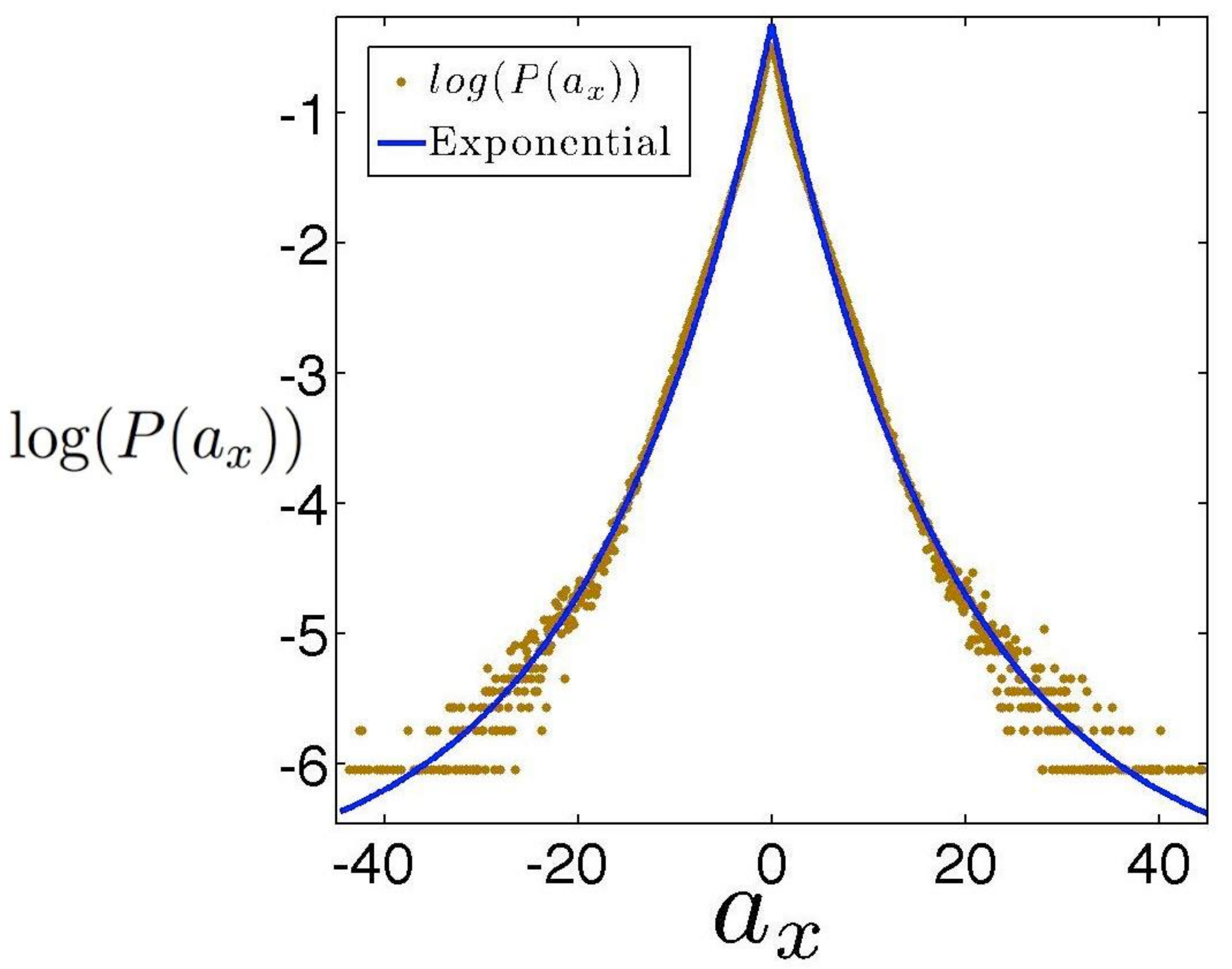}}
\subfigure[]{\includegraphics[totalheight=0.165\textheight,]{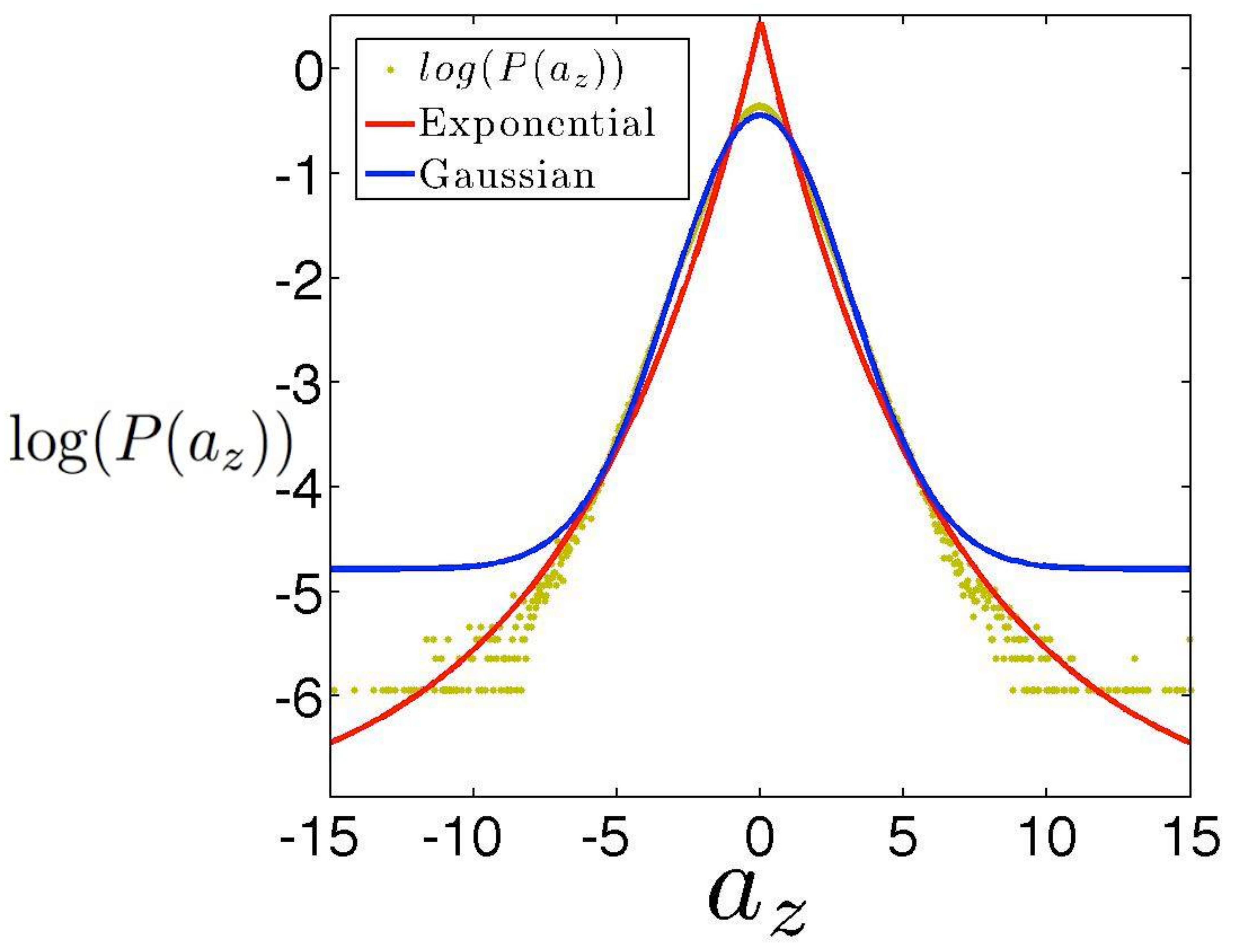}}
\subfigure[]{\includegraphics[totalheight=0.165\textheight,]{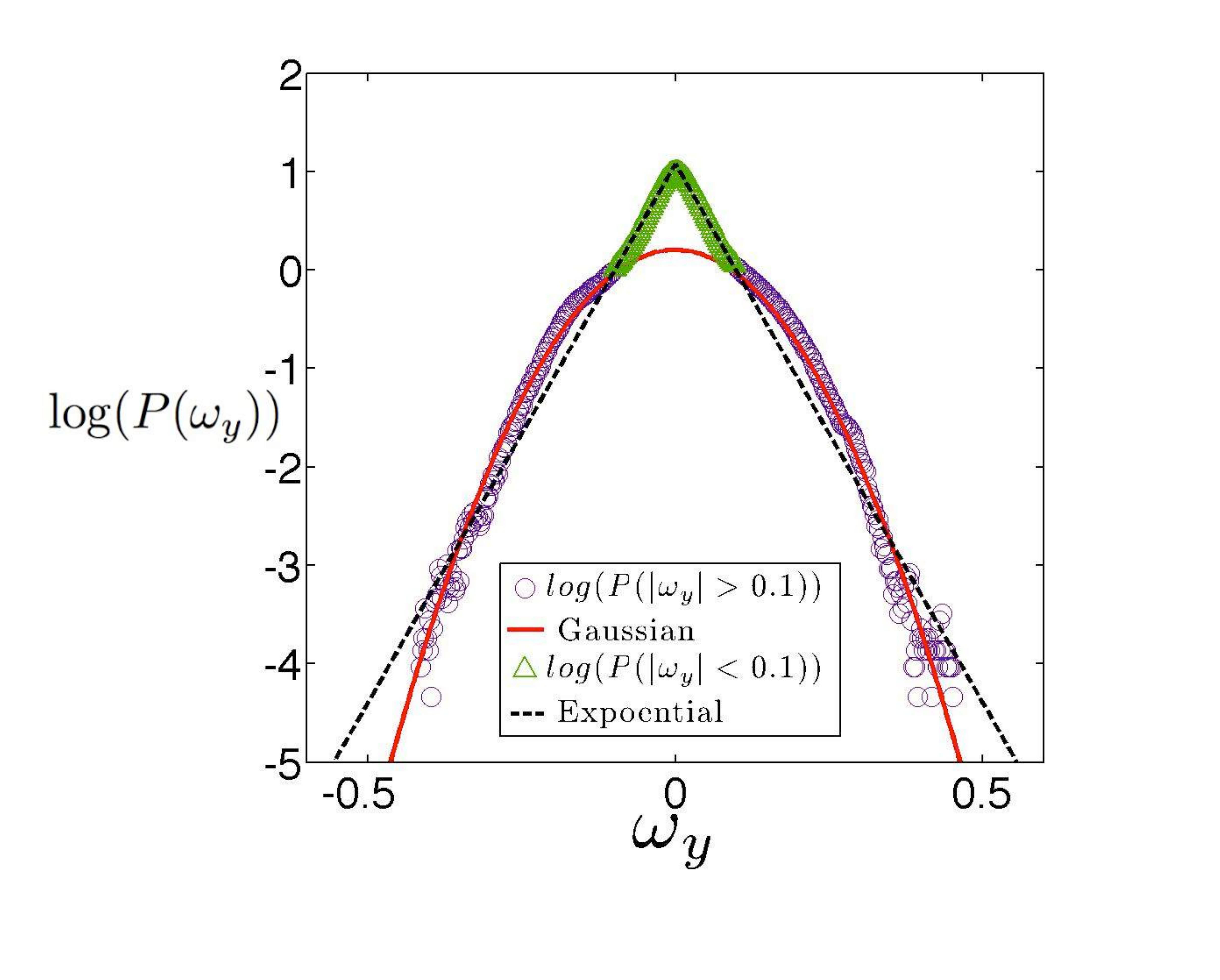}}
\caption{The probability curve fits:(\emph{a}) the exponential fit on $P(a_x)$ at $Re = 0.6$ and $\phi = 0.15$ (\emph{b}) the exponential and Gaussian fits on $P(a_z)$ for $Re = 0.6$ and $\phi = 0.25$ (\emph{c}) The exponential fit for the tip of $P(\omega_y)$ and the Gaussian distribution for the shoulder of $P(\omega_y)$ at $Re = 0.6$ and $\phi = 0.1$.}
 \label{fig:FITS}
\end{figure}

We consider in this section the acceleration of particles, as well as the force and torque on the particles.  {\color{red} Probability density functions of linear velocities have been discussed in KM08.} In a Stokes flow, the forces and torques are vanishingly small, but finite hydrodynamic forces and torques, of mean zero, are generated at finite $Re$, and these are related to the linear and angular accelerations of the particles.  These fluctuating quantities are of interest to understand the statistical physics of these materials more fully.  Hence, we sample and report the probability density function (PDF) of linear and angular accelerations and angular velocities. The scales used for the linear and angular accelerations are $a \dot{\gamma}^{2}$ and $a^{2} \dot{\gamma}^{2}$, respectively.  The angular velocity is scaled by $\dot{\gamma}^{-1}$.  We compare the distributions with the Gaussian and exponential probability distributions using curve fitting. The Gaussian and exponential distributions follow the general form \newline

\begin{equation}
 P \propto exp\left[ - b |\alpha|^{c}\right]
\end{equation}
where $\alpha$ is the quantity of interest (linear acceleration, angular acceleration, or angular velocity) and $b$ is a fitting parameter. The distribution type is fixed by whether $c=1$ for an exponential or $c=2$ for a Gaussian distributions. We present and discuss the fittings for a select set of data in figure ~\ref{fig:FITS}, to which we refer for cases requiring more discussion. \newline

\begin{figure}
\centering
\subfigure[]{\includegraphics[totalheight=0.16\textheight,]{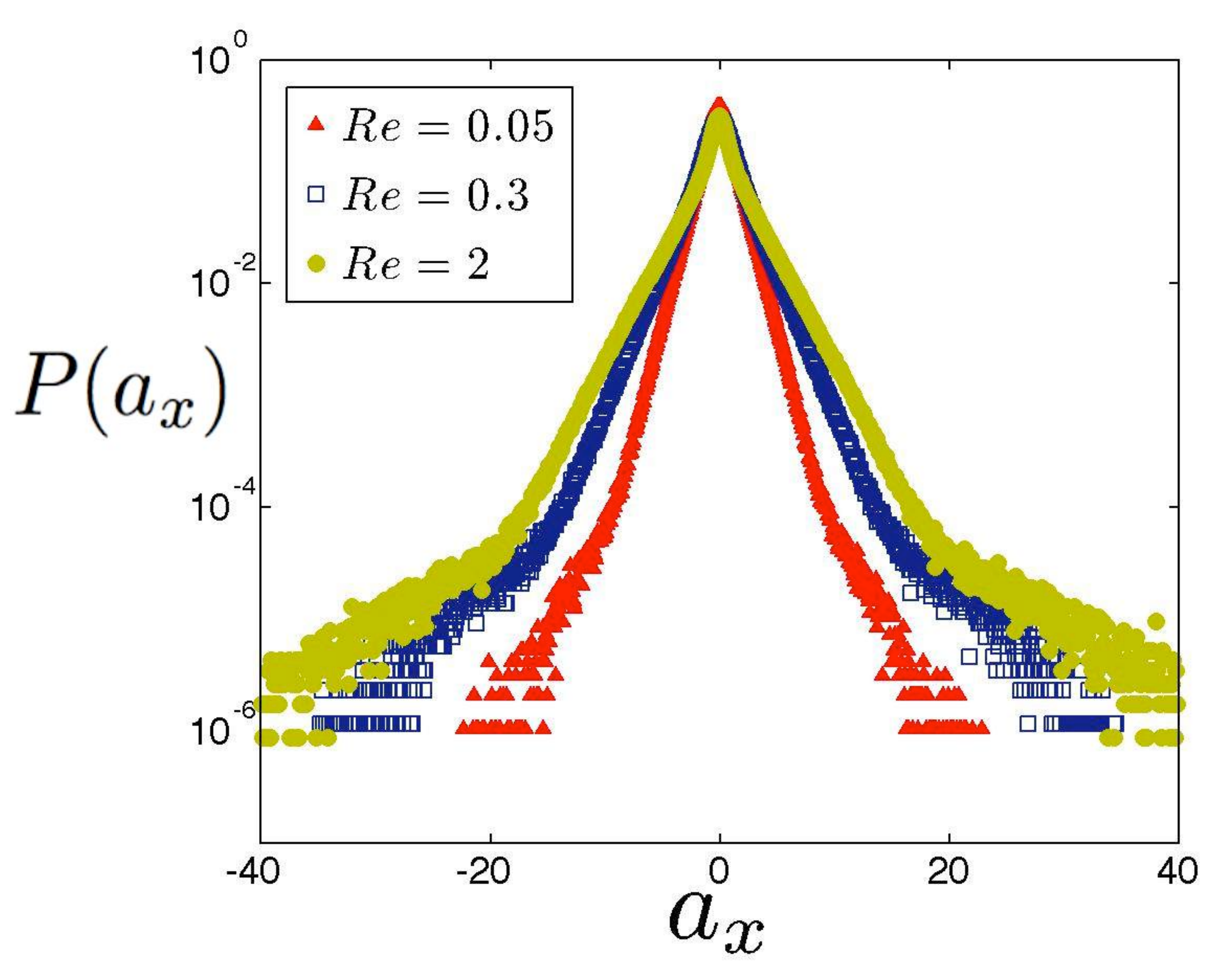}}
\subfigure[]{\includegraphics[totalheight=0.16\textheight,]{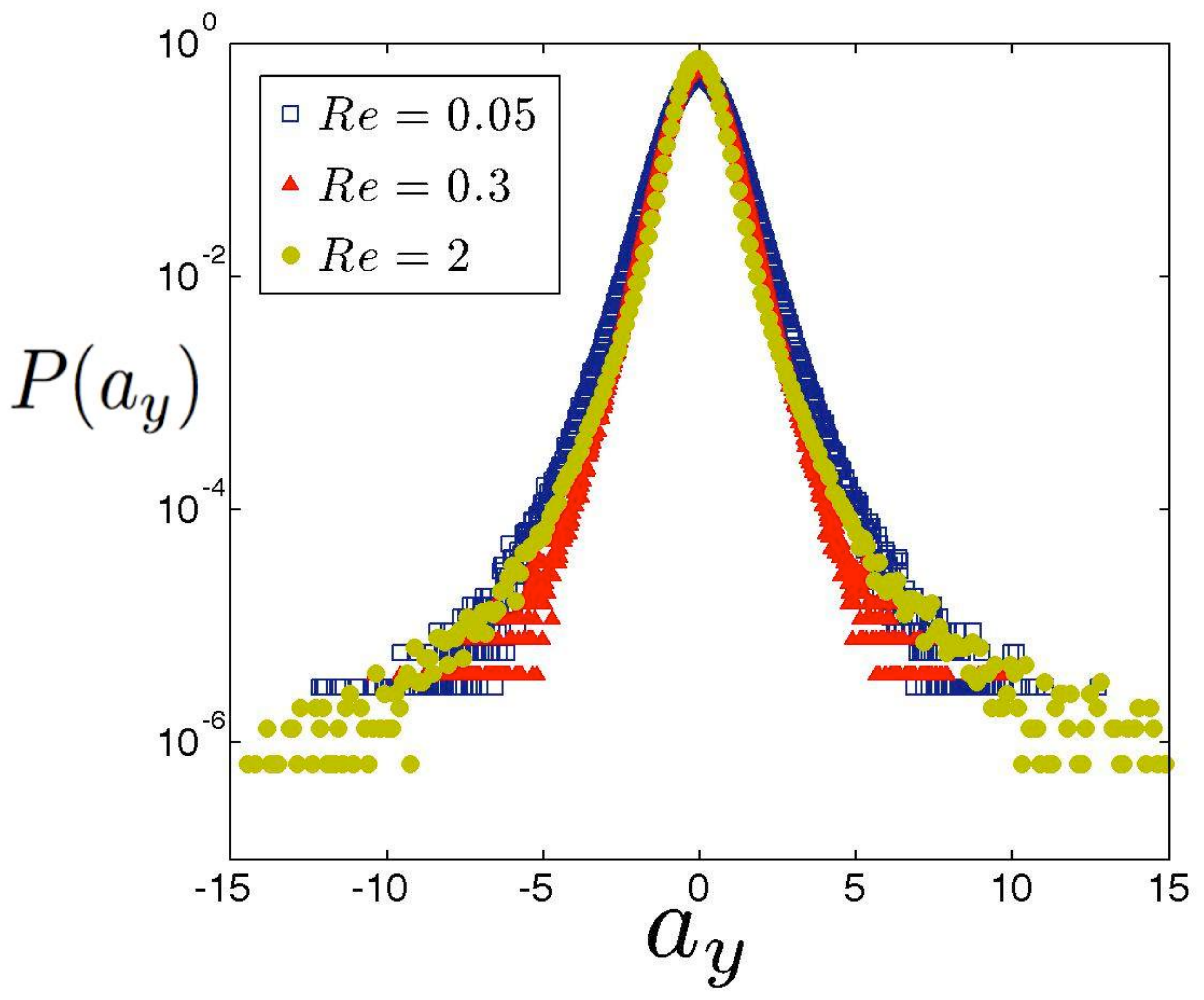}}
\subfigure[]{\includegraphics[totalheight=0.16\textheight,]{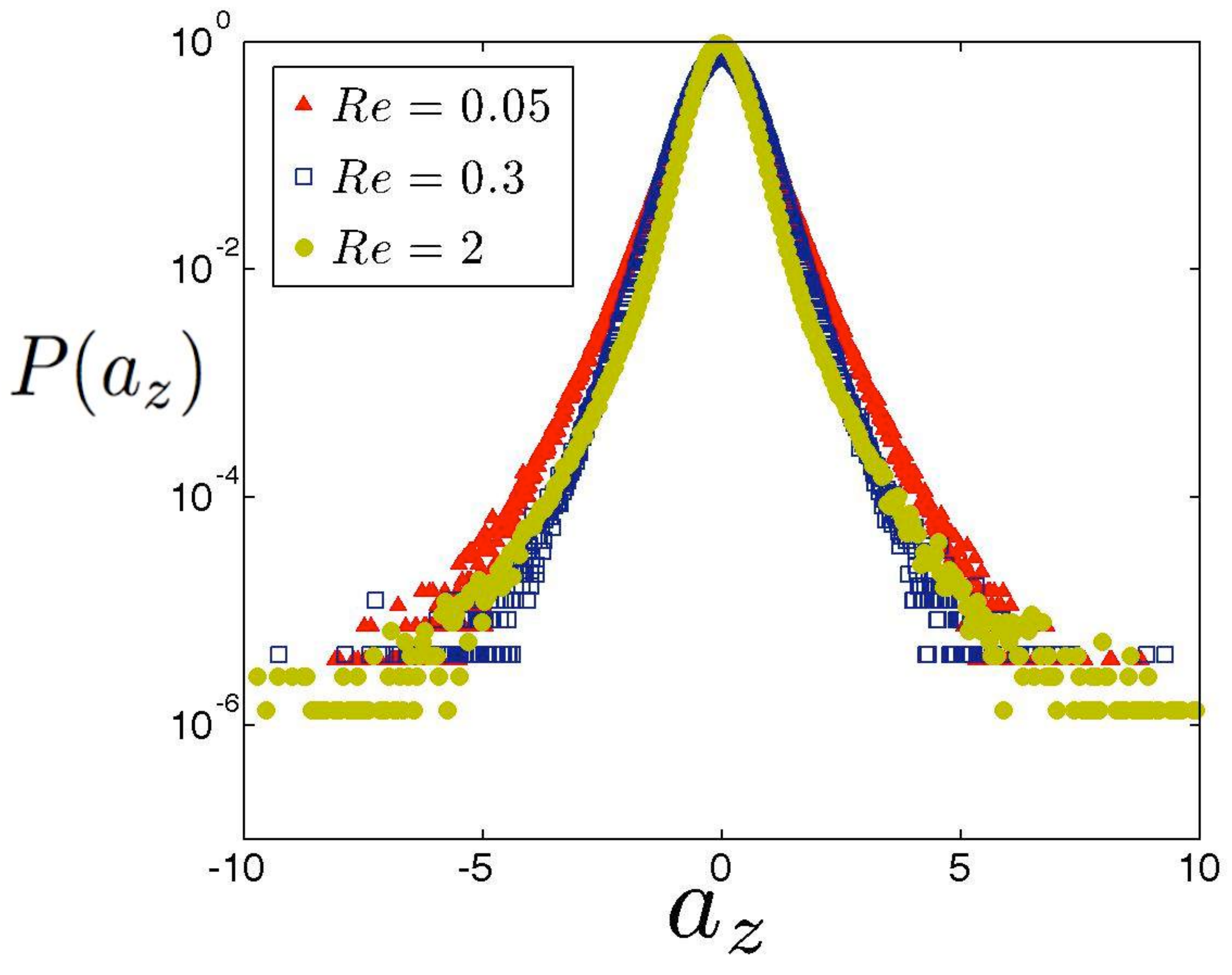}}
\subfigure[]{\includegraphics[totalheight=0.16\textheight,]{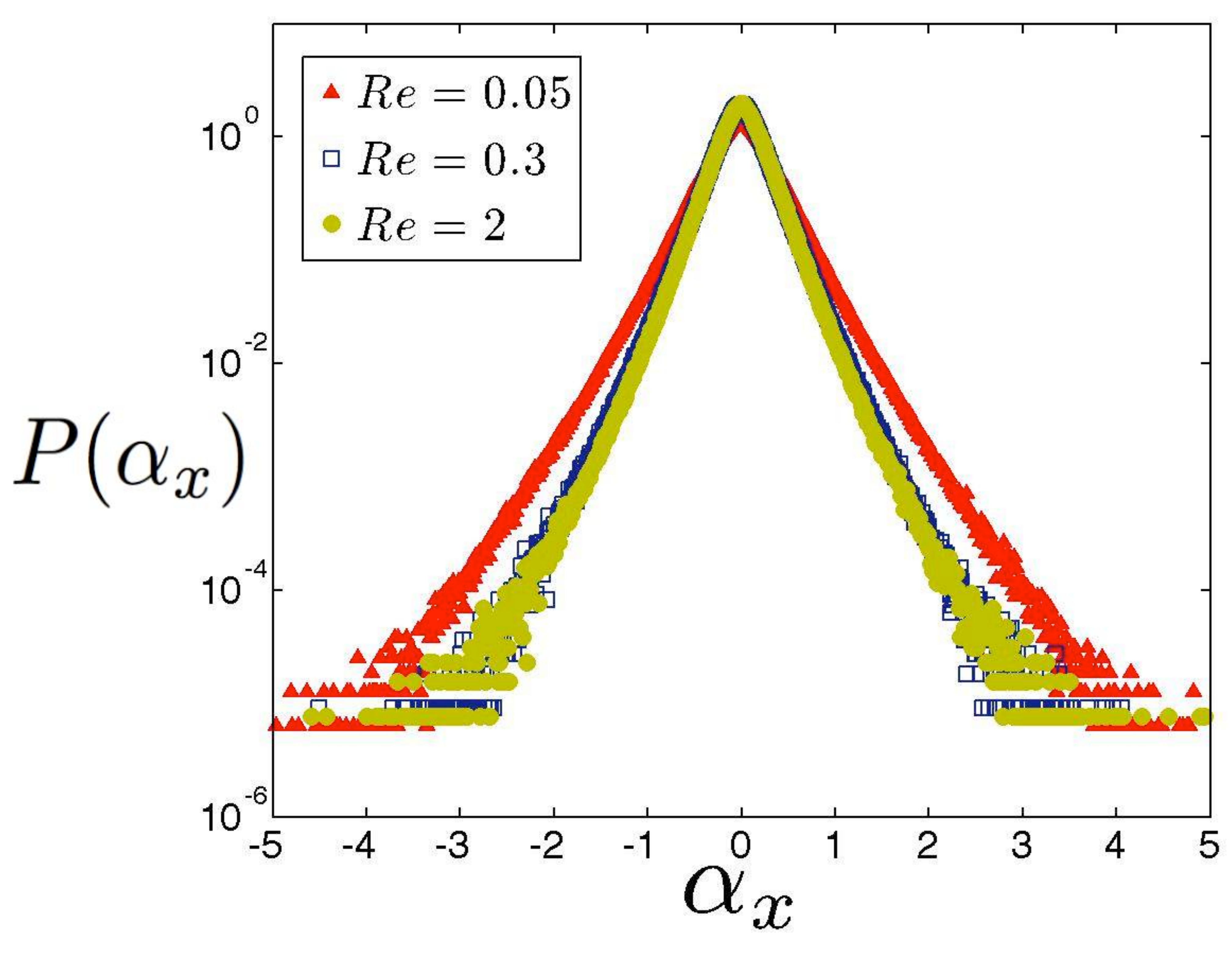}}
\subfigure[]{\includegraphics[totalheight=0.16\textheight,]{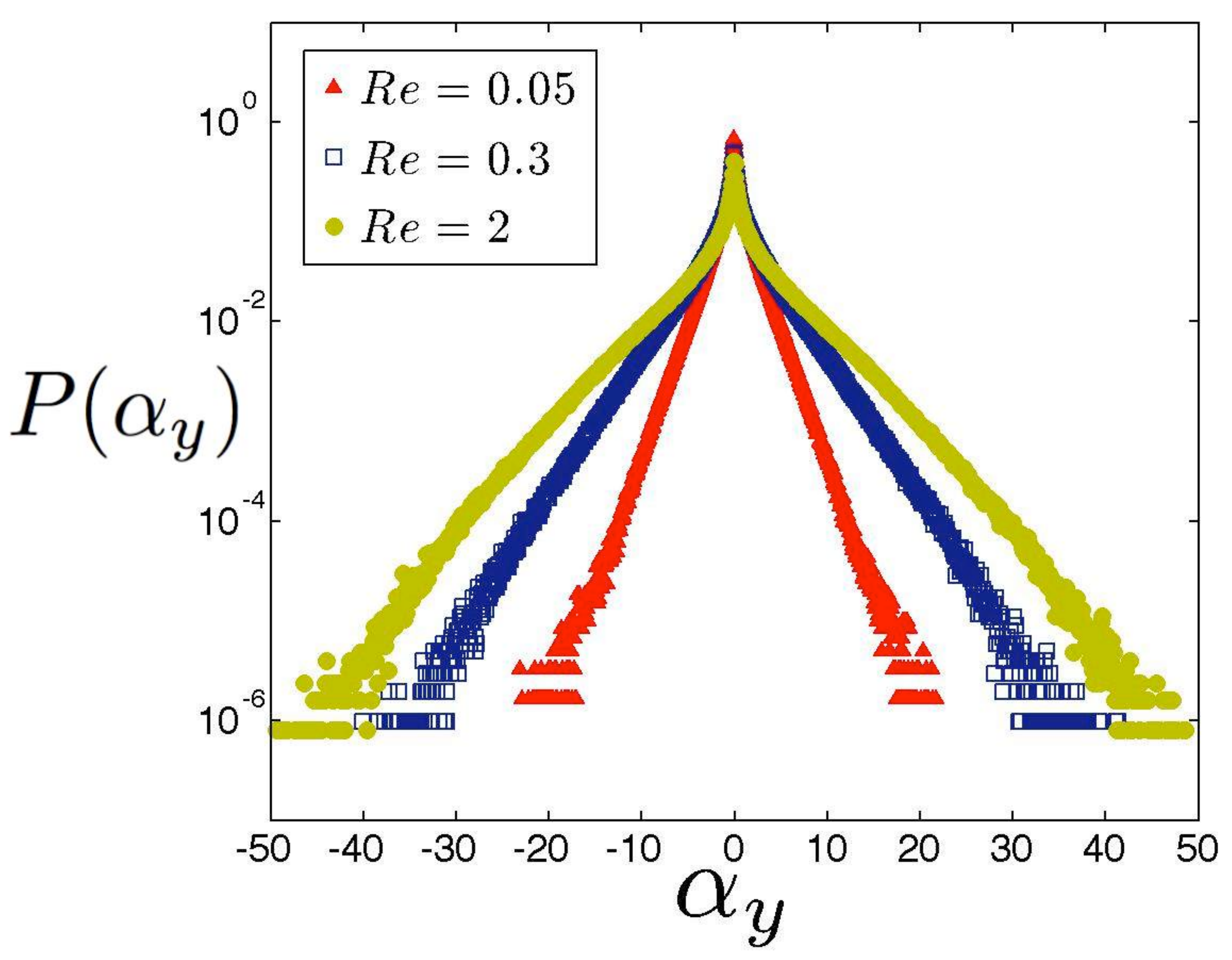}}
\subfigure[]{\includegraphics[totalheight=0.16\textheight,]{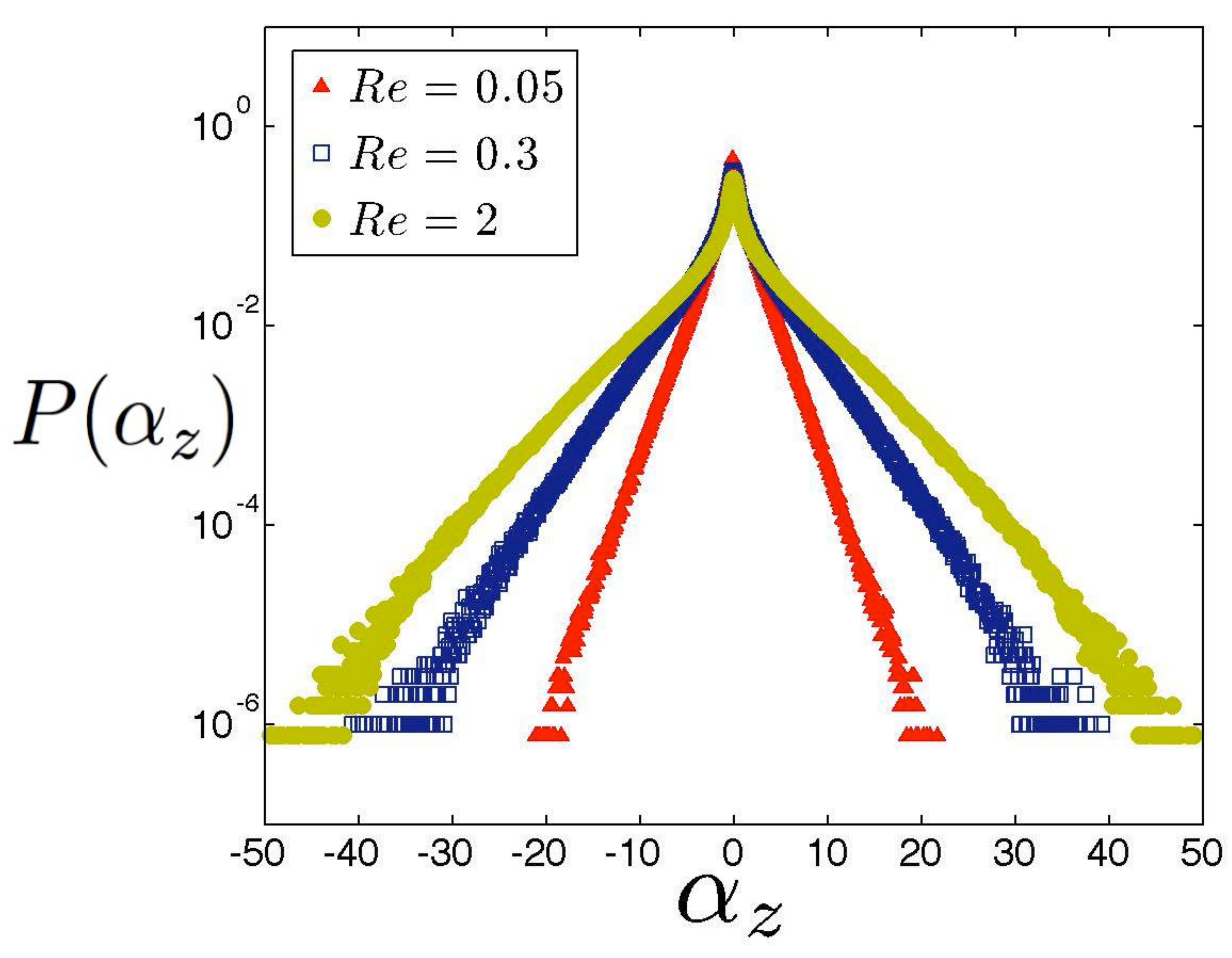}}
\caption{Probability density functions of (\emph{a}) $a_x$, (\emph{b}) $a_y$, (\emph{c}) $a_z$, (\emph{d}) $\alpha_x$, (\emph{e}) $\alpha_y$ and (\emph{f}) $\alpha_z$, for several $Re$, at $\phi = 0.15$. }

 \label{fig:PDFRE}
\end{figure}

We first consider the effect of inertia on the distribution of linear and angular accelerations for $\phi = 0.15$.  We observe in figure ~\ref{fig:PDFRE} (\emph{a}) that increasing $Re$ results in a broader distribution around the average of $\langle a_x\rangle = 0$.  For all values of $Re$, $a_x$ is exponentially distributed. Figure ~\ref{fig:FITS} (\emph{a}) shows the quality of exponential fit for the PDF of $a_x$ for $\phi = 0.15$ and $Re = 0.6$. We observe in figure ~\ref{fig:PDFRE} (\emph{a}) that the width of the PDF has two distinct broadenings. For all $Re$, the first increase of the width occurs at a normalized acceleration of $|a_x| \simeq 2$. For $|a_x| \gtrsim 2$ the influence of inertia becomes pronounced and the distributions separate. The curves broaden again at larger $a_x$.  The point of this second broadening depends on $Re$.  Figure ~\ref{fig:PDFRE} (\emph{b}) and (\emph{c}) display the distributions of $a_y$ and $a_z$, and we see that $Re$ does not influence these distributions significantly. In order to compare the exponential and Gaussian distributions, we show in figure  ~\ref{fig:FITS} (\emph{b}) the curve fittings for $a_z$ at $Re = 0.6$ and $\phi = 0.25$. The exponential distribution predicts an excessively sharp peak at $a_z = 0$, while the width of the PDF is broader than the exponential distribution in the range  $ 1\lesssim |a_z| \lesssim 5$. If we discard the low probability data for $|a_z| \geqslant 5$, the Gaussian distribution properly fits the remaining part of the PDF. This pattern is observed for both $a_y$ and $a_z$ at all $\phi$ and $Re$. \newline

Figure  ~\ref{fig:PDFRE} (\emph{d})-(\emph{f}) shows the PDF of angular acceleration for different $Re$ at $\phi = 0.15$. An exponential distribution is the better fit relative to a Gaussian for the PDF curves. The maximum in the PDF curves of $\alpha_y$ and $\alpha_z$ forms a cusp which causes deviation from exponential distribution.  We observe in figures ~\ref{fig:PDFRE} (\emph{e}) and (\emph{f}) that the width of the PDF of $\alpha_y$ and $\alpha_z$ increases noticeably with $Re$. The PDFs of $\alpha_y$ and $\alpha_z$ span a similar range of magnitudes and the curves are similar; $\alpha_x$ covers a significantly smaller range of values. \newline

\begin{figure}
\centering
\subfigure[]{\includegraphics[totalheight=0.16\textheight,]{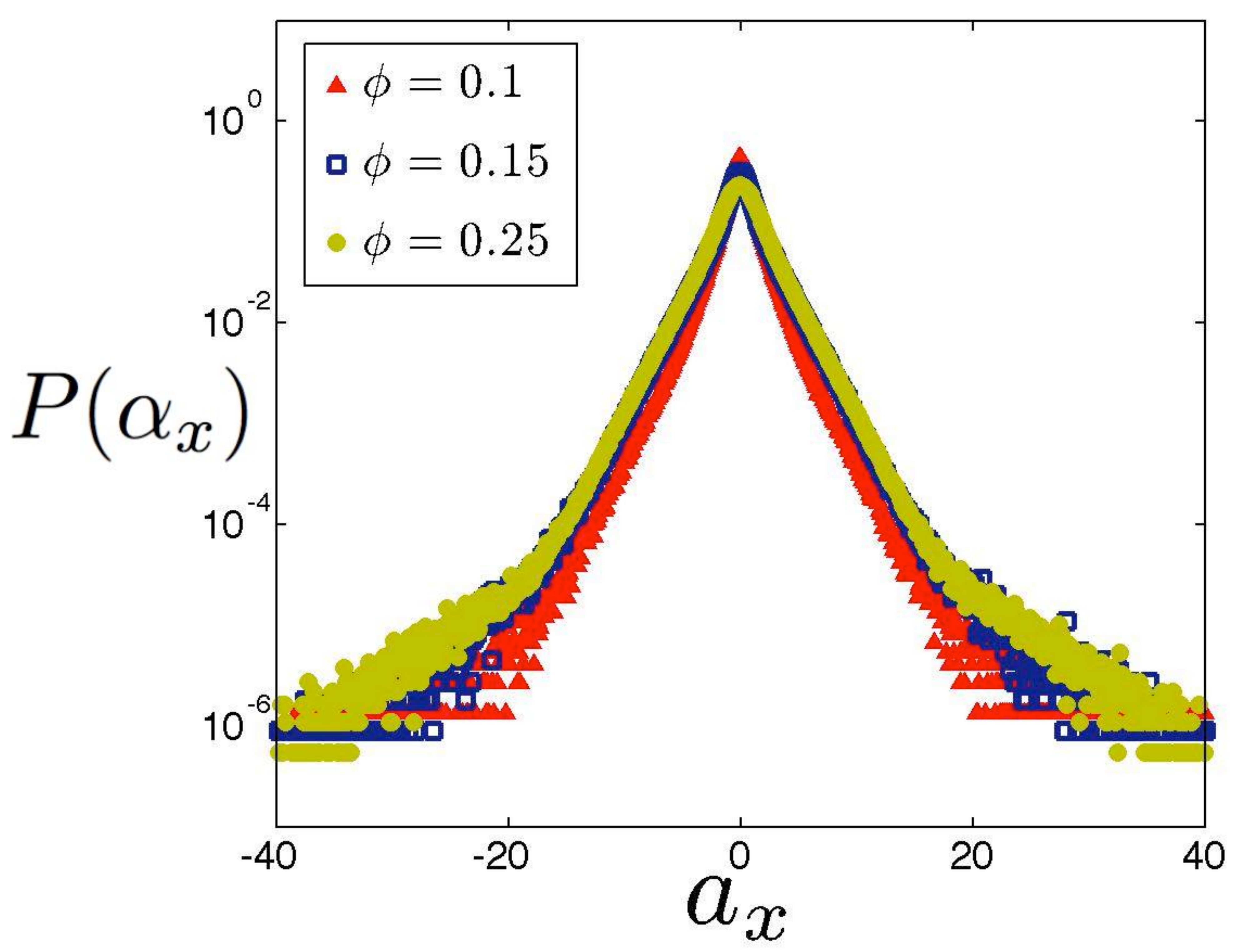}}
\subfigure[]{\includegraphics[totalheight=0.16\textheight,]{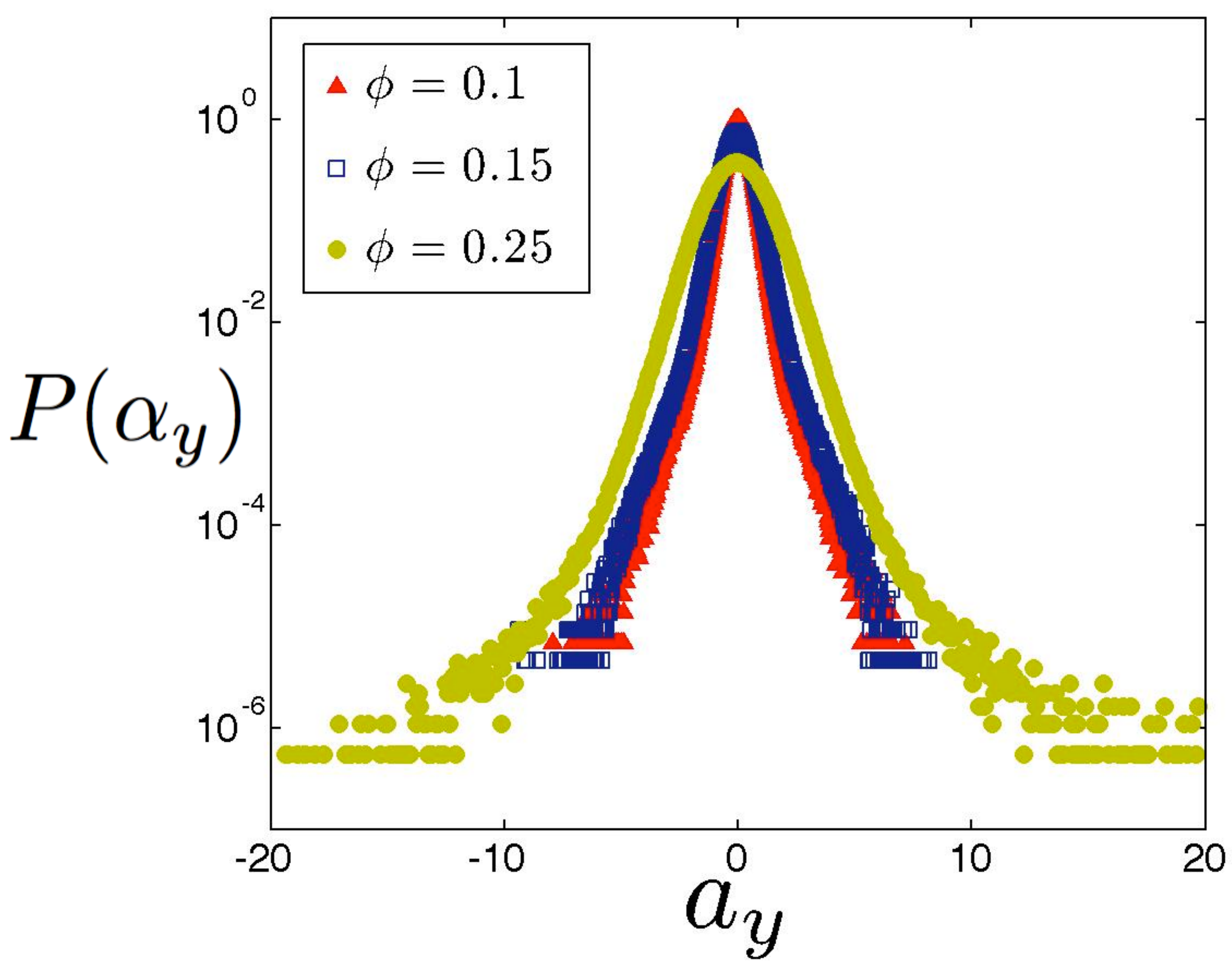}}
\subfigure[]{\includegraphics[totalheight=0.16\textheight,]{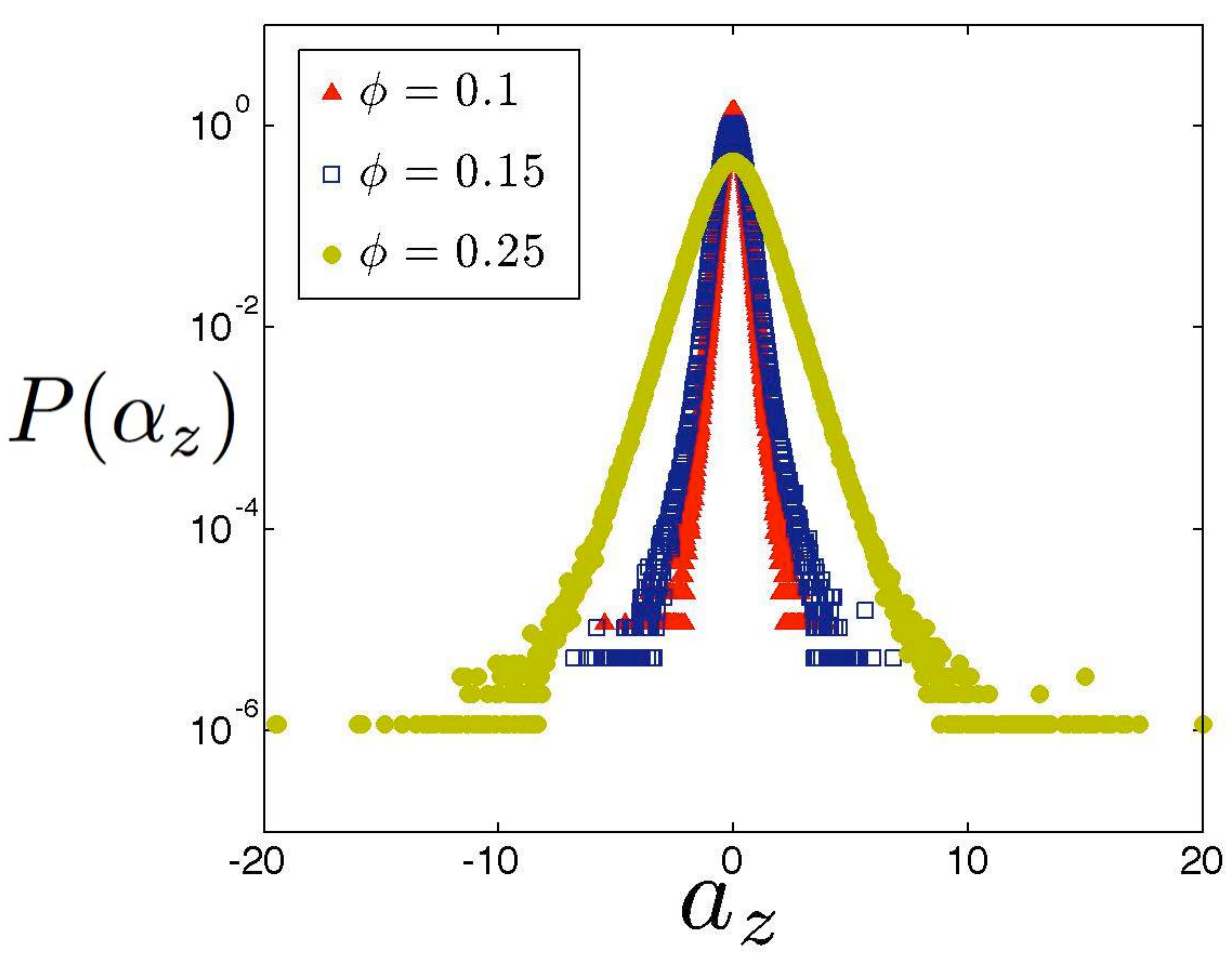}}
\subfigure[]{\includegraphics[totalheight=0.16\textheight,]{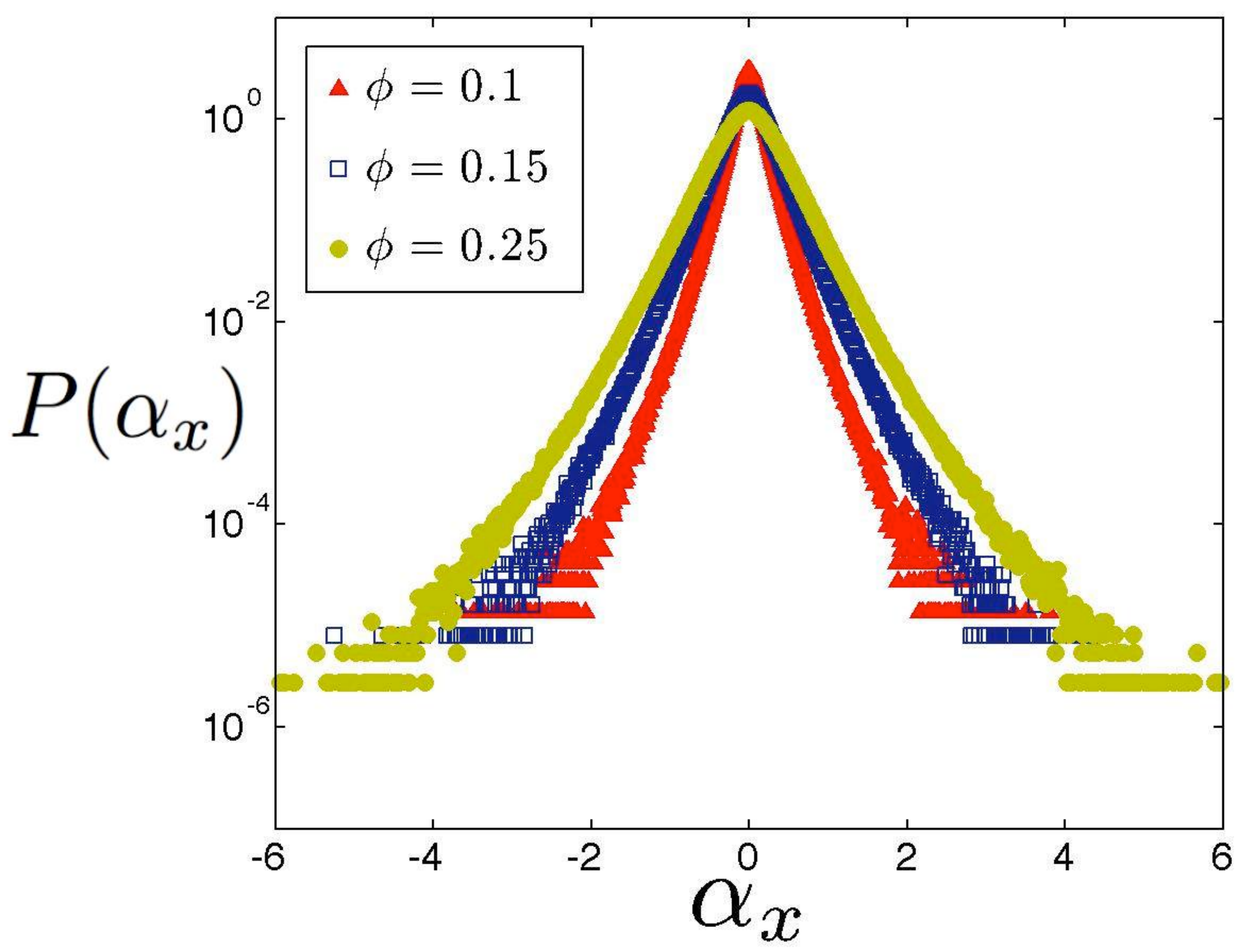}}
\subfigure[]{\includegraphics[totalheight=0.16\textheight,]{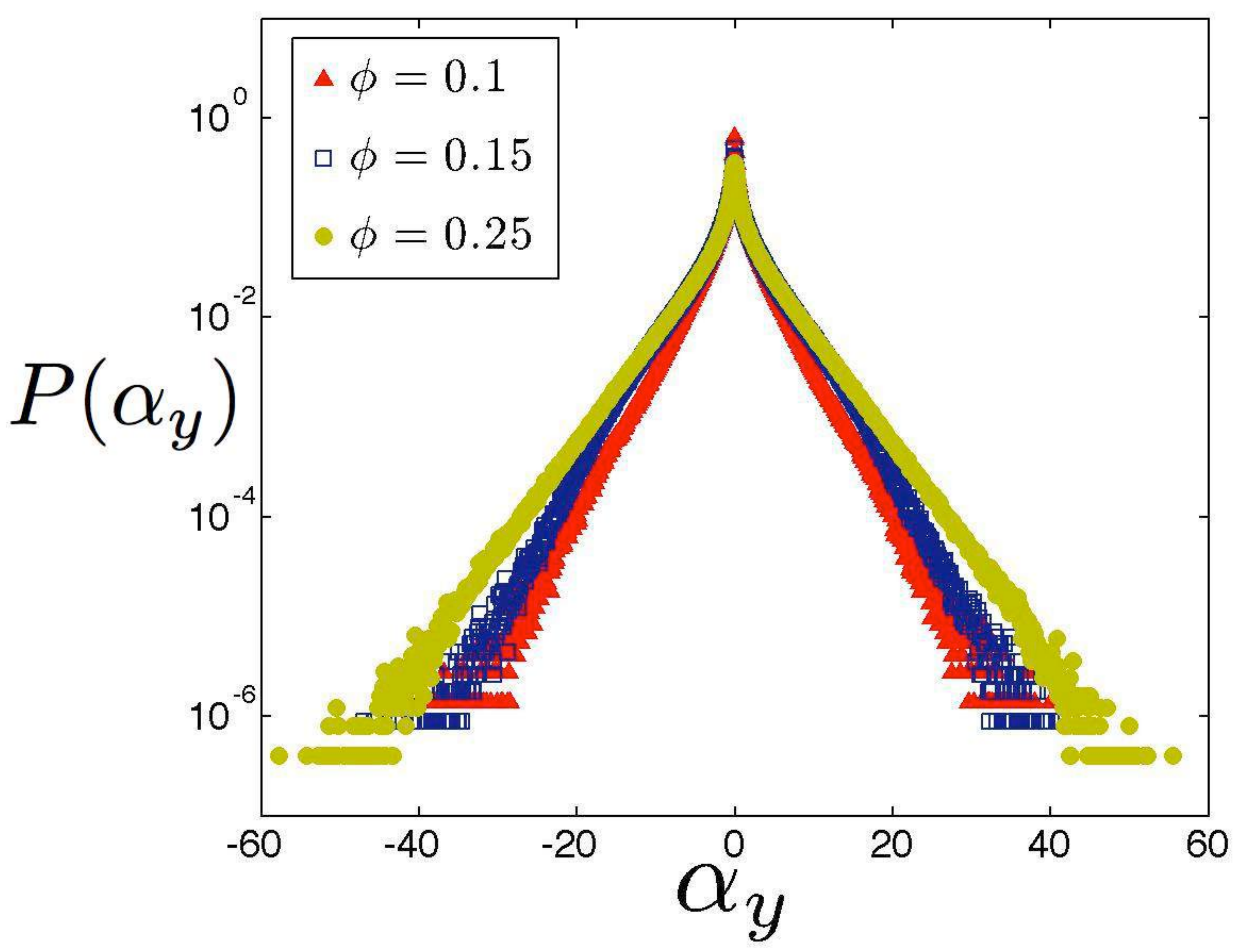}}
\subfigure[]{\includegraphics[totalheight=0.16\textheight,]{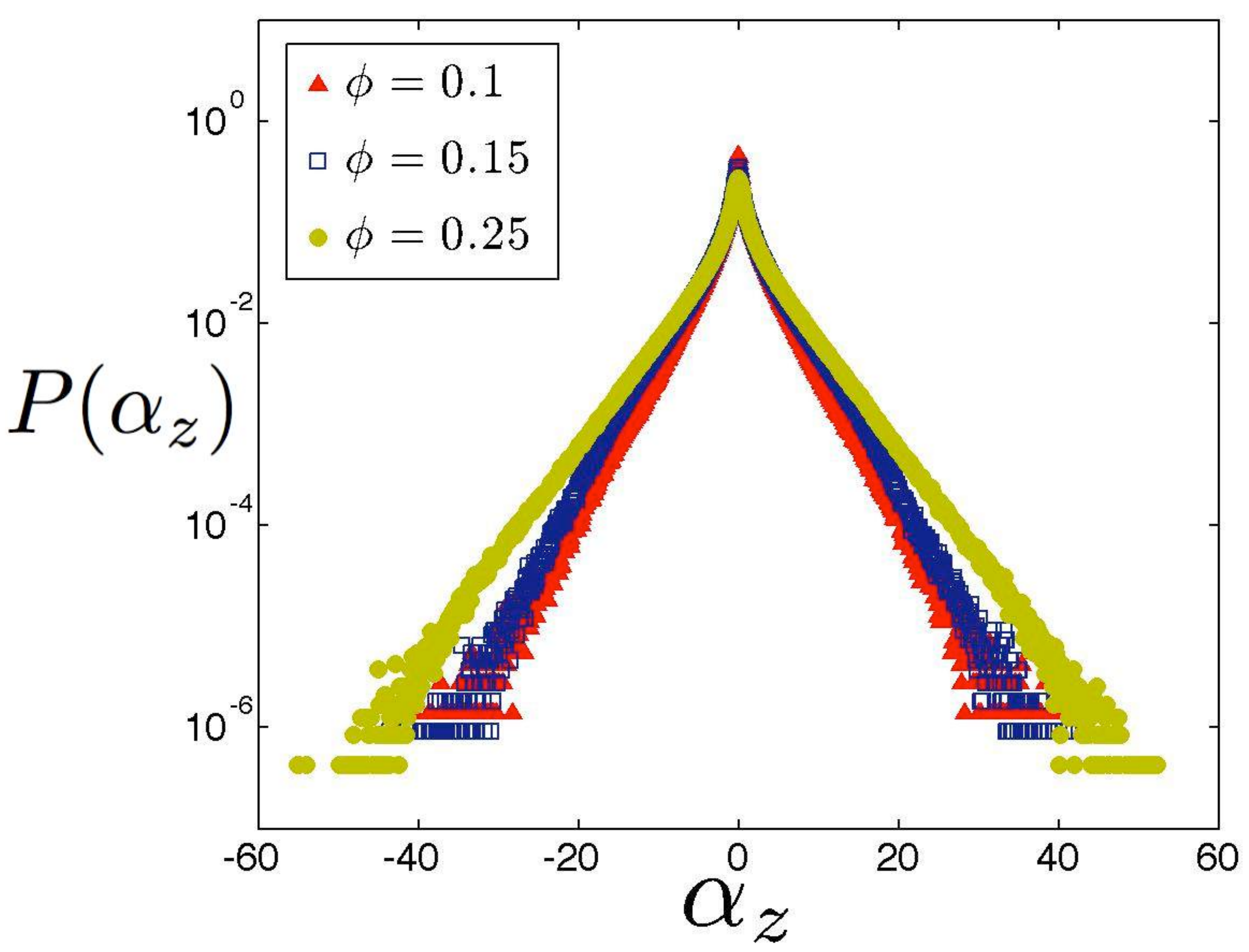}}
\caption{The effect of $\phi$ on the PDF of (\emph{a}) $a_x$,  (\emph{b}) $a_y$, (\emph{c}) $a_z$, (\emph{d}) $\alpha_x$, (\emph{e}) $\alpha_y$ and (\emph{f}) $\alpha_z$  at $Re = 0.6$.}
 \label{fig:S-Dist-PHI}
\end{figure}

In figure ~\ref{fig:S-Dist-PHI} we present the PDFs of linear and angular accelerations at varying volume fractions, for $Re = 0.6$.  For better visibility, we demonstrate the PDF at three volume fractions. Figure ~\ref{fig:S-Dist-PHI} depicts the influence of $\phi$ on the PDF of linear accelerations. As demonstrated in figure  ~\ref{fig:FITS} (\emph{a}),(\emph{b}), the best fit for $a_x$ is an exponential distribution and by discarding the spreading tails from the PDF of $a_y$ and $a_z$, their PDFs can be represented by the Gaussian distribution. The width of the PDFs of linear accelerations increases with increasing $\phi$,  with this effect less pronounced for $a_x$.  The PDFs of the angular accelerations are exponential and their width increases with increasing $\phi$.  It is observed in figures ~\ref{fig:S-Dist-PHI} (\emph{e}), (\emph{f}) that there is a pronounced similarity between the PDFs of $\alpha_y$ and $\alpha_z$. \newline

\begin{figure}
\centering
\subfigure[]{\includegraphics[totalheight=0.15\textheight,]{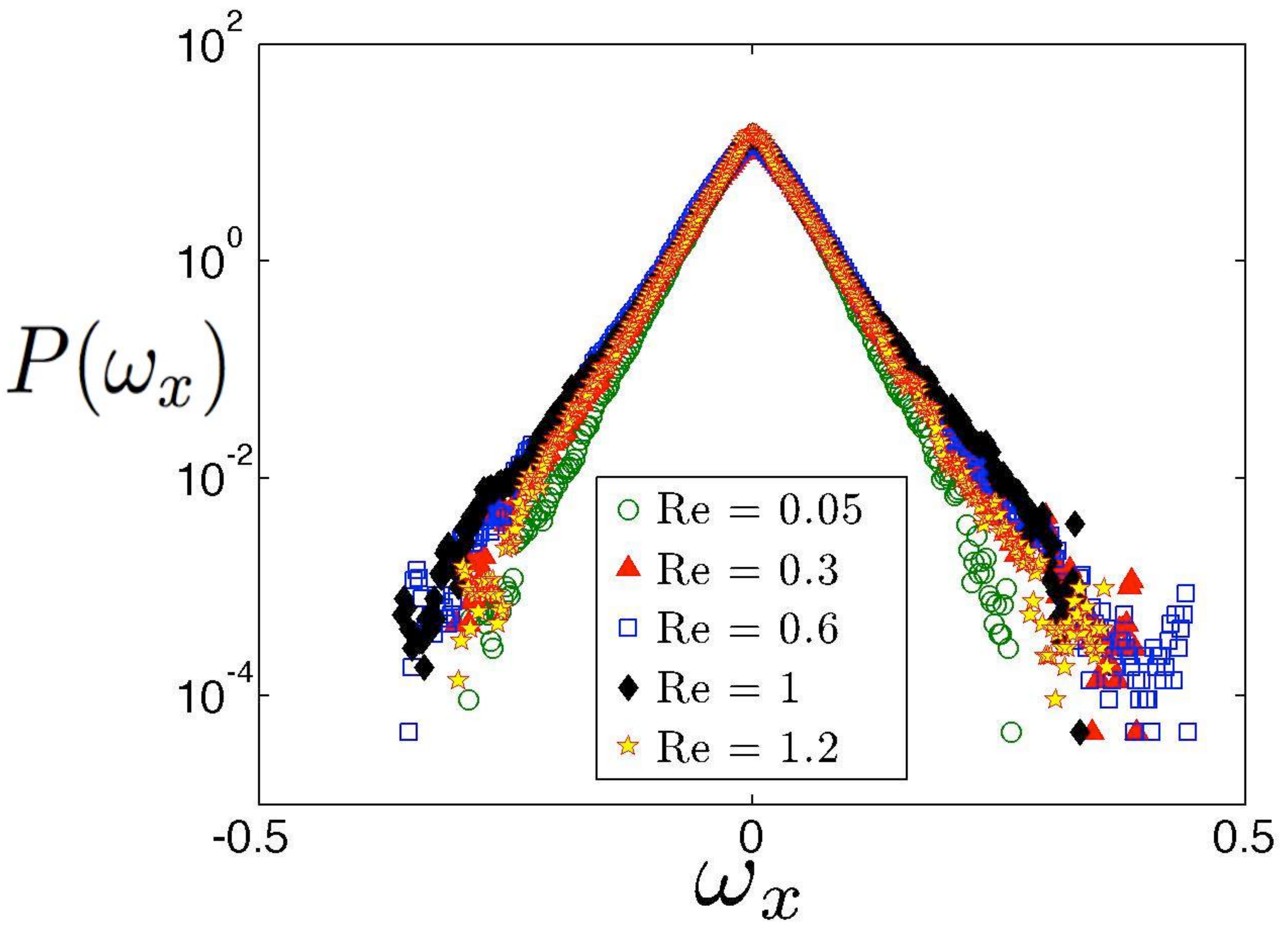}}
\subfigure[]{\includegraphics[totalheight=0.15\textheight,]{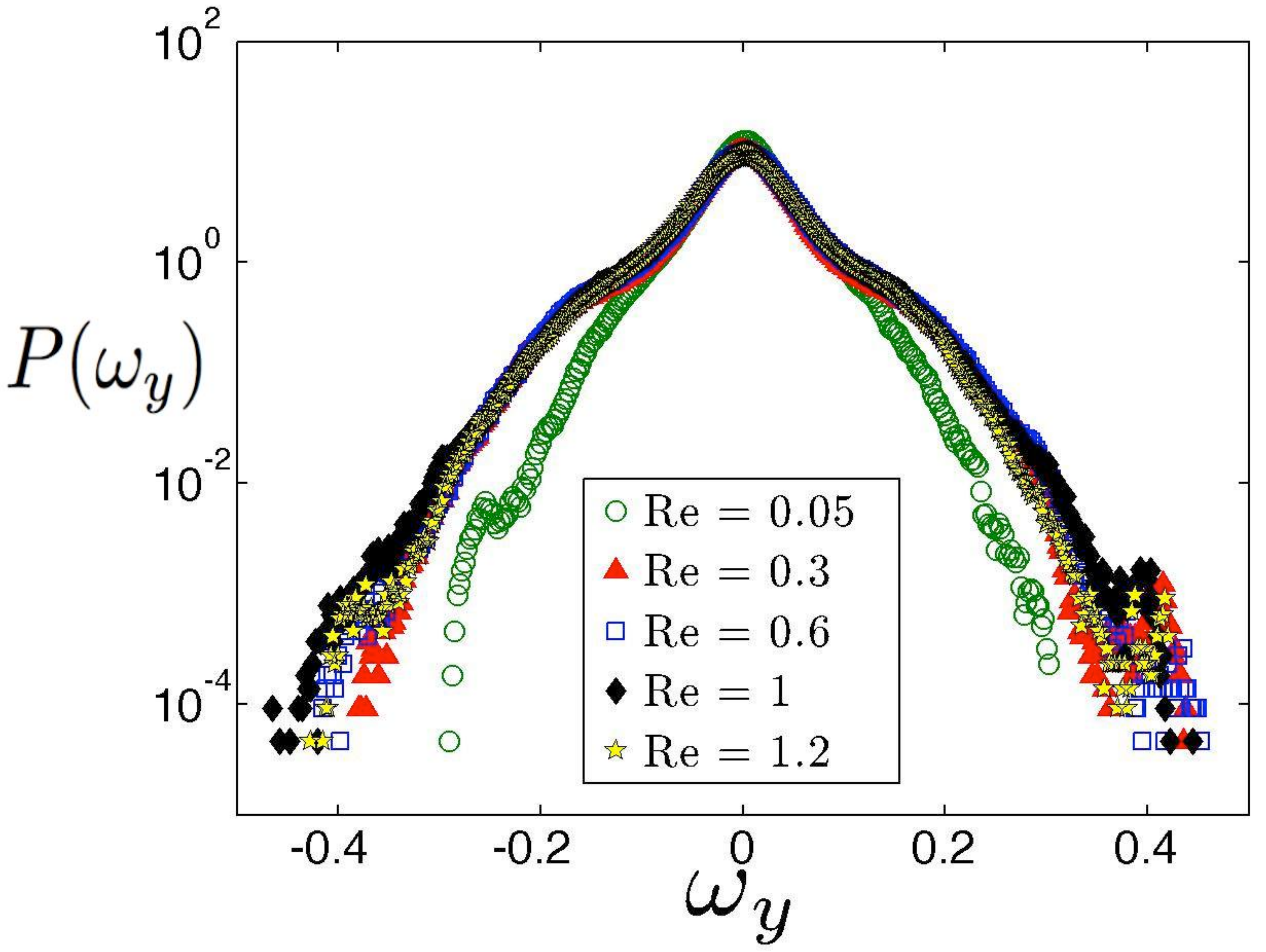}}
\subfigure[]{\includegraphics[totalheight=0.15\textheight,]{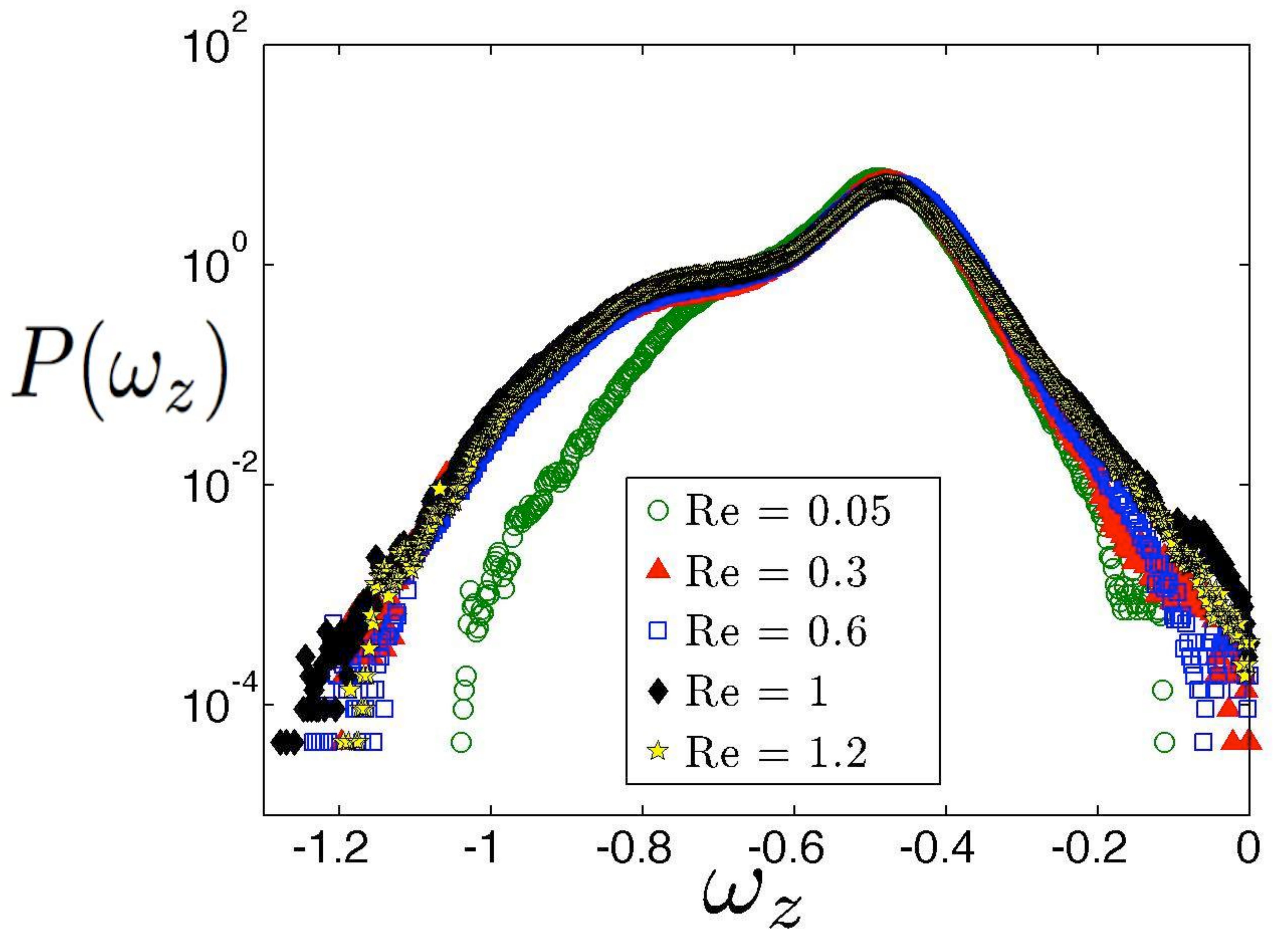}}
\subfigure[]{\includegraphics[totalheight=0.15\textheight,]{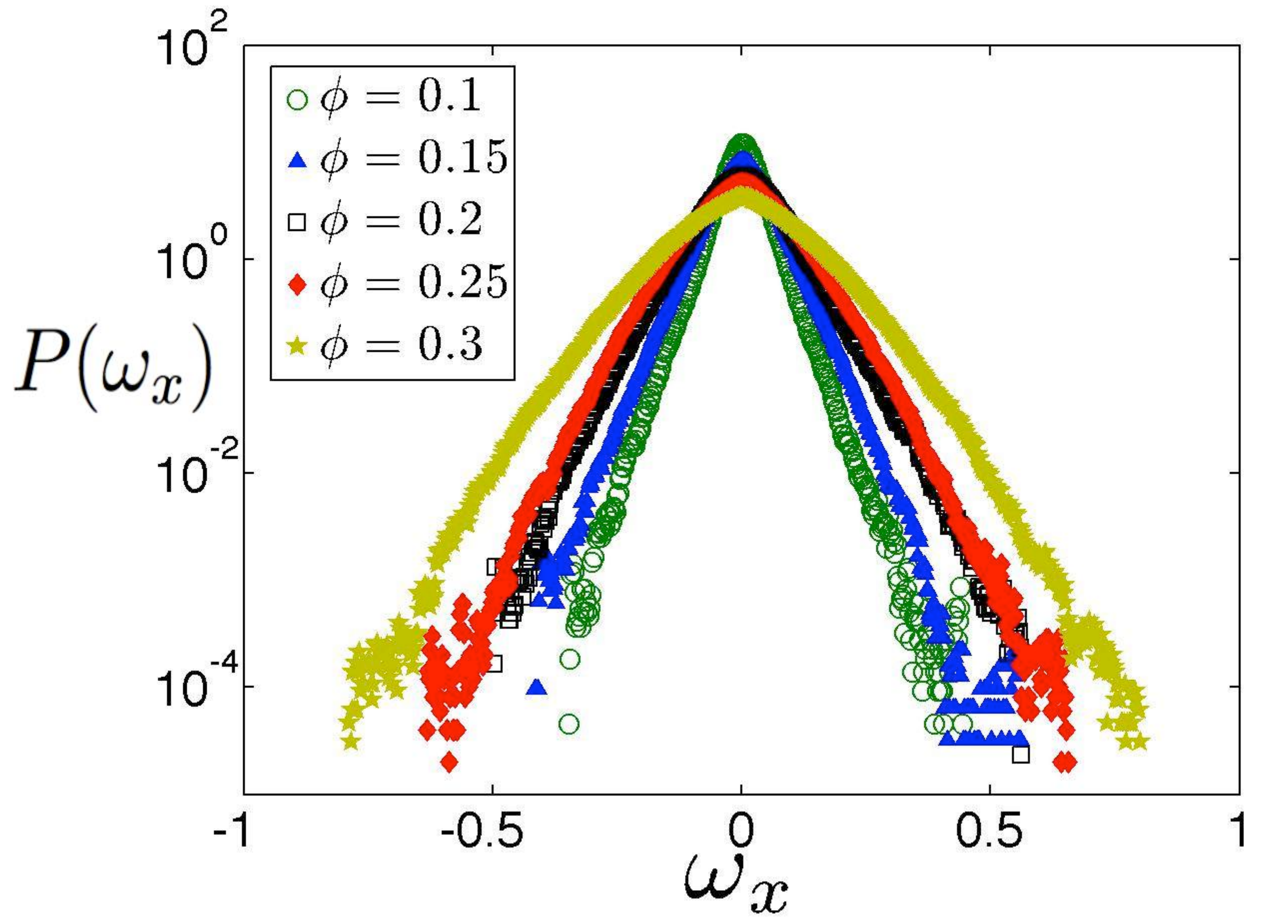}}
\subfigure[]{\includegraphics[totalheight=0.15\textheight,]{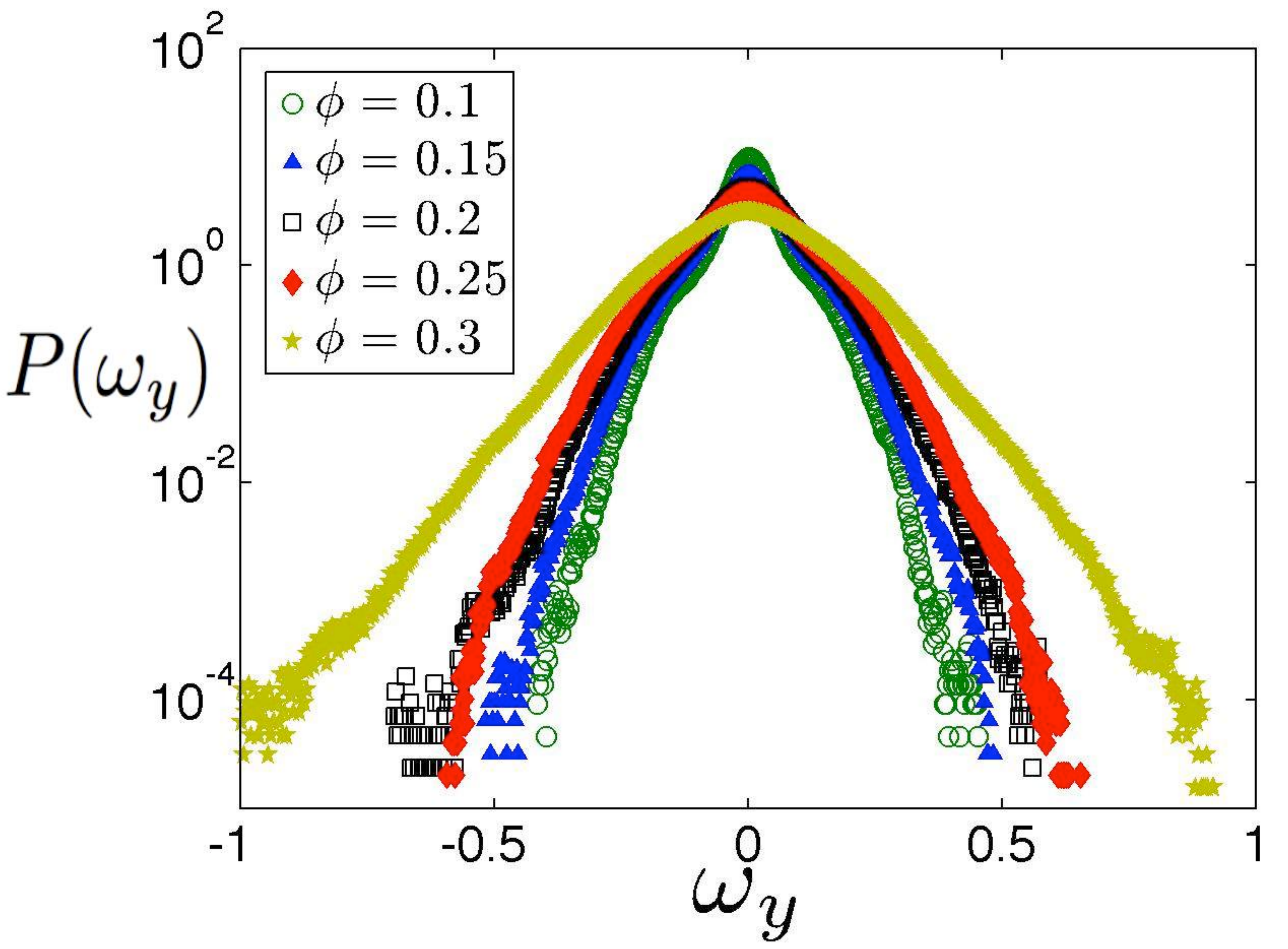}}
\subfigure[]{\includegraphics[totalheight=0.15\textheight,]{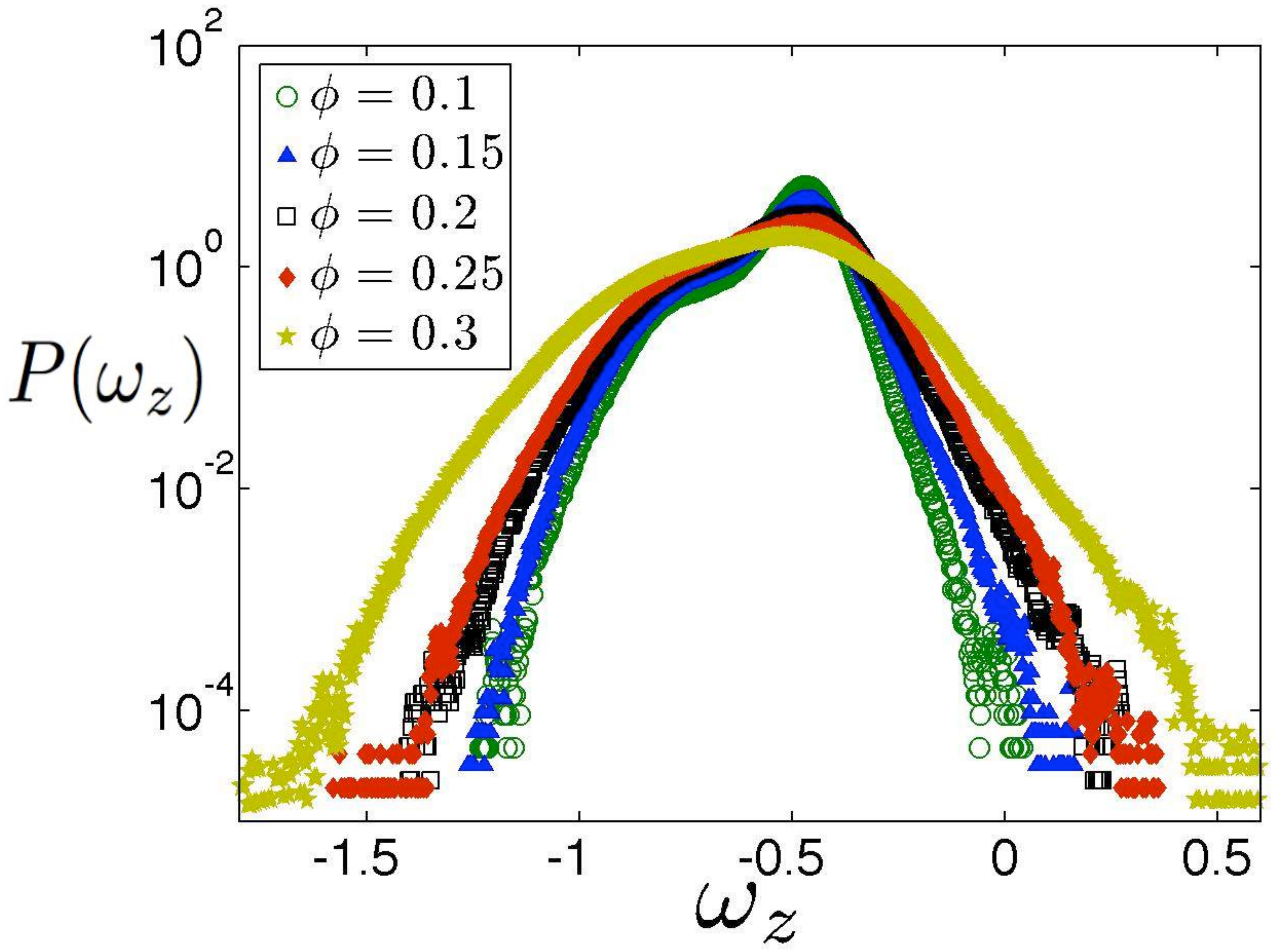}}
\caption{The probability density function of angular velocity and its variation with $Re$ and $\phi$. Varying $Re$ at $\phi = 0.1$ for (\emph{a}) $\omega_x$, (\emph{b}) $\omega_y$, and (\emph{c}) $\omega_z$.Varying $\phi$ at $Re=0.6$ for (\emph{d}) $\omega_x$, (\emph{e}) $\omega_y$, and (\emph{f}) $\omega_z$. Angular velocity is normalized by $\dot{\gamma}$.}
 \label{fig:Omega}
\end{figure}

 We provide the PDF of $\omega$ and its variation with $Re$ and $\phi$ in figure  ~\ref{fig:Omega}. Recall that an isolated sphere in Stokes flow has $\omega_z = -0.5\dot{\gamma}$ and this decreases with $Re$, but only slightly for the range of $Re \leqslant 2$ studied here. The volume fraction in figures  ~\ref{fig:Omega} (\emph{a})-(\emph{c}) is $0.1$ and the $Re$ in figures  ~\ref{fig:Omega} (\emph{d})-(\emph{f}) is $0.6$. Figure ~\ref{fig:Omega} (\emph{a}) provides the PDF of $\omega_x$ and its variation with $Re$. $\omega_x$ at all values of $\phi$ is exponentially distributed. The effect of $\phi$ on the PDF of $\omega_x$ is displayed in figure ~\ref{fig:Omega} (\emph{d}) where we observe the increase of the width of the distribution with increasing $\phi$.  Figures ~\ref{fig:Omega} (\emph{b}) and ~\ref{fig:Omega} (\emph{e}) depict the effect of $Re$ and $\phi$ on the PDF of $\alpha_y$ respectively.  We find $\omega_x$ at all values of $\phi$ to be well-described by an exponential distribution. The effect of $\phi$ on the PDF of $\omega_x$ is displayed in figure ~\ref{fig:Omega} (\emph{d}), showing an increase of the width of the distribution with increasing $\phi$.  Figure ~\ref{fig:Omega} (\emph{b}) and ~\ref{fig:Omega} (\emph{e}) show the effect of $Re$ and $\phi$ on the PDF of $\alpha_y$, respectively.  We can split the PDFs of $\omega_y$ into \emph{tip} and  \emph{shoulder} regions. The tip forms around the average ($\langle \omega_y\rangle = 0$) which is followed by a widening of the PDF at the shoulder zone. We exhibit the curve fits of the Gaussian and exponential distributions on the PDF of $\omega_y$ in figure ~\ref{fig:FITS} (\emph{c}).  It can be observed that the tip forms exponential distribution but the Gaussian distribution is a better fit for the shoulder.  We see in figure ~\ref{fig:Omega} (\emph{b}) that $Re$ does not influence the tip. By increasing $Re$ from $0.05$ to $0.3$, the shoulder zone broadens but further increase of $Re$ has negligible influence.  Figure  ~\ref{fig:Omega} (\emph{e}) shows that increasing $\phi$ flattens the tip of the distribution and results in a wider distribution.  Because particles near the walls have been excluded from the calculations, the large difference can not be due to layering of the particles near the walls.\newline

Figures  ~\ref{fig:Omega} $(c)$ and $(f)$ illustrate the distribution of values of $\omega_z$. The form of the PDF of $\omega_z$ is more complicated that that of $\omega_x$ or $\omega_y$; in particular it is asymmetric about the mean, and thus is not well-represented by either the Gaussian or exponential distributions. The PDF exhibits a peak around $\omega_z = -0.5 \dot{\gamma}$ (or -0.5 when made dimensionless by $\dot{\gamma}$ as shown in the figure), although the mean is found to deviate from this value. This peak is observed for different values of $Re$, and we find that $Re$ does not have much influence on the shape of the PDF near the peak; however, for $\omega_z \lesssim -0.5$, the width of the PDF at $Re = 0.05$ is clearly narrower than for higher $Re$. Figure  ~\ref{fig:Omega} $(f)$ shows that with increasing $\phi$ the peak of the PDF flattens, but $\omega_z$ remains asymmetrically distributed around the average.  The PDF of $\omega_z$ in  Stokes flow for $\phi \geqslant 0.05$ has previously been shown (Drazer \emph{et al.} 2004) from Stokesian Dynamics simulations to have an asymmetric distribution around the average, and to take on mean value $\langle\omega_z\rangle\ne -0.5$. The distribution of $\omega_z$ in Stokes flow has a more smooth shape and does not exhibit a localized peak, and this shape of the PDF is approached for finite $Re$ at $\phi = 0.3$. The average value of $\omega_z$ decreases (tends to become more negative) with increasing $\phi$ and $Re$, while the standard deviation is larger for higher $\phi$ and $Re$. We give the mean and the standard deviation of $\omega_z$ distributions for various $\phi$ and $Re$ in table ~\ref{tab:mstd}. \newline

\begin{table}
\begin{center}
\def~{\hphantom{0}}
\begin{tabular}{lccc}
Re & \hspace{30 pt} $\phi$ & \hspace{30 pt} Mean & \hspace{30 pt}  Standard deviation \\
0.05 & \hspace{30 pt}   0.1 & \hspace{30 pt} -0.5105 & \hspace{30 pt} 0.092\\
         & \hspace{30 pt}  0.15 & \hspace{30 pt} -0.5132 & \hspace{30 pt} 0.108\\  
         & \hspace{30 pt}  0.2 & \hspace{30 pt} -0.525 & \hspace{30 pt} 0.1413\\  
         & \hspace{30 pt}  0.25 & \hspace{30 pt} -0.5272 & \hspace{30 pt} 0.161\\  
         & \hspace{30 pt}  0.3  & \hspace{30 pt} -0.5554 & \hspace{30 pt} 0.2026\\ 
         \\
0.6    & \hspace{30 pt}  0.1 & \hspace{30 pt} -0.5159 & \hspace{30 pt} 0.1191\\ 
          & \hspace{30 pt}  0.15 & \hspace{30 pt} -0.5217 & \hspace{30 pt} 0.1371\\  
          & \hspace{30 pt}  0.2 & \hspace{30 pt} -0.5367 & \hspace{30 pt} 0.1609\\
          & \hspace{30 pt}  0.25 & \hspace{30 pt} -0.5447 & \hspace{30 pt} 0.1744\\
          & \hspace{30 pt}  0.3 & \hspace{30 pt} -0.5828 & \hspace{30 pt} 0.2184\\
          \\
 1       & \hspace{30 pt}  0.1 & \hspace{30 pt} -0.5275 & \hspace{30 pt} 0.1297\\
          & \hspace{30 pt}  0.15 & \hspace{30 pt} -0.5279 & \hspace{30 pt} 0.1342\\
          & \hspace{30 pt}  0.2 & \hspace{30 pt} -0.5415 & \hspace{30 pt} 0.157\\
          & \hspace{30 pt}  0.25 & \hspace{30 pt} -0.5611 & \hspace{30 pt} 0.1866\\
          & \hspace{30 pt}  0.3 & \hspace{30 pt} -0.5826 & \hspace{30 pt} 0.2122\\
\end{tabular}
\caption{The mean and standard deviation of $\omega_z$ for $Re = 0.05$, $0.6$ and $1$ at various volume fractions.}
\label{tab:mstd}
\end{center}
\end{table}

\section{Conclusion}

We have studied the simple-shear properties of suspensions of neutrally buoyant spherical particles at a range of solid fraction, $\phi$, and particle-scale Reynolds number, $Re$. Employing the lattice-Boltzmann method (LBM) for suspensions in a wall-bounded periodic domain, we have studied the microstructure and rheological properties of the suspensions for $0.005 \leqslant Re \leqslant 5$ and $0.1 \leqslant \phi \leqslant 0.35$. The distributions of hydrodynamic force and torque have also been investigated.

Inertia leads to an increase of the pair distribution function, $g(\boldsymbol{r})$, at contact. The influence of inertia diminishes at larger volume fractions as the excluded volume effects are dominant. By studying the three dimensional structure of the pair distribution function near contact,  we have demonstrated that increasing $Re$ leads to accumulation of $g(\boldsymbol{r})$ on the plane of shear. Because particles are driven together more strongly by shearing at larger $\phi$, the contact value of $g(\boldsymbol{r})$ is larger at higher $\phi$ and the distribution is more homogeneous at contact. \newline

To study the rheological properties, the first and second normal stress differences ($N_{1}$ and $N_{2}$), the particle pressure $\Pi$, and the viscosity ($\mu$) of suspensions at various $\phi$ and $Re$ have been computed. By employing the formulation of suspension stress developed by Batchelor (1970), we differentiate the contributions of stresslet, acceleration stress and Reynolds stress on the total particle contribution to the suspension stress.  The stresslet-generated $N_{1}$ is negative for all $Re$ and $\phi$ with higher magnitude at larger $Re$ and $\phi$.  $N_{2}$ is an increasing function of $Re$ up to a critical $\phi$ beyond which, $N_{2}$ starts to decrease. The critical volume fraction corresponding to the maximum $N_{2}$ depends on $Re$ and varies in the range $0.15\leqslant\phi\leqslant0.25$, with larger critical $\phi$ for higher $Re$. At $Re = 5$ inertia amplifies the effect of excluded volume and the pattern of $N_2$ versus $Re$ is reversed. Increase of either $Re$ or $\phi$ results in larger particle pressure. The viscosity measured by computing the shear component of the particle stress tensor $\Sigma_{12}^p$ and the shear stress on the walls did not show significant change with $Re$.  \newline

Through investigation of the individual components of the stresslet tensor, we have found that $S_{11}$ is the dominant component. Studying the average distribution of normal components of stresslet tensor on the particle surface reveals that at significant particle-scale inertia, a compressive stress develops along the flow direction. This effect is amplified at larger volume fractions.  We propose an interpretation of this result based on the average pair trajectories.  Both open and reversing trajectories push the particles toward each other in the compressional region of the shear flow and generate compressive stress on the particle surfaces. \newline

This work has shown how various mechanisms contribute to the particle stress up to $Re=O(1)$.  The stresslet and Reynolds stress contribute significantly, while the acceleration stress is found to be negligible. We have found that $N_{1}$ is dominated by the stresslet, but Reynolds stress influences $N_{2}$ especially for $Re \ge 1$, where its magnitude is comparable with the stresslet. \newline
                                                                                                                                                                                                                                                                                                                                                                                                                                                                                                         
One aspect of the difference between finite inertia and Stokes flow suspensions is that particles in an inertial suspension experience a local force and torque due to interaction with one another.   Particle interactions cause fluctuations which are associated at finite $Re$ with these forces and torques.  To provide greater insight to the statistical physics of sheared suspensions, we have presented the probability density functions of linear and angular accelerations of the particles, and their dependence on $\phi$ and $Re$.  While the average linear and angular acceleration is zero for all $Re$ and $\phi$, there is a distribution around the average.  We have measured the distribution of angular velocity ($\omega$) of the particles, finding as in Drazer \emph{et al.} (2004) that there is a deviation of the mean value of $\omega_z$ from the dilute Stokes-flow result of $\langle \omega \rangle = -0.5$ to more negative values.  This result is a many-body effect, as finite inertia will cause a single isolated particle to rotate with less negative values (Mikulencak \& Morris 2004).     \newline

\section*{Acknowlegdement}
This work was supported by NSF grant CBET 0847271. This research was also supported, in part, by a grant of computer time from the City University of New York High Performance Computing Center under NSF Grants CNS-0855217, CNS-0958379, and ACI-1126113..  We are grateful to Professor A. J. C. Ladd of the University of Florida for providing the initial lattice-Boltzmann code.

\bibliographystyle{jfm}

\end{document}